\overfullrule=0pt
\newcount\mgnf  %ingrandimento
\mgnf=0

\ifnum\mgnf=0
\def\openone{\leavevmode\hbox{\ninerm 1\kern-3.3pt\tenrm1}}%
\def\*{\vglue0.3truecm}\fi
\ifnum\mgnf=1
\def\openone{\leavevmode\hbox{\ninerm 1\kern-3.63pt\tenrm1}}%
\def\*{\vglue0.5truecm}\fi
\ifnum\mgnf=0
\magnification=\magstep0
\hsize=15truecm\vsize=23truecm
\parindent=4.pt\baselineskip=0.45cm
\font\titolo=cmbx12
\font\titolone=cmbx10 scaled\magstep 2
\font\cs=cmcsc10
%\font\css=cmcsc7
\font\ottorm=cmr8

\font\euftw=eufm10
\font\msytw=msbm10

\font\msytwww=msbm7
\font\indbf=cmbx10 scaled\magstep1

\fi
\ifnum\mgnf=2
   \magnification=\magstep0\hoffset=0.truecm
   \hsize=15truecm\vsize=24.truecm
   \baselineskip=18truept plus0.1pt minus0.1pt \parindent=0.9truecm
   \lineskip=0.5truecm\lineskiplimit=0.1pt      \parskip=0.1pt plus1pt
\font\titolo=cmbx12 scaled\magstep 1
\font\titolone=cmbx10 scaled\magstep 3
\font\cs=cmcsc10 scaled\magstep 1
\font\ottorm=cmr8 scaled\magstep 1

\font\euftw=eufm10 scaled\magstep1
\font\msytw=msbm10 scaled\magstep1

\font\msytwww=msbm7 scaled\magstep1

\font\indbf=cmbx10 scaled\magstep2
\fi

\global\newcount\numsec\global\newcount\numapp
\global\newcount\numfor\global\newcount\numfig\global\newcount\numsub
\numsec=0\numapp=0\numfig=1
\def\veroparagrafo{\number\numsec}\def\veraformula{\number\numfor}
\def\veraappendice{\number\numapp}\def\verasub{\number\numsub}
\def\verafigura{\number\numfig}

\def\section(#1,#2){\advance\numsec by 1\numfor=1\numsub=1%
\SIA p,#1,{\veroparagrafo} %
\write15{\string\Fp (#1){\secc(#1)}}%
\write16{ sec. #1 ==> \secc(#1)  }%
\hbox to \hsize{\titolo\hfill \number\numsec. #2\hfill%
\expandafter{\alato(sec. #1)}}\*}

\def\appendix(#1,#2){\advance\numapp by 1\numfor=1\numsub=1%
\SIA p,#1,{A\veraappendice} %
\write15{\string\Fp (#1){\secc(#1)}}%
\write16{ app. #1 ==> \secc(#1)  }%
\hbox to \hsize{\titolo\hfill Appendix A\number\numapp. #2\hfill%
\expandafter{\alato(app. #1)}}\*}

\def\senondefinito#1{\expandafter\ifx\csname#1\endcsname\relax}

\def\SIA #1,#2,#3 {\senondefinito{#1#2}%
\expandafter\xdef\csname #1#2\endcsname{#3}\else
\write16{???? ma #1#2 e' gia' stato definito !!!!} \fi}

\def \Fe(#1)#2{\SIA fe,#1,#2 }
\def \Fp(#1)#2{\SIA fp,#1,#2 }
\def \Fg(#1)#2{\SIA fg,#1,#2 }

\def\etichetta(#1){(\veroparagrafo.\veraformula)%
\SIA e,#1,(\veroparagrafo.\veraformula) %
\global\advance\numfor by 1%
\write15{\string\Fe (#1){\equ(#1)}}%
\write16{ EQ #1 ==> \equ(#1)  }}

\def\etichettaa(#1){(A\veraappendice.\veraformula)%
\SIA e,#1,(A\veraappendice.\veraformula) %
\global\advance\numfor by 1%
\write15{\string\Fe (#1){\equ(#1)}}%
\write16{ EQ #1 ==> \equ(#1) }}

\def\getichetta(#1){Fig. \verafigura%
\SIA g,#1,{\verafigura} %
\global\advance\numfig by 1%
\write15{\string\Fg (#1){\graf(#1)}}%
\write16{ Fig. #1 ==> \graf(#1) }}

\def\etichettap(#1){\veroparagrafo.\verasub%
\SIA p,#1,{\veroparagrafo.\verasub} %
\global\advance\numsub by 1%
\write15{\string\Fp (#1){\secc(#1)}}%
\write16{ par #1 ==> \secc(#1)  }}

\def\etichettapa(#1){A\veraappendice.\verasub%
\SIA p,#1,{A\veraappendice.\verasub} %
\global\advance\numsub by 1%
\write15{\string\Fp (#1){\secc(#1)}}%
\write16{ par #1 ==> \secc(#1)  }}

\def\Eq(#1){\eqno{\etichetta(#1)\alato(#1)}}
\def\eq(#1){\etichetta(#1)\alato(#1)}
\def\Eqa(#1){\eqno{\etichettaa(#1)\alato(#1)}}
\def\eqa(#1){\etichettaa(#1)\alato(#1)}
\def\eqg(#1){\getichetta(#1)\alato(fig. #1)}
\def\sub(#1){\0\palato(p. #1){\bf \etichettap(#1)\hskip.3truecm}}
\def\asub(#1){\0\palato(p. #1){\bf \etichettapa(#1)\hskip.3truecm}}

\def\equv(#1){\senondefinito{fe#1}$\clubsuit$#1%
\write16{eq. #1 non e' (ancora) definita}%
\else\csname fe#1\endcsname\fi}
\def\grafv(#1){\senondefinito{fg#1}$\clubsuit$#1%
\write16{fig. #1 non e' (ancora) definito}%
\else\csname fg#1\endcsname\fi}
\def\secv(#1){\senondefinito{fp#1}$\clubsuit$#1%
\write16{par. #1 non e' (ancora) definito}%
\else\csname fp#1\endcsname\fi}

\def\equ(#1){\senondefinito{e#1}\equv(#1)\else\csname e#1\endcsname\fi}
\def\graf(#1){\senondefinito{g#1}\grafv(#1)\else\csname g#1\endcsname\fi}
\def\secc(#1){\senondefinito{p#1}\secv(#1)\else\csname p#1\endcsname\fi}
\def\sec(#1){{\S\secc(#1)}}

\def\BOZZA{
\def\alato(##1){\rlap{\kern-\hsize\kern-1.2truecm{$\scriptstyle##1$}}}
\def\palato(##1){\rlap{\kern-1.2truecm{$\scriptstyle##1$}}}
}

\def\alato(#1){}
\def\galato(#1){}
\def\palato(#1){}

{\count255=\time\divide\count255 by 60 \xdef\hourmin{\number\count255}
        \multiply\count255 by-60\advance\count255 by\time
   \xdef\hourmin{\hourmin:\ifnum\count255<10 0\fi\the\count255}}

\def\oramin{\hourmin }

\def\data{\number\day/\ifcase\month\or gennaio \or febbraio \or marzo \or
aprile \or maggio \or giugno \or luglio \or agosto \or settembre
\or ottobre \or novembre \or dicembre \fi/\number\year;\ \oramin}
\setbox200\hbox{$\scriptscriptstyle \data $}
\footline={\rlap{\hbox{\copy200}}\tenrm\hss \number\pageno\hss}

\let\a=\alpha \let\b=\beta  \let\g=\gamma     \let\d=\delta  \let\e=\varepsilon
\let\z=\zeta  \let\h=\eta   \let\th=\vartheta \let\k=\kappa   \let\l=\lambda
\let\m=\mu    \let\n=\nu    \let\x=\xi        \let\p=\pi      \let\r=\rho
\let\s=\sigma \let\t=\tau        \let\c=\chi
   \let\o=\omega 
 \let\D=\Delta     \let\L=\Lambda  
           
\let\O=\Omega 

\def\\{\hfill\break} \let\==\equiv

\let\io=\infty 

\let\0=\noindent

\def\ie{\hbox{\it i.e.\ }}
\let\dpr=\partial 
\let\bs=\backslash
\def\defin{{\buildrel def\over=}}
\def\tende#1{\,\vtop{\ialign{##\crcr\rightarrowfill\crcr
 \noalign{\kern-1pt\nointerlineskip}
 \hskip3.pt${\scriptstyle #1}$\hskip3.pt\crcr}}\,}
\def\otto{\,{\kern-1.truept\leftarrow\kern-5.truept\to\kern-1.truept}\,}
\def\fra#1#2{{#1\over#2}}
\def\Pf{{\rm Pf}\,}

\def\PP{{\cal P}}\def\EE{{\cal E}}\def\MM{{\cal M}}\def\VV{{\cal V}}
\def\HH{{\cal H}}
\def\TT{{\cal T}}\def\NN{{\cal N}}\def\BB{{\cal B}}
\def\RR{{\cal R}}\def\LL{{\cal L}}
\def\DD{{\cal D}}\def\SS{{\cal S}}

\def\der{{\rm d}}
\def\T#1{{#1_{\kern-3pt\lower7pt\hbox{$\widetilde{}$}}\kern3pt}}
\def\VVV#1{{\underline #1}_{\kern-3pt
\lower7pt\hbox{$\widetilde{}$}}\kern3pt\,}
\def\W#1{#1_{\kern-3pt\lower7.5pt\hbox{$\widetilde{}$}}\kern2pt\,}

\def\lis{\overline}
 
  \def\sign{{\rm sign}\,}
\def\indica{\leaders \hbox to 0.5cm{\hss.\hss}\hfill}
\def\guida{\leaders\hbox to 1em{\hss.\hss}\hfill}
\mathchardef\oo= "0521

\def\pp{{\bf p}}\def\qq{{\bf q}}\def\xx{{\bf x}}
\def\yy{{\bf y}}\def\kk{{\bf k}}\def\nn{{\bf n}}
\def\dd{{\bf d}}\def\uu{{\bf u}}
 \def\bP{{\bf P}}
\def\tt{{\bf t}}
\def\ss{{\underline \sigma}}\def\oo{{\underline \omega}}
\def\aa{{\underline \alpha}}
\def\un{{\underline \nu}}\def\ul{{\underline \lambda}}

\def\qed{\raise1pt\hbox{\vrule height5pt width5pt depth0pt}}

\def\indic{\hbox{\raise-2pt \hbox{\indbf 1}}}
\def\bk#1#2{\bar\kk_{#1#2}}

\def\virg{\quad,\quad}
\def\bT{{\bf T}}

\def\MMM{\hbox{\euftw M}}
\def\RRR{\hbox{\msytw R}}

 \def\ZZZ{\hbox{\msytw Z}}
 \def\zzz{\hbox{\msytwww Z}}

%%% INSERIMENTO FIGURE ( se si usa DVIPS )
%
% Se si vuole utilizzare delle macro postscript personali, contenute
% nel file ini.ps, togliere il commento alla riga seguente
%\special{header=ini.pst}
%
% Il comando seguente inserisce una scatola contenente #3 in modo che
% l'angolo superiore sinistro occupi la posizione (#1,#2)
%
\def\ins#1#2#3{\vbox to0pt{\kern-#2 \hbox{\kern#1 #3}\vss}\nointerlineskip}
%
% Il comando seguente crea una scatola di dimensioni #1x#2 contenente
% il disegno descritto in #4.ps;
% in questo disegno si possono introdurre delle stringhe usando \ins
% e mettendo le istruzioni relative nell'argomento #3.
% Il file #4.ps contiene le istruzioni postscript, che devono essere scritte
% presupponendo che l'origine sia nell'angolo inferiore sinistro della
% scatola, mentre per il resto l'ambiente grafico e' quello standard.
% #5 deve essere della forma \eqg("nome simbolico").
%
% Le istruzioni postscript possono essere inserite nel file che contiene
% l'istruzione \insertplot, racchiudendole fra le istruzioni \initfig{#4}
% e \endfig; inoltre ogni riga deve cominciare con "write13<" e deve finire
% con ">". In questo modo si crea il file #4.ps relativo alla figura.
%
\newdimen\xshift \newdimen\xwidth \newdimen\yshift

\def\insertplot#1#2#3#4#5{\par%
\xwidth=#1 \xshift=\hsize \advance\xshift by-\xwidth \divide\xshift by 2%
\yshift=#2 \divide\yshift by 2%
\line{\hskip\xshift \vbox to #2{\vfil%
#3 \includegraphics{#4.ps}}\hfill \raise\yshift\hbox{#5}}}

\def\initfig#1{%
\catcode`\%=12\catcode`\{=12\catcode`\}=12
\catcode`\<=1\catcode`\>=2
\openout13=#1.ps}

\def\endfig{%
\closeout13
\catcode`\%=14\catcode`\{=1
\catcode`\}=2\catcode`\<=12\catcode`\>=12}

\def\insertplotttt#1#2#3#4{\par%
\xwidth=#1 \xshift=\hsize \advance\xshift by-\xwidth \divide\xshift by 2%
\yshift=#2 \divide\yshift by 2%
\line{\hskip\xshift \vbox to #2{\vfil%
\ifnum\driver=0 #3
%
\special{ps: plotfile #4.ps}
\fi
\ifnum\driver=1 #3 \includegraphics{#4.ps}\fi
\ifnum\driver=2 #3
\ifnum\mgnf=0\special{#4.ps 1. 1. scale} \fi
\ifnum\mgnf=1\special{#4.ps 1.2 1.2 scale}\fi
\fi }\hfill}}

%%%%%   FIGURE fig51.ps

\initfig{fig51}
\write13<%!>
\write13<%%BoundingBox 0 0 300 150>
\write13<gsave .5 setlinewidth 40 20 260 {dup 0 moveto 140 lineto} for stroke
grestore>
\write13</punto { gsave  % uso: x1 y1 punto>
\write13<2 0 360 newpath arc fill stroke grestore} def>
\write13<40 75 punto>
\write13<60 75 punto>
\write13<80 75 punto>
\write13<100 75 punto 120 68 punto 140 61 punto 160 54 punto 180 47 punto 200
40 punto>
\write13<220 33 punto 240 26 punto 260 19 punto>
\write13<120 82.5 punto>
\write13<140 90 punto>
\write13<160 80 punto>
\write13<160 100 punto>
\write13<180 110 punto>
\write13<180 70 punto>
\write13<200 60 punto>
\write13<200 120 punto>
\write13<220 110 punto>
\write13<220 50 punto>
\write13<240 100 punto>
\write13<240 60 punto>
\write13<120 50 punto>
\write13<260 20 punto>
\write13<240 40 punto>
\write13<240 50 punto>
\write13<260 70 punto>
\write13<200 80 punto>
\write13<260 90 punto>
\write13<260 110 punto>
\write13<220 130 punto>
\write13<40 75 moveto 100 75 lineto 140 90 lineto 200 120 lineto 220 130
lineto>
\write13<200 120 moveto 240 100 lineto 260 110 lineto>
\write13<240 100 moveto 260 90 lineto>
\write13<140 90 moveto 180 70 lineto 200 80 lineto>
\write13<180 70 moveto 220 50 lineto 260 70 lineto>
\write13<220 50 moveto 240 40 lineto>
\write13<220 50 moveto 240 50 lineto>
\write13<100 75 moveto 260 20 lineto>
\write13<100 75 moveto 120 50 lineto stroke>
\write13<grestore>
\endfig

\openin14=\jobname.aux \ifeof14 \relax \else
\input \jobname.aux \closein14 \fi
\openout15=\jobname.aux
\footline={\tenrm\hss \number\pageno\hss}
%\BOZZA
{\baselineskip=12pt
\centerline{\titolone Anomalous universality in the 
Anisotropic Ashkin--Teller model}
\vskip1.truecm
\centerline{\titolo{A. Giuliani%
\footnote{${}^{\ast}$}{\ottorm Partially supported by 
NSF Grant DMR 01--279--26;
Dipartimento di Fisica, Universit\`a 
di Roma ``La Sapienza'', P.zzale A. Moro, 2, I-00185, Roma; 
and INFN, sezione di Roma1; e--mail: 
alessandro.giuliani@roma1.infn.it}, V. Mastropietro%
\footnote{${}^{\ast\ast}$}{\ottorm Dipartimento di Matematica, Universit\`a 
di Roma ``Tor
Vergata'', Via della Ricerca Scientifica, I-00133, Roma; e--mail: 
mastropi@mat.uniroma2.it}}}
\vskip.5truecm
\0{\cs Abstract.}
{\it The Ashkin--Teller (AT) model
is a generalization of Ising 2--d 
to a four states spin model; it can be written in the form 
of two Ising layers (in general 
with different couplings)
interacting via a four--spin interaction.
It was conjectured long ago (by Kadanoff and Wegner, 
Wu and Lin, Baxter and others) that
AT has in general two critical points, and that universality holds,
in the sense that the critical exponents are the same as in the Ising model,
except when the couplings of the two Ising layers are equal (isotropic case).
We obtain an explicit expression for the specific 
heat from which we prove this conjecture in the 
weakly interacting case 
and we locate precisely the critical points.
We find the somewhat unexpected feature that,
despite universality holds for the specific heat, 
nevertheless nonuniversal critical indexes
appear: for instance the distance between 
the critical points rescales with an anomalous exponent 
as we let the couplings of the two Ising layers
coincide (isotropic limit); and so does the 
constant in front of the logarithm in the specific heat.
Our result also explains
how the crossover from universal to nonuniversal 
behaviour is realized.}
}
\*
\section(1,Introduction)
\*
\sub(1.1) {\it Historical introduction.}
Ashkin and Teller [AT] introduced 
their model as a generalization
of the Ising model to a four 
component system; in each site of a bidimensional
lattice there is a spin which can take four values,
and only nearest neighbor spins interact.
The model can be also considered a generalization
of the four state Potts model to which
it reduces for a suitable choice of the parameters.

A very convenient representation
of the Ashkin Teller model is 
in terms of Ising spins [F];
one associates with each site of the square
lattice two spins variables, $\s^{(1)}_\xx$ and 
$\s^{(2)}_\xx$;
the partition function is given by 
$Z_{\L_M}=\sum_{\s^{(1)},\s^{(2)}} e^{-H_{\L_M}}$, where 
$$\eqalign{&H_{\L_M}(\s^{(1)},\s^{(2)})=J^{(1)}H_I(\s^{(1)})+
J^{(2)}H_I(\s^{(2)})+\l V(\s^{(1)},\s^{(2)})=\sum_{\xx\in\L_M}
H_\xx^{AT}\;,\cr
&H_I(\s^{(j)})=-\sum_{\xx\in\L_M} [\s^{(j)}_{\xx}\s^{(j)}_{\xx+\hat e_1}+
\s^{(j)}_{\xx}\s^{(j)}_{\xx+\hat e_0}]\;,\cr
&V(\s^{(1)},\s^{(2)})=-
\sum_{\xx\in\L_M}[\s^{(2)}_{\xx}\s^{(2)}_{\xx+\hat e_0}
\s^{(1)}_{\xx}\s^{(1)}_{\xx+\hat e_0}+
\s^{(2)}_{\xx}\s^{(2)}_{\xx+\hat e_1}
\s^{(1)}_{\xx}\s^{(1)}_{\xx+\hat e_1}]\;,\cr}\Eq(1.1)$$
where $H_I$ is the Ising model hamiltonian,
$\hat e_1$, $\hat e_0$ are the unit vectors $\hat e_1=(1,0)$,
$\hat e_0=(0,1)$ and $\L_M$ is a square subset of $\ZZZ^2$ of side $M$. 
The free energy 
and the specific heat are given by
$$f=\lim_{M\to\io}{1\over M^2}\log Z_{\L_M}\virg 
C_{v}=\lim_{M\to\io}{1\over M^2}\sum_{\xx,\yy\in\L_M}<H_\xx^{AT}
H_\yy^{AT}>_{\L_M,T}\;,\Eq(cv)$$
where $<\cdot>_{\L_M,T}$ denotes
the truncated expectation w.r.t. the Gibbs distribution
with Hamiltonian \equ(1.1). 
The case $J^{(1)}=J^{(2)}$ is called {\it isotropic}.
For $\l=0$ the model
reduces to two independent Ising models
and it has 
{\it two critical points} if $J^{(1)}\not= J^{(2)}$;
it was conjectured by Kadanoff and Wegner [K][KW]
and later on by
Wu and Lin [WL] that the AT model
has in general two critical points 
also when $\l\not=0$,
except when the model is isotropic.

The {\it isotropic} case was studied
by Kadanoff [K] who, 
by scaling theory, conjectured a relation between 
the critical exponents of {\it isotropic} AT
and those of the {\it Eight vertex} model, 
which had been solved by Baxter and has nonuniversal indexes. 
Further evidence for the validity 
of Kadanoff's prediction was given by [PB]
(using second order renormalization group arguments)
and by [LP][N] (by a heuristic mapping of both
models into the massive Luttinger model describing 
one dimensional interacting fermions
in the continuum).
Indeed non universal
critical behaviour in the specific heat
in the isotropic AT model, for small
$\l$, has been rigorously established in [M1].

The {\it anisotropic} case is much less understood.
As we said, it is believed that 
there are {\it two} critical points, 
contrary to what happens in the isotropic case.
Baxter [Ba] conjectured that  
"presumably" {\it universality} holds at the critical points
for $J^{(1)}\not= J^{(2)}$
(\ie the critical indices are the same as in the Ising model),
except when $J^{(1)}=J^{(2)}$ when the two critical
points coincide and nonuniversal behaviour is found.
Since the 1970's, the anisotropic AT model 
was studied by various approximate or numerical methods:
Migdal--Kadanoff Renormalization Group [DR],
Monte Carlo Renormalization group [Be],
finite size scaling [Bad]; such results give evidence of
the fact that, far away from the isotropic point, 
AT has two critical points and belongs to the same universality class
of Ising; however they do not give informations
about the precise relative location of the critical points
and the critical behaviour of the specific heat 
when $J^{(1)}$ is close to $J^{(2)}$.
The problem of how 
the crossover from universal to nonuniversal
behaviour is realized in the isotropic limit
remained for years completely unsolved, 
even at a heuristic level.

We will study the anisotropic Ashkin--Teller model
by writing the partition function and the specific heat
as Grassmann integrals corresponding to a $d=1+1$
{\it interacting} fermionic theory; 
this is possible because the Ising model
can be reformulated as a {\it free fermions} model 
(see [SML][H][S] or [ID]). One can then
take advantage from the theory
of Grassmann integrals for 
weakly interacting $d=1+1$ fermions, which is quite
well developed, starting from [BG1]
(see also [BG][GM] or [BM] for extensive
reviews). Fermionic RG methods for classical spin models
have been already applied in [PS] to the
Ising model perturbed by a four spin
interaction, proving a {\it universality} result
for the specific heat; and in [M1] to prove a {\it
nonuniversality} result for the 8 vertex or the isotropic
AT model. By such techniques one can develop a 
perturbative expansion, convergent up to the critical points,
uniformly in the parameters.

\sub(1.1a) {\it Main results.}
We find convenient to introduce the variables 
$t^{(j)}=\tanh J^{(j)}$, $j=1,2$ and
$$t={t^{(1)}+t^{(2)}\over 2}
\virg u={t^{(1)}-t^{(2)}\over 2}\;\Eq(1.1aa).$$
The parameter $u$ measures the {\it anisotropy}
of the system. We consider then the free
energy or the specific heat as functions of $t,u,\l$.

If $\l=0$, AT is exactly
solvable, because the Hamiltonian \equ(1.1)
is the sum of two indipendent Ising model
Hamiltonians. From the Ising model exact solution
[O][SML][MW] one finds that $f$
is analytic for all $t,u$ except for
$$t=t_c^\pm=\sqrt2-1\pm|u|\Eq(1.aa)$$
and for $t$ close to $t_c^\pm$
the specific heat $C_v$ has a logarithmic 
divergence: $C_v\simeq -C\log|t-t_c^\pm|$, where 
$C>0$ and $\simeq$ means that the
ratio of both sides tends to 1 as $t\to t_c^\pm$.

We consider the case in which $\l$
is small with respect to $\sqrt2-1$ and 
we distinguish two 
regimes.\\
\01) If $u$ is much bigger than $\l$ (so that the unperturbed 
critical points are well separated)
we find that the presence of
$\l$ just changes by a small amount 
the location of the critical points, \ie we find that 
the critical points have the form $t^\pm_c=\sqrt2-1+O(\l)\pm|u|
\big(1+O(\l)\big)$; moreover the asymptotic behaviour
of $C_v$ at criticality remains essentially  
unchanged: $C_v\simeq-C\log|t-t^\pm_c|$.
\\
\02) When $u$ is small compared to $\l$ the interaction
has a more dramatic effect.
We find that the system has still only two critical points $t^\pm_c(\l,u)$;
their center $(t^+_c+t_c^-)/2$ is just shifted by $O(\l)$
from $\sqrt2-1$, as in item (1);
however their relative location 
scales, as $u\to 0$, 
with an ``anomalous critical exponent'' $\h(\l)$,
continously varying with $\l$: more precisely 
we find that $t^+_c-t_c^-=O(|u|^{1+\h})$,
where $\h$ is analytic in $\l$ near $\l=0$ and
$\h=-b \l +O\big(\l^2\big)$, $b>0$. In particular the relative 
location of the critical points as a function of the anisotropy parameter $u$
with $\l$ fixed and small has a different qualitative behaviour,
depending on the sign of $\l$, see Fig 1.

%%%%%%%%%%%%%%%%%%%%%%%%%%%%%%%%%%%%%%%%%%%%%%%%%%%%%%%%%%%%%%%%%%%%%%%%%%%%%
%%%%%%%%%%%%%%%%%%%%%%%%%%%%%%%%FIGURA 1%%%%%%%%%%%%%%%%%%%%%%%%%%%%%%%%%%%%%
%%%%%%%%%%%%%%%%%%%%%%%%%%%%%%%%%%%%%%%%%%%%%%%%%%%%%%%%%%%%%%%%%%%%%%%%%%%%%

\insertplot{300pt}{110pt}%
{\ins{170pt}{-24pt}{$u$}
\ins{44pt}{83pt}{$t^+_c-t^-_c$}
\ins{151pt}{18pt}{$\l<0$}
\ins{107pt}{60pt}{$\l>0$}
\ins{110pt}{25pt}{$\l=0$}
}%
{tc}{}
\vskip1.5cm
\line{\vtop{\line{\hskip2.5truecm\vbox{\advance\hsize by -5.1 truecm
\0{\ottorm FIG 1. 
The qualitative behaviour of ${\scriptstyle 
t^+_c(\l,u)-t_c^-(\l,u)}$ as a function of 
${\scriptstyle u}$ for two different values of ${\scriptstyle \l}$ 
(in arbitrary units). The graphs are (qualitative) plots of ${\scriptstyle 
2|u|^{1+\h}}$, with ${\scriptstyle \h\simeq -b\l}$, ${\scriptstyle b>0}$. 
} \hfill} }}}
\vskip.5cm

%%%%%%%%%%%%%%%%%%%%%%%%%%%%%%%%%%%%%%%%%%%%%%%%%%%%%%%%%%%%%%%%%%%%%%%%%%%%%%
%%%%%%%%%%%%%%%%%%%%%%%%%%%%%%%%%%%%%%%%%%%%%%%%%%%%%%%%%%%%%%%%%%%%%%%%%%%%%%
 
For $t\to t_c^\pm(\l,u)$
the specific heat $C_v$ has still a logarithmic 
divergence but, for all $u\not=0$, the constant in front of the log
is $O(|u|^{\h_c})$, where $\h_c$ is analytic in $\l$ for small $\l$ and $\h_c
=a \l+O\big(\l^2\big)$, $a\not=0$. The logarithmic behaviour is found only
in an extremely small region around
the critical points; outside this region, $C_v$ varies as $t\to t_c^\pm(\l,u)$
according to a power law behaviour with nonuniversal 
exponent.
The conclusion is that, for all $u\not=0$,
there is universality for the 
specific heat (which diverges with the same exponent as in the Ising model);
nevertheless {\it nonuniversal critical indexes} appear in the theory,
in the difference between the critical points and in the constant 
in front of the logarithm in the specific heat.
One can speak of {\it anomalous universality}
as the specific heat diverges at criticality
as in Ising, but the isotropic limit $u\to 0$ 
is reached with nonuniversal
critical indices. 

%The presence of nonuniversal critical exponents implies that
%the anisotropic AT model is not analytically close
%to the Ising model uniformly in $u$
%(for instance the ratio 
%$C_v(\l)/C_v(0)$, which is $O(|u|^{\h_c})$ 
%for $|u|<<|\l|$, cannot be written  
%as a power series in $\l$ uniformly convergent in $u$), 
%even though is in the
%same universality class (at least for what concerns the
%critical behaviour of the specific heat).\\

With the notations introduced above
and calling $D$ a sufficiently small $O(1)$ interval 
(\ie with amplitude independent of $\l$)
centered around $\sqrt2-1$, 
we can express our main result as follows. 
\*
\0{\bf Main Theorem.}
{\it 
There exists $\e_1$ such that, for 
$t\pm u\in D$, $j=1,2$, 
and $|\l|\le \e_1$, one can define two functions
$t_c^\pm(\l,u)$ with the following properties:
$$t_c^\pm(\l,u)=\sqrt2-1+\n^*(\l)
\pm|u|^{1+\h}\big(1+F^\pm(\l,u)\big)\;,
\Eq(1.5)$$
where
$|\n^*(\l)|\le c|\l|$,
$|F^\pm(\l,u)|\le c|\l|$,
for some positive constant $c$ and 
$\h=\h(\l)$ is an analytic function of $\l$ s.t.
$\h(\l)=-b \l +O(\l^2)$, $b>0$, and:
\\ 
\01) the free energy $f(t,u,\l)$
and the specific heat $C_v(t,u,\l)$ in \equ(cv)
are analytic in the region $t\pm u\in D$, $|\l|\le \e_1$ and 
$t\not=t_c^\pm(\l,u)$;\\
\02) in the same region of parameters,
the specific heat can be written as:
$$\eqalign{&C_v=-C_1
\D^{2\h_c}\log{\big|t-t_c^-\big|\big|t-t_c^+\big|
\over \D^2}
+C_2{1-\D^{2\h_c}\over \h_c}+C_3\cr
%&{\rm where:}\qquad \cr
}\Eq(1.6)$$
where: 
$\D^2\ \defin (t-\lis t_c)^2+(u^2)^{1+\h}$ and 
$\lis t_c\defin (t^+_c+t^-_c)/ 2$;
the exponent $\h_c=\h_c(\l)=a \l +O(\l^2)$,
$a\not =0$, is analytic in $\l$; 
the functions $C_j=C_j(\l,t,u)$, $j=1,2,3$, are
bounded above and below by $O(1)$ constants; finally $C_1-C_2$
vanishes for $\l=u=0$.}
\*
\0{\bf Remarks}\\
\01) The key hypothesis 
for the validity of Main Theorem is the 
smallness of $\l$. When $\l=0$ the critical points
correspond to $t\pm u=\sqrt2-1$: hence for simplicity
we restrict $t\pm u$ in a sufficiently small $O(1)$
interval around $\sqrt2-1$. A possible explicit choice for $D$, 
convenient for our proof, 
could be $D=[{3(\sqrt2-1)\over 4},{5(\sqrt2-1)\over 4}]$.  
Our technique would allow us to prove the above theorem, at the cost of a 
lengthier discussion, 
for any $t^{(1)},t^{(2)}>0$: of course in that case we should 
distinguish different regions of parameters and treat in a different
way the cases of low or high temperature or the case of big anisotropy
(\ie the cases $t<<\sqrt2-1$ or $t>>\sqrt2-1$ or $|u|>>1$).\\
\02) \equ(1.6) shows how the crossover from universal to nonuniversal behaviour
is realized. When $u\not=0$ only the first term in \equ(1.6) can be singular
in correspondence of the two critical points; it has a logarithmic singularity 
(as in the Ising model) with a constant $O(\D^{2\h_c})$ in front. However
the logarithmic term dominates on the second one only if $t$ varies 
inside an extremely small region $O(|u|^{1+\h}e^{-a/|\l|})$, $a>0$,
around the 
critical points. Outside such region the power law behaviour 
corresponding to the second addend in \equ(1.6) dominates. When $u\to 0$ one
recovers the power law decay found in [M1] for the isotropic case. See Fig 2.

%%%%%%%%%%%%%%%%%%%%%%%%%%%%%%%%%%%%%%%%%%%%%%%%%%%%%%%%%%%%%%%%%%%%%%%%%%%%%
%%%%%%%%%%%%%%%%%%%%%%%%%%%%%%%%FIGURA 2%%%%%%%%%%%%%%%%%%%%%%%%%%%%%%%%%%%%%
%%%%%%%%%%%%%%%%%%%%%%%%%%%%%%%%%%%%%%%%%%%%%%%%%%%%%%%%%%%%%%%%%%%%%%%%%%%%%

\insertplot{300pt}{90pt}%
{%\ins{30pt}{70pt}{$C_v$}
%\ins{113pt}{30pt}{$\l>0$}
%\ins{130pt}{-30pt}{$\l=0$}
%\ins{270pt}{-53pt}{$\l<0$}
%\ins{290pt}{-115pt}{$t-\lis t_c$}
%\ins{158pt}{68pt}{$t^+_c$}
%\ins{115pt}{56pt}{$(\ )$}
%\ins{155.5pt}{56pt}{$(\ )$}
%\ins{234pt}{47pt}{$t$}
%\ins{234pt}{-2pt}{$t$}
%\ins{130pt}{78pt}{$4|u|^{1+\h}$}
%\ins{130pt}{46pt}{$2|u|^{1+\h}$}
%\ins{43pt}{40pt}{$|u|^{1+\h}\Big\{$}
%\ins{43pt}{60pt}{$|u|^{1+\h}\Big\{$}
%\ins{138pt}{18pt}{$\lis t_c$}
%\ins{134pt}{-5pt}{$|D|$}
}%
{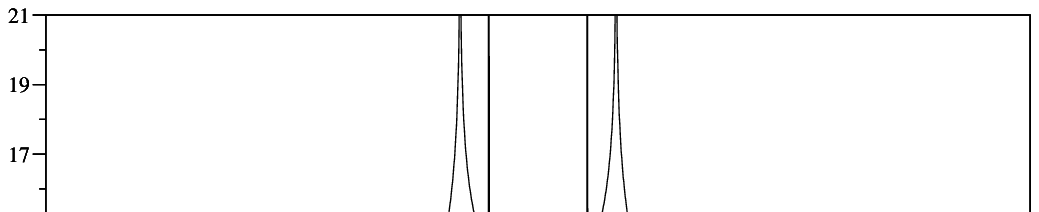}{}
\vskip6.18cm
\line{\vtop{\line{\hskip1.3truecm\vbox{\advance\hsize by -2.0 truecm
\0{\ottorm FIG 2. The qualitative behaviour of ${\scriptstyle C_v}$ 
as a function of ${\scriptstyle t-\lis t_c}$, 
where ${\scriptstyle \lis t_c=(t^+_c+t^-_c)/2}$. The three graphs are 
plots of \equ(1.6), with ${\scriptstyle C_1=C_2=1}$, ${\scriptstyle C_3=0}$, 
${\scriptstyle u=0.01}$, 
${\scriptstyle \h=\h_c=0.1,0,-0.1}$ respectively; the central curve 
corresponds to ${\scriptstyle\l=0}$, the upper one to ${\scriptstyle\l>0}$
and the lower to ${\scriptstyle\l<0}$.
} \hfill} }}}
\vskip.5cm

%%%%%%%%%%%%%%%%%%%%%%%%%%%%%%%%%%%%%%%%%%%%%%%%%%%%%%%%%%%%%%%%%%%%%%%%%%%%%%
%%%%%%%%%%%%%%%%%%%%%%%%%%%%%%%%%%%%%%%%%%%%%%%%%%%%%%%%%%%%%%%%%%%%%%%%%%%%%%

\03) By the result of item (1) of Main Theorem, $C_v$ is analytic in
$\l,t,u$ outside the critical line. This is not appearent 
from \equ(1.6), because $\D$ is non analytic in $u$ at $u=0$ (of course the
bounded functions $C_j$ are non analytic in $u$ also, in a suitable
way compensating the non analyticity of $\D$). We get to \equ(1.6)
by interpolating two different asymptotic behaviours of $C_v$ in the regions
$|t-\lis t_c|<2|u|^{1+\h}$ and $|t-\lis t_c|\ge 2|u|^{1+\h}$ and the
non analyticity of $\D$ is introduced ``by hands'' by our estimates
and it is not intrinsic for $C_v$. \equ(1.6) is simply a convenient 
way to describe the crossover between different
critical behaviours of $C_v$.\\
\04) We do not study the free energy directly at $t=t_c^\pm(\l,u)$,
therefore in order to show that
$t=t_c^\pm(\l,u)$ is a critical point we must study 
some thermodynamic property like 
the specific heat by evaluating
it at $t\ne t_c^\pm(\l,u)$ and $M=\io$ and then verify that it has a
singular behavior as $t\to t_c^\pm$. The case $t$ precisely equal to
$t_c^\pm$ cannot be discussed at the moment 
with our techniques, in spite of the uniformity
of our bounds as $t\to t_c^\pm$. The reason is that 
we write the AT partition function as a sum
of 16 different partition functions, differing
for boundary terms. Our estimates on each single term 
are uniform up to the critical point;
however, in order to show that the free energy computed with one of the 
16 terms is the same as the complete free energy,
we need to stay at $t\ne t_c^\pm$: in this case boundary terms are suppressed
as $\sim e^{-\k M|t-t_c^\pm|}$, $\k>0$, as $M\to\io$. If we stay exactly
at the critical point cancellations between the 16 terms can be present
(as it is well known already from the Ising model exact solution [MW])
and we do not have control on the behaviour of the free energy, as 
the infinite volume limit is approached.\\

\sub(1.1b) {\it Strategy of the proof.}
It is well known that the free energy and the specific heat of
the Ising model can be expressed as a sum of {\it Pfaffians} [MW]
which can be equivalently written, 
see [ID][S], 
as {\it Grassmann functional integrals}, see for instance App A of [M1] or 
\S 4 of [GM]
for the basic definitions of Grassmann variables and Grassmann integration. 
The formal action of the Ising model in terms 
of Grassmann variables $\psi,\lis\psi$ has the form 
$$\eqalign{&%S(\psi,\lis\psi)=
\sum_\xx{t\over 4}\Big[
\psi_\xx(\partial_1-i\partial_0)\psi_\xx+
\lis\psi_\xx(\partial_1+i\partial_0)\lis\psi_\xx-2
i\lis\psi_\xx(\partial_1+\dpr_0)\psi_\xx\Big]
+i(\sqrt{2}-1-t)\lis\psi_\xx\psi_\xx\;,\cr}\Eq(isi)$$
where $\dpr_j$ are discrete derivatives. $\psi$ and $\lis\psi$ are called 
{\it Majorana} fields, see [ID], because of an analogy with
relativistic Majorana fermions. 
They are massive, because of the presence
of the last term in \equ(isi); criticality corresponds
to the massless case ($t=\sqrt2-1$). 
If $\l=0$ the free energy and specific
heat can be written as sum of Grassmann integrals describing {\it two}
kinds of Majorana fields, with masses $m^{(1)}=t^{(1)}-\sqrt2+1$
and $m^{(2)}=t^{(2)}-\sqrt2+1$.
The critical points are obtained by choosing 
one of the two fields massless
(in the isotropic case $t^{(1)}=t^{(2)}$ and the two fields
become massless togheter).

If $\l\not=0$ again the free energy and the specific heat
can be written as Grassmann integrals, but the Majorana
fields are {\it interacting} with a short range potential.
By performing a suitable change
of variables, the partition function can be written, see \S 2 and \S 3, as 
a sum of terms $\Xi_{AT}^{\g_1,\g_2}$ ($\g_1,\g_2$ label 
different boundary conditions) of the form
$$\Xi_{AT}^{\g_1,\g_2}=\int P(d\psi) e^{-\VV^{(1)}(\sqrt{Z_1}\psi)}\virg
P(d\psi)=\DD\psi\ e^{-Z_1(\psi^+,A\psi)}\;,\Eq(Xi)$$
where: $\psi=\{\psi^+_{\o,\xx},\psi^{-}_{\o,\xx}\}_{\o=\pm 1}$ 
are elements of a Grassmann algebra; $\DD\psi$ is a symbol for 
the Grassmann integration; $\VV^{(1)}$ is a short range interaction, 
sum of monomials in $\psi$
of any degree, whose quartic term is weighted by a constant $\l_1=O(\l)$;
and $Z_1(\psi^+,A\psi)$ has the form:
$$Z_1\sum_{\xx,\o}
\psi^+_{\o,\xx}(\dpr_1-i\o\dpr_0)\psi^-_{\o,\xx}
-i\o\s_1\psi^+_{\o,\xx}\psi^-_{-\o,\xx}+
i\o\m_1\psi^\a_{\o,\xx}
\psi^\a_{-\o,-\xx}-\b_1\psi^\a_{\o,\xx}(\dpr_1-i\o\dpr_0)\psi^\a_{\o,\xx}
\Eq(lab1)$$
with $\s_1=O(t-\sqrt2+1)+O(\l)$, $\m_1,\b_1=O(u)$
(in particular in the isotropic case the terms proportional to $\m_1$ and 
$\b_1$ are absent).
If $\l=0$, $\s_1=(m^{(1)}+m^{(2)})/2$
and $\m_1=(m^{(2)}-m^{(1)})/2$. 
$\psi^\pm$ are called {\it Dirac} fields, because of an analogy with
relativistic Dirac fermions; they are 
combinations of the Majorana variables $\psi^{(j)},\lis\psi^{(j)}$, $j=1,2$,
associated with the two Ising layers
in \equ(1.1): hence the description in terms 
of Dirac variables mixes intrinsically the two Ising models
and will be useful in a range of momentum scale
in which the two layers appear to be essentially equal.

One can compute $\Xi_{AT}^{\g_1,\g_2}$ by expanding $e^{-\VV^{(1)}(
{\sqrt Z_1}\psi)}$
in Taylor series and integrating term by term the Grassmann monomials;
since the propagators of $P(d\psi)$ (\ie the elements
of $A^{-1}$, see \equ(Xi), \equ(lab1)) diverge for $\kk={\bf 0}$
and $\s_1\pm\m_1=0$ 
in the infinite volume limit $M\to\io$, 
the series can converge uniformly in $M$ only in a region
outside $|\s_1\pm\m_1|\le c$, for some $c$, \ie in the thermodynamic
limit it can converge 
only far from the critical points.

Since we are interested in the critical behaviour of the system, 
we set up a more complicated procedure to evaluate 
the partition function, based on (Wilsonian)
Renormalization Group (RG).
The first step is to decompose the integration $P(d\psi)$ as a product of
independent integrations: $P(d\psi)=\prod_{h=-\io}^{1} P(d\psi^{(h)})$,
where the momentum space propagator corresponding to $P(d\psi^{(h)})$ is
not singular, but $O(\g^{-h})$, for $M\to\io$, $\g$ being a fixed {\it
scaling parameter} larger than $1$. This decomposition is realized by
slicing in a smooth way the momentum space, so that $\psi^{(h)}$, if $h\le
0$, depends only on the momenta between $\g^{h-1}$ and $\g^{h+1}$. 
We compute the Grassmann integrals defining the partition function
by iteratively integrating
the fields $\psi^{(1)},\psi^{(0)},\ldots$, see \S 4.
After each integration step we rewrite the partition function in a way 
similar to \equ(Xi), with the quadratic form $Z_1(\psi^+,A\psi)$ replaced by 
$Z_h(\psi^+, A^{(h)}\psi)$, which has the same structure of \equ(lab1), 
with $Z_h,\s_h,\m_h$
replacing $Z_1,\s_1,\m_1$; the structure of $Z_h(\psi^+, A^{(h)}
\psi)$ is preserved because of
symmetry properties, guaranteeing that many other possible 
quadratic ``local'' terms are indeed vanishing, 
or {\it irrelevant} in a RG sense.
The interaction $\VV^{(1)}$ is replaced by an {\it effective
action} $\VV^{(h)}$, $h\le 0$, 
given by a sum of monomials of $\psi$ of arbitrary
order, with kernels decaying in real space on scale $\g^{-h}$; in particular
the quartic term is weighted by a coupling constant $\l_h$ and
the kernels of $\VV^{(h)}$ are {\it analytic functions} 
of $\{\l_h,\ldots,\l_1\}$,
if $\l_k$ are small enough, $k\ge h$, and $|\s_k|\g^{-k},|\m_k|\g^{-k}\le 1$
(say -- the constant $1$ could be replaced by any other constant $O(1)$).

In this way the problem of finding good bounds for $\log\Xi_{AT}$ 
is reformulated
into the problem of controlling the size of $\l_h,\s_h,\m_h$, $h\le 0$,
under the RG iterations. 
We use a crucial property,
called {\it vanishing of Beta function}, 
to prove that actually, if $\l$ is small enough, $|\l_h|\le 2|\l_1|$
(recall that $\l_1=O(\l)$). The possibility of controlling the flow of 
$\l_h$ is 
the main reason for describing the system in terms of Dirac variables.
For $\s_h,\m_h,Z_h$, we find that, under RG iterations, they 
evolve as: $\s_h\simeq\s_1 \g^{b_2\l h}$, $\m_h\simeq \m_1 \g^{-b_2\l h}$, 
$Z_h\simeq\g^{-b_1\l^2h}$. Note in particular that $Z_h$ grows exponentially
with an exponent $O(\l^2)$; this is connected with the presence of
``critical indexes'' in the correlation functions, which means that
their long distance behaviour is qualitatively changed by the interaction.

We perform the iterative integration descrided above up to a
scale $h^*_1$ such that $(|\s_{h^*_1}|+|\m_{h^*_1}|)\g^{-h^*_1}=O(1)$, 
in such a way that $(|\s_{h}|+|\m_{h}|)\g^{-h}\le O(1)$, for all
$h\ge h^*_1$ and convergence of the kernels
of the effective potential can be guaranteed by our estimates. 
In the range of scales $h\ge h^*_1$ the flow
of the effective coupling constant $\l_h$ is essentially the same 
as for the isotropic AT model [M1] (since $|\m_h|\g^{-h}$ is small the
iteration ``does not see'' the anisotropy and 
the system seems to behave
as if there was just one critical point) and nonuniversal critical indexes are 
generated (they appear in the flows of $\s_h,\m_h$ and $Z_h$), following 
the same mechanism of the isotropic case.

We note that after the integration of $\psi^{(1)},\ldots,\psi^{(h^*_1+1)}$,
we can still reformulate the problem in terms of the original 
Majorana fermions
$\psi^{(1, \le h^*_1)}$, $\psi^{(2, \le h^*_1)}$
associated with the two Ising models in \equ(1.1).
On scale $h^*_1$ their masses are deeply changed 
w.r.t. $t^{(1)}-\sqrt2+1$ and $t^{(2)}-\sqrt2+1$:
they are given by $m_{h^*_1}^{(1)}=
|\s_{h^*_1}|+|\m_{h^*_1}|$
and $m_{h^*_1}^{(2)}=|\s_{h^*_1}|-|\m_{h^*_1}|$. Note that
the condition $|\s_{h^*_1}|+|\m_{h^*_1}|=O(\g^{h^*_1})$ 
implies that the field $\psi^{(1,\le h^*_1)}$ is massive on scale $h^*_1$
(so that the Ising layer with $j=1$ is ``far from criticality'' 
on the same scale). This implies
that we can integrate (without any multiscale decomposition)
the massive Majorana
field $\psi^{(1,\le h^*_1)}$, obtaining
an effective theory
of a single Majorana field with mass $|\s_{h^*_1}|-|\m_{h^*_1}|$, which 
can be arbitrarly small.
The integration of the scales $\le h^*_1$, see \S 6, is done again by
a multiscale decomposition similar to the one just described; an
important feature is however that there are no more quartic
marginal terms, because the anticommutativity
of Grassmann variables forbids local quartic monomials
of a single Majorana fermion. The problem is essentially
equivalent to the study of a single perturbed Ising model
with ``upper'' cutoff on momentum space $O(\g^{h^*_1})$ and mass 
$|\s_{h^*_1}|-|\m_{h^*_1}|$. The flow of the effective mass and
of $Z_h$ is non anomalous in this regime: in particular
the mass of Majorana field is just shifted by $O(\l\g^{h^*_1})$ from
$|\s_{h^*_1}|-|\m_{h^*_1}|$. Criticality 
is found when the effective mass on scale $-\io$ is vanishing;
the values of $t,u$ for which this happens
are found by solving a non trivial implicit function problem.

Finally, see \S 7, we define a similar expansion
for the specific heat and 
we compute its asymptotic behaviour arbitrarily near the
critical points.

Technically it is an interesting
feature of this problem that there are two regimes in which the
system must be described in terms of different fields:
a first one in which the natural variables are Dirac Grassmann variables,
and a second one in which they are Majorana; note that
the scale separating the two
regimes is dynamically generated by the RG iterations (and of course 
its precise location is not crucial and $h^*_1$ can be modified in $h^*_1+n$,
$n\in\ZZZ$, without qualitatively affecting the bounds).
%something similar is believed to happen in BCS theory in which
%in the first regime one uses fermionic variables, in the second
%bosonic ones. 

\*
\section(2,Fermionic representation)

\sub(2.1) The partition function $\Xi_{I}^{(j)}=\sum_{\s^{(j)}}
\exp\{-J^{(j)}H_I(\s^{(j)})\}$ of the Ising model can
be written as a Grassmann integral; this is a classical result, mainly
due to [LMS][Ka][H][MW][S]. In Appendix A1, starting from a formula 
obtained in [MW], we prove that
$$\Xi_I^{(j)}=(-1)^{M^2}{(2\cosh  J^{(j)})^{M^2} \over
2}\sum_{\e,\e'=\pm} \int \prod_{\xx\in\L_M} dH^{(j)}_\xx
d\lis H^{(j)}_\xx d V^{(j)}_\xx d\lis V^{(j)}_\xx
(-1)^{\d_{\g}} e^{S^{(j)}_\g(t^{(j)})}
\Eq(2.1)$$
where $j=1,2$ denotes the lattice, $\g= (\e,\e')$ and $\d_\g$ is
$\d_{+,+}=1,\d_{+,-}=\d_{-,+}=\d_{-,-}=2$ and, if $t^{(j)}=\tanh J^{(j)}$,
$$\eqalign{S^{(j)}_\g(t^{(j)})&=
t^{(j)}\sum_{\xx\in\L_M} \left[\lis H^{(j)}_{\xx} H^{(j)}_{\xx+\hat e_1}+
\lis V^{(j)}_{\xx} V^{(j)}_{\xx+\hat e_0}\right]+\cr
&+\sum_{\xx\in\L_M}\left[\lis H^{(j)}_{\xx} H^{(j)}_{\xx}+
\lis V^{(j)}_{\xx}
V^{(j)}_{\xx}+\lis V^{(j)}_{\xx} \lis H^{(j)}_{\xx}+
V^{(j)}_{\xx} \lis H^{(j)}_{\xx}+
H^{(j)}_{\xx} \lis V^{(j)}_{\xx}+
V^{(j)}_{\xx} H^{(j)}_{\xx}\right]\;,\cr}\Eq(2.2)$$
where $H^{(j)}_\xx,\lis H^{(j)}_\xx,V^{(j)}_\xx,
\lis V^{(j)}_\xx$ are {\it Grassmann variables} verifying different
boundary conditions depending on the label $\g=(\e,\e')$ which is not
affixed explicitly, to simplify the notations, \ie
$$\eqalign{
&\lis H^{(j)}_{\xx+M\hat e_0}=\e\lis H^{(j)}_{\xx}
\virg\lis H^{(j)}_{\xx+M\hat e_1}=\e'\lis H^{(j)}_{\xx}\cr
&H^{(j)}_{\xx+M\hat e_0}=\e H^{(j)}_{\xx}\virg
H^{(j)}_{\xx+\hat e_1}=\e' H^{(j)}_{\xx}\cr}\virg\e,\e'=\pm\Eq(2.3)$$
and identical definitions are set for the variables $V^{(j)},{\lis
V}^{(j)}$; we shall say that $\lis H^{(j)}, H^{(j)},\lis V^{(j)},V^{(j)}$ 
satisfy $\e$--periodic
($\e'$--periodic) boundary conditions in vertical (horizontal) 
direction. \\

\sub(2.2) By expanding in power series $\exp\{-\l V\}$, we see that 
the partition function of the model \equ(1.1) is
$$\eqalign{\Xi_{AT}&=\sum_{\s^{(1)},\,\s^{(2)}}
e^{-J^{(1)}H_I(\s^{(1)})}e^{-J^{(2)}
H_I(\s^{(2)})}e^{-\l V(\s^{(1)},\,\s^{(2)})}
=\cr
&=(\cosh \l)^{2M^2}\sum_{\s^{(1)},\,\s^{(2)}}
e^{- J^{(1)}H_I(\s^{(1)})- J^{(2)}
H_I(\s^{(2)})}\cdot\cr
&\cdot\prod_{\xx\in\L_M}
\left(1+\hat\l 
\s^{(1)}_{\xx}\s^{(1)}_{\xx+\hat e_1}
\s^{(2)}_{\xx}\s^{(2)}_{\xx+\hat e_1}\right)\left(1+\hat\l
\s^{(1)}_{\xx}\s^{(1)}_{\xx+\hat e_0}
\s^{(2)}_{\xx}\s^{(2)}_{\xx+\hat e_0}\right)\;,\cr}\Eq(2.8)$$
where $\hat\l=\tanh \l$. The r.h.s. of \equ(2.8) can be rewritten as:
$$\Xi_{AT}=\Bigg[\prod_{\xx\in\L_M\atop i=0,1}
\Big(1+\hat\l{\dpr^2\over\dpr 
J^{(1)}_{\xx,\xx+\hat e_i}\dpr J^{(2)}_{\xx,\xx+\hat e_i}}\Big)\Bigg]
\Xi_{I}^{(1)}(\{J^{(1)}_{\xx,\xx'}\})\Xi_{I}^{
(2)}(\{J^{(2)}_{\xx,\xx'}\})\Big|_{
\{J^{(j)}_{\xx,\xx'}\}=\{J^{(j)}\}}\;,\Eq(2.9)$$
where $\Xi_I^{(j)}(\{J^{(j)}_{\xx,\xx'}\})$ is the partition function 
of an Ising model in which the couplings are allowed to depend on the 
bonds (the coupling associated to the n.n. bond $(\xx,\xx')$ on the lattice
$j$ is called $J^{(j)}_{\xx,\xx'}$).
Using for $\Xi_{I}^{(1)}(\{J^{(1)}_{\xx,\xx'}\})$ an expression similar 
to \equ(2.1), we find that we can
express $\Xi_{AT}$ as a sum of  sixteen partition functions labeled
by $\g_1,\g_2=(\e_1,\e_1'),(\e_2,\e_2')$ (corresponding to
choosing each $\e_j$ and $\e_j'$ as $\pm$):
$$\Xi_{AT}= {1\over 4} (\cosh \l)^{2M^2}\sum_{\g_1,\g_2}
(-1)^{\d_{\g_1}+\d_{\g_2}}Z_{AT}^{\g_1,\g_2}\;,\Eq(2.10)$$
each of which is given by a functional integral
$$\eqalign{&\Xi_{AT}^{\g_1,\g_2}=
\big[4(1+\hat\l t^{(1)} t^{(2)})\big]^{M^2}\prod_{j=1}^2
(\cosh J^{(j)})^{M^2}(-1)^{M^2}
\cdot\cr
&\qquad\cdot
\int\prod_{\xx\in\L_M}^{j=1,2} dH^{(j)}_\xx d
\lis H^{(j)}_\xx d V^{(j)}_\xx
d\lis V^{(j)}_\xx\, e^{S^{(1)}_{\g_1}(t_\l^{(1)})+S^{(2)}_{\g_2}
(t_\l^{(2)})+V_\l}\;,\cr}
\Eq(2.11)$$
where, if we define
$$\eqalign{&
\l^{(j)}={\hat\l \big[t(1-t^2+u^2)+(-1)^ju(1+t^2-u^2)\big]\over 1+\hat\l 
(t^2-u^2)}
\;,\cr}\Eq(2.13)$$
we have that $t_\l^{(j)}$, $j=1,2$, 
is given by $t_\l^{(j)}=t^{(j)}+\l^{(j)}$ and 
$V_\l$ by:   
$$V_\l=\sum_{\xx\in\L_M} 
\l \left(\lis H^{(1)}_{\xx} H^{(1)}_{\xx+\hat e_1}
\lis H^{(2)}_{\xx} H^{(2)}_{\xx+\hat e_1}+ \lis V^{(1)}_\xx V^{(1)}_{
\xx+\hat e_0} \lis V^{(2)}_{\xx} V^{(2)}_{\xx+\hat e_0}\right)
\;,\qquad
\l={\l^{(1)}\l^{(2)}\over \hat\l (t^2-u^2)}\;.\Eq(2.15)$$
\\
\sub(2.3) 
From now on, we shall study in detail only the partition function 
$\Xi_{AT}^{-}\defin \Xi_{AT}^{(-,-),(-,-)}$, \ie the partition function
in which all Grassmannian variables verify antiperiodic boundary
conditions (see \equ(2.3)). We shall see in \sec(an) below that, 
if $(\l, t, u)$ {\it does not belong} to the {\it critical surface}, which 
is a suitable 2--dimensional subset of $[-\e_1,\e_1]\times D\times [-{|D|\over
2},{|D|\over 2}]$ which we will 
explicitely determine in \sec(6.1), 
the partition function $\Xi_{AT}^{\g_1,\g_2}$ divided by 
$\Xi_I^{(1)\g_1}\Xi_I^{(2)\g_2}$
is exponentially insensitive to boundary conditions as $M\to\io$.

As in [M1] we find convenient to perform the following change of
variables, $\a=\pm$, $\o=\pm 1$:
$$\eqalign{&{1\over \sqrt2}\sum_{j=1,2}(-i\a)^{j-1}\big(\lis H_\xx^{(j)}
+i\o H^{(j)}_\xx\big)=e^{i\o\p/4}\big(\psi^\a_{\o,\xx}-\c^\a_{\o,\xx}\big)\cr
&{1\over \sqrt2}\sum_{j=1,2}(-i\a)^{j-1}\big(\lis V_\xx^{(j)}
+i\o V^{(j)}_\xx\big)=\psi^\a_{\o,\xx}+\c^\a_{\o,\xx}\;.\cr}\Eq(2.19)$$
Let $\kk\in D_{-,-}$, where $D_{-,-}$ is the set of $\kk$'s 
such that $k=2\p/M(n_1+1/2)$ and $k_0=2\p/M(n_0+1/2)$, where
$-[M/2]\le n_0,n_1\le [(M-1)/2]$, $n_0,n_1\in\ZZZ$. The 
Fourier transform of the Grassmanian fields $\phi^\a_{\o,\xx}$, $\phi=\psi,\c$,
is given by $\hat\phi^\a_{\o,\kk}\defin\sum_{\xx\in\L_M} 
e^{-i\a\kk\xx}\phi^\a_{\o,\xx}$.\\

With the above definitions,
it is straightforward algebra to verify that the final expression is:
$$\Xi_{AT}^-=e^{-E M^2}\int P(d\psi)
P(d\chi)e^{Q(\psi,\c)+V(\psi,\c)}\;,\Eq(2.26)$$
where: $E$ is a suitable constant; $Q(\psi,\c)$ collects the quadratic
terms of the form $\psi^{\a_1}_{\o_1,\xx_1}\c^{\a_2}_{\o_2,\xx_2}$;
$V(\psi,\c)$ is the quartic interaction (it is equal to $V_\l$, see 
\equ(2.15), in terms of the $\psi^\pm_\o$, $\c^\pm_\o$ variables); 
$P(d\phi)$, $\phi=\psi,\c$, is:
$$\eqalign{
&P(d\phi)=\NN^{-1}_\phi\prod_{\kk\in D_{-,-}} \prod_{\o=\pm 1}
d\phi^{+}_{\o,\kk}
d\phi^{-}_{\o,\kk}
\exp\Bigl\{-{t_\l\over 4M^2}\sum_{\kk\in D_{-,-}}
{\bf\Phi_\kk^{+,T}} A_\phi(\kk){\bf\Phi}_\kk\Bigr\}\;,\cr
&A_\phi(\kk)=
\left( \matrix{i\sin k+\sin k_0 & -i \s_\phi(\kk)
& -{\m\over 2}(i\sin k+\sin k_0) & i \m(\kk)\cr
i \s_\phi(\kk) & i\sin k-\sin k_0& -i \m(\kk)& 
-{\m\over 2}(i\sin k-\sin k_0) \cr
-{\m\over 2}(i\sin k+\sin k_0)& 
i \m(\kk) & i\sin k+\sin k_0 & -i \s_\phi(\kk)\cr
-i \m(\kk)& -{\m\over 2}(i\sin k-\sin k_0)&i \s_\phi(\kk)&i\sin k-\sin k_0\cr}
\right)\cr}\Eq(2.27)$$
where
$${\bf\Phi^{+,T}}_\kk=(\hat\phi^+_{1,\kk},
\hat\phi^+_{-1,\kk},\hat\phi^-_{1,-\kk},
\hat\phi^-_{-1,-\kk})\virg
{\bf\Phi^{T}}_\kk=(\hat\phi^-_{1,\kk},\hat\phi^-_{-1,\kk},
\hat\phi^+_{1,-\kk},\hat\phi^+_{-1,-\kk})
\;,\Eq(2.28)$$
$\NN_\phi$ is chosen in such a way that $\int P(d\phi)=1$ and, if 
we define $t_\l\defin(t_\l^{(1)}+t_\l^{(2)})/ 2$,
$u_\l\defin 
(t_\l^{(1)}-t_\l^{(2)})/ 2$, for $\phi=\psi,\c$ we have: 
%&c(\kk)\defin \cos k_0+\cos k-2\virg d(\kk)\defin 
%(u_\l/t_\l)(2-\cos k-\cos k_0)\;,\cr}
%\Eq(2.250a)$$
%%
%we have 
%
$$\eqalign{&\s_\phi(\kk)=2\Bigl(1+{\pm\sqrt 2+1 \over 
t_\l}\Bigr)+\cos k_0+\cos k-2\virg \mu(\kk)=-(u_\l/t_\l)(\cos k+\cos k_0)
\;.\cr}\Eq(2.29)$$
In the first of \equ(2.29) the $-$ ($+$) sign corresponds to $\phi=\psi$ 
($\phi=\c$). The parameter $\m$ in \equ(2.27) is given by 
$\m\defin\m({\bf 0})$.

It is convenient to split the $\sqrt2-1$ appearing in the definition of 
$\s_{\psi}(\kk)$ as:
$$\sqrt2-1=(\sqrt2-1+{\n\over 2})-{\n\over 2}\defin t_\psi-{\n\over 2}
\;,\Eq(nu)$$
where $\n$ is a parameter 
to be properly chosen later as a function of $\l$,
in such a way that the average location of the critical points
will be given by $t_\l=t_\psi$; in other words 
$\n$ has the role of a {\it counterterm} fixing 
the middle point of the critical temperatures. The splitting \equ(nu)
induces the following splitting of $P(d\psi)$:
$$P(d\psi)=P_\s(d\psi)e^{-\n F_\n(\psi)}\virg
F_\n(\psi)\defin {1\over 2M^2}\sum_{\kk,\o}
(-i\o)\hat\psi^+_{\o,\kk}\hat\psi^-_{-\o,\kk}\;,\Eq(ppsi)$$
where $P_\s(d\psi)$
is given by \equ(2.27) with $\phi=\psi$ and 
$\s\defin 2(1-t_\psi/t_\l)$ replacing $\s_\psi({\bf 0})$.

\*
\sub(2.6)  {\it Integration of the $\chi$ variables}.
The propagators $<\phi_{\xx,\o}^\s\phi_{\yy,\o'}^{\s'}>$ 
of the fermionic integration $P(d\phi)$ verify
the following bound, for some $A,\k>0$:
$$|<\phi_{\xx,\o}^\s\phi_{\yy,\o'}^{\s'}>|\le A e^{-\k \bar m_\phi
|\xx-\yy|}\;,\Eq(b)$$
where $\bar m_\phi$ is the minimum between
$|m^{(1)}_{\phi}|$ and $|m^{(2)}_{\phi}|$ and, for $j=1,2$, 
$m^{(j)}_{\phi}$ is given by $m^{(j)}_{\phi}\defin
2(t_\l^{(j)}-t_\phi)/t_\l$, $j=1,2$.
Note that both $m^{(1)}_{\c}$
and $m^{(2)}_{\c}$ are $O(1)$.
This suggests to integrate first the $\chi$ variables.\\

After the integration of the $\c$ variables we shall rewrite \equ(2.26)
as
$$\Xi^-_{AT}=e^{-M^2 E_1}\int P_{Z_1,\s_1,\m_1,C_1}(d\psi)
e^{-\VV^{(1)}(\sqrt{Z_1}\psi)}\;,
\qquad \VV^{(1)}(0)=0\;,\Eq(2.36)$$
where $C_1(\kk)\=1$, $Z_1=t_\psi$, $\s_1={\s/(1-{\s\over 2})}$,
$\m_1={\m/(1-{\s\over 2})}$
and $P_{Z_1,\s_1,\m_1,C_1}(d\psi)$ is the exponential 
of a quadratic form:
$$\eqalign{& P_{Z_1,\s_1,\m_1,C_1}(d\psi)=\NN_1^{-1}
\prod_{\kk\in D_{-,-}}^{\o=\pm 1}
d\psi^{+}_{\o,\kk}
d\psi^{-}_{\o,\kk}
\exp\Bigl[-{1 \over 4 M^2}\sum_{\kk\in D_{-,-}}
Z_1 C_1(\kk)
\Psi_\kk^{+,T} A_\psi^{(1)}(\kk)\Psi_\kk\Bigr]\;,\cr
&A_\psi^{(1)}(\kk)=\left( \matrix{M^{(1)}(\kk)& N^{(1)}(\kk)\cr
N^{(1)}(\kk)&M^{(1)}(\kk)\cr}\right)\cr
&M^{(1)}(\kk)=\left(\matrix{i \sin k+\sin k_0+a^+_1(\kk)& 
-i\left(\s_1+c_1(\kk)\right)\cr
i\left(\s_1+c_1(\kk)\right) & i \sin k-\sin k_0+a^-_1(\kk) 
\cr}\right)\cr
&N^{(1)}(\kk)=\left(\matrix{b^+_1(\kk)&i\left(\m_1+d_1(\kk)\right)\cr
-i\left(\mu_1+d_1(\kk)\right)& b^-_1(\kk)\cr}
\right)\;,\cr}\Eq(2.38)$$
where $\NN_1$ is chosen 
in such a way that $\int P_{Z_1,\s_1,\m_1,C_1}(d \psi)=1$.
Moreover $\VV^{(1)}$ is the {\it interaction}, which can be expressed as a sum
of monomials in $\psi$ of arbitrary order:
$$\VV^{(1)}(\psi)=\sum_{n=1}^\io
\sum_{\kk_1,\ldots,\kk_{2n}\atop\aa,\oo}
\prod_{i=1}^{2n}\hat\psi^{\a_i(\le 1)}_{\o_i,\kk_i}
\widehat W_{2n,\aa,\oo}^{(1)}(\kk_1,\ldots,\kk_{2n-1})
\d(\sum_{i=1}^{2n}\a_i\kk_i)\Eq(vv1)$$
and $\d(\kk)=\sum_{\nn\in\zzz^2}\d_{\kk,2\p\nn}$. The constant $E_1$ in 
\equ(2.36), the functions $a^\pm_1,b^\pm_1,c_1,d_1$ in \equ(2.38)
and the kernels $\widehat W_{2n,\aa,\oo}^{(1)}$
in \equ(vv1) have the properties described in the following Theorem, 
proved in Appendix \secc(a1). 
Note that from now on we will consider all functions
appearing in the theory as functions of $\l,\s_1,\m_1$ (of course
$t$ and $u$ can be analytically and elementarily expressed in terms 
of $\l,\s_1,\m_1$). We shall also assume $|\s_1|,|\m_1|$ bounded by some $O(1)$
constant.
Note that if $t\pm u$ belong
to a sufficiently small interval $D$ centered around $\sqrt2-1$, 
as assumed in the hypothesis of the Main Theorem in \sec(1), then of course 
$|\s_1|,|\m_1|\le c_1$ for a suitable constant $c_1$ (in 
particular, if $D$ is chosen as in Remark (1) following the Main Theorem, 
we find $|\s_1|\le 1+O(\e_1)$ and $|\m_1|\le 2+O(\e_1)$).
\\

{\bf Theorem 2.1} {\it Assume that $|\s_1|,|\m_1|\le c_1$
for some constant $c_1>0$.
There exist a constant
$\e_1$ such that,
if $|\l|,|\n|\le \e_1$, 
then $\Xi^-_{AT}$ can be written as in \equ(2.36), \equ(2.38), \equ(vv1),
where:\\ 
\01) $E_1$ is an $O(1)$ constant;\\
\02) $a^\pm_1(\kk), b^\pm_1(\kk)$ are analytic odd
functions of $\kk$ and $c_1(\kk),d_1(\kk)$ real analytic  
even functions of $\kk$; in a neighborhood of $\kk= {\bf 0}$,  
$a_1^\pm(\kk)=O(\s_1\kk)
+O(\kk^3)$, $b_1^{\pm}(\kk)=O(\m_1\kk)+O(\kk^3)$,
$c_1(\kk)=O(\kk^2)$ and $d_1(\kk)=O(\m_1\kk^2)$;\\ 
\03) the determinant 
$|\det A_\psi(\kk)|$ can be bounded above and below by some
constant times $\big[(\s_1-\m_1)^2+|c(\kk)|\big]\big[(\s_1+\m_1)^2+|c(\kk)|
\big]$ 
and $c(\kk)=\cos k_0+\cos k-2$;\\ 
\04) $\widehat W_{2n,\aa,\oo}^{(1)}$ are analytic functions of
$\kk_i,\l,\n,\s_1,\m_1$, $i=1,\ldots,2n$ and, for some constant $C$,
$$|\widehat W_{2n,\aa,\oo}^{(1)}(\kk_1,\ldots,\kk_{2n-1})|
\le M^2 C^n|\l|^{\max\{1,n/2\}}\;;
\Eq(2.40)$$
\04--a) the terms in \equ(2.40) with $n=2$ can be written as
$$\eqalign{
&L_1\sum_{\kk_1,\ldots,\kk_4} \hat\psi^{+}_{1,\kk_1}
\hat\psi^{+}_{-1,\kk_2}\hat\psi^{-}_{-1,\kk_3}\hat\psi^{-}_{1,
\kk_4}\d(\kk_1+\kk_2-\kk_3-\kk_4)+
\cr &+\sum_{\kk_1,\ldots,\kk_4}\sum_{\aa,\oo}
\widetilde W_{4,
\underline\a,\underline\o}(\kk_1,\kk_2,\kk_3)
\hat\psi^{\a_1}_{\o_1,\kk_1}\hat\psi^{\a_2}_{\o_2,\kk_2}
\hat\psi^{\a_3}_{\o_3,\kk_3}\hat\psi^{\a_4}_{\o_4,\kk_4}
\d(\sum_{i=1}^4\a_i\kk_i)\;,\cr}\Eq(8.29)$$
where $L_1$ is real and $\widetilde W_{4,
\underline\a,\underline\o}(\kk_1,\kk_2,\kk_3)$ vanishes at
$\kk_1=\kk_2=\kk_3=\left({\p\over M},{\p\over
M}\right)$;\\
\04--b)
the term in \equ(2.40) with $n=1$ can be written as:
$$\eqalign{
&{1\over 4}\sum_{\o,\a=\pm}\
\sum_\kk \Bigl[S_1(-i\o)\hat\psi^{+}_{\o,\kk}\hat
\psi^{-}_{-\o,\kk}+ M_1(i\o)\hat\psi^{\a}_{\o,\kk}\hat\psi^{\a}_{
-\o,-\kk}+F_1(i\sin k+\o\sin k_0)\hat\psi^{\a}_{\o,\kk}\hat\psi^{\a
}_{\o,-\kk}+\cr
&+G_1(i\sin k+\o\sin k_0)\hat\psi^{+}_{\o,\kk}\hat\psi^{-
}_{\o,\kk}\Bigr]
+\sum_{\kk}\sum_{\aa,\oo}\widetilde
W_{2,\underline\a,\underline\o}(\kk)
\hat\psi^{\a_1}_{\o_1,\kk}
\hat\psi_{\o_2,-\a_1\a_2\kk}^{\a_{2}}\cr}\Eq(8.30)$$
where: $\widetilde
W_{2,\underline\a,\underline\o}(\kk)$ is $O(\kk^2)$ in a neighborhood 
of $\kk={\bf 0}$; $S_1, M_1, F_1,G_1$ are real analytic functions of
$\l,\s_1,\m_1,\nu$ s.t. $F_1=O(\l\m_1)$ and
$$\eqalign{&L_1=l_1+O(\l\s_1)+O(\l\m_1)\virg
S_1=s_1+\g n_1+O(\l\s_1^2)+O(\l\m_1^2)\cr
&M_1=m_1+O(\l\m_1\s_1)+O(\l\m_1^3)\virg
G_1=z_1+O(\l\s_1)+O(\l\m_1)\cr}\Eq(8.30a)$$
with $s_1=\s_1 f_1$, $m_1=\m_1 f_2$ and
$l_1, n_1, f_1, f_2, z_1$ independent of $\s_1,\m_1$;
moreover $l_1=\l/Z_1^2+O(\l^2)$, 
$f_1,f_2=O(\l)$, $\g n_1=\n/Z_1+c^\n_1\l+O(\l^2)$, for some $c^\n_1$ 
independent of $\l$, and $z_1=O(\l^2)$.}

\*
\0{\bf Remark.} 
The meaning of Theorem 2.1 is that after the integration of the $\chi$
fields we are left with a fermionic integration similar
to \equ(2.27) up to corrections which are at least $O(\kk^2)$,
and an effective interaction containing terms with any number
of fields. 

{\it A priori} many 
bilinear terms with kernel $O(1)$ or $O(\kk)$ with respect to
$\kk$ near $\kk={\bf 0}$ 
could be generated by the $\chi$--integration
besides the ones originally present in \equ(2.27); however
{\it symmetry considerations restrict drastically the number
of possible bilinear terms} $O(1)$ or $O(\kk)$. 
Only one new term of the form
$\sum_\kk(i\sin k+\o\sin k_0)\hat\psi^{\a}_{\o,\kk}\hat\psi^{\a}_{\o,-\kk}$
appears, which is ``dimensionally'' {\it marginal}
in a RG sense; however it is weighted by a constant 
$O(\l\m_1)$ and this will improve its 
``dimension'', so that it will result to be {\it irrelevant},
see \sec(4.2) below.\\

\*
\section(4,Integration of the {$\psi$} variables: first regime)

\sub(4.1) {\it Multiscale analysis}.
From the bound on $\det A^{(1)}_\psi(\kk)$ described in Theorem 2.1, 
we see that the $\psi$ fields have a mass 
given by $\min\{|\s_1-\m_1|,|\s_1+\m_1|\}$, which can be arbitrarly small; 
their integration in the
infrared region (small $\kk$) needs a multiscale analysis.
We introduce a {\sl scaling parameter} $\g>1$ which will be used to
define a geometrically growing sequence of length scales
$1,\g,\g^2,\ldots$, \ie of geometrically decreasing momentum
scales $\g^h,\,h=0,-1,-2,\ldots$
Correspondingly we introduce 
$C^\io$ compact support functions $f_h(\kk)$ $h\le 1$, 
with the following properties: if $|\kk|\defin\sqrt{\sin^2 k+\sin^2 k_0}$, 
when $h\le 0$, $f_h(\kk) = 0$ for $|\kk| <\g^{h-2}$ or $|\kk|
>\g^{h}$, and $f_h(\kk)= 1$, if $|\kk| =\g^{h-1}$; 
$f_1(\kk)=0$ for $|\kk|\le\g^{-1}$ and $f_1(\kk)=1$ for $|\kk|\ge 1$;
furthermore:
$$1=\sum_{h=h_{M} }^1 f_h(\kk)\virg {\rm where:}\qquad h_M=\min
\{h:\g^{h}> \sqrt2\sin{\p\over M}\}
\;,\Eq(4.2)$$
and $\sqrt2\sin(\p/ M)$ is the smallest momentum allowed by
the antiperiodic boundary conditions, \ie $\sqrt2\sin(\p/ M)=\min_{\kk\in 
D_{-,-}}|\kk|$.

The purpose is to perform the integration of \equ(2.38) over the 
fermion fields in an iterative way.
After each iteration we shall be left with a ``simpler''
Grassmannian integration to perform: if
$h=1,0,-1,\ldots,h_M$, we shall write
$$\Xi^-_{AT}=\int P_{Z_h,\s_h,\m_h,C_h}(d\psi^{(\le h)}) \, e^{-\VV^{(h)}
(\sqrt{Z_h}\psi^{(\le h)})-M^2 E_h}\;,\quad \VV^{(h)}(0)=0\;,\Eq(4.3)$$
where the quantities $Z_h$,
$\s_h$, $\m_h$, $C_h$, $P_{Z_h,\s_h,\m_h,C_h}(d\psi^{(\le h)})$, 
$\VV^{(h)}$ and $E_h$ have to be defined recursively and the result of
the last iteration will be $\Xi_{AT}^-=e^{-M^2 E_{-1+h_M}}$, 
\ie the value of the partition function. 

$P_{Z_h,\s_h,\m_h,C_h}(d\psi^{(\le h)})$ is defined by \equ(2.38) in which 
we replace $Z_1,\s_1,\m_1,a^\o_1, b^\o_1,c_1,d_1,C_1(\kk)$ with
$Z_h,\s_h,\m_h,a^\o_h, b^\o_h,c_h,d_h,C_h(\kk)$, where
$C_h(\kk)^{-1}=\sum_{j=h_{M}}^hf_j(\kk)$. Moreover
$$\eqalign{\VV^{(h)}(\psi)&= \sum_{n=1}^\io{1\over M^{2n}}
\sum_{\kk_1,\ldots,\kk_{2n-1},\atop \aa,\oo}
\prod_{i=1}^{2n}\hat\psi^{\a_i(\le h)}_{\o_i,\kk_i}
\widehat W_{2n,\aa,\oo}^{(h)}
(\kk_1,\ldots,\kk_{2n-1})\d(\sum_{i=1}^{2n}\a_i\kk_i)\defin\cr
&\defin
\sum_{n=1}^\io
\sum_{\xx_1,\ldots,\xx_{2n},\atop \ss,\underline j,\oo,\aa}
\prod_{i=1}^{2n}\partial^{\s_i}_{j_i}\psi^{\a_i(\le h)}_{\o_i,\xx_i}
W_{2n,\ss,\underline j,\aa,\oo}^{(h)}(\xx_1,\ldots,\xx_{2n})\;,\cr}\Eq(4.6)$$
where in the last line $j_i=0,1$, $\s_i\ge 0$ and 
$\dpr_j$ is the forward discrete derivative in the $\hat e_j$ direction.

Note that the field $\psi^{(\le h)}$, whose propagator is given by the 
inverse of $Z_h C_h(\kk)A^{(h)}_\psi$, 
has the same support of $C_h^{-1}(\kk)$, that is 
on a strip of width $\g^h$ around the singularity $\kk={\bf 0}$. The field
$\psi^{(\le 1)}$ coincides with the field $\psi$ of previous section, so that
\equ(2.36) is the same as \equ(4.3) with $h=1$.  

It is crucial for the following to think 
$\widehat W_{2n,\aa,\oo}^{(h)}$, $h\le 1$, as functions of the 
variables $\s_k(\kk),\m_k(\kk)$,
$k=h, h+1,\ldots,0,1$, $\kk\in D_{-,-}$. The iterative construction below
will inductively imply that the dependence on these variables 
is well defined (note that for $h=1$
we can think the kernels of $\VV^{(1)}$ as functions of $\s_1,\m_1$, see 
Theorem 2.1). 

\*
\sub(4.2) {\it The localization operator.}
We now begin to describe the iterative construction leading to \equ(4.3).
The first step consits in defining a {\it localization} operator $\LL$ 
acting on the kernels of $\VV^{(h)}$, in terms of which we shall rewrite
$\VV^{(h)}=\LL\VV^{(h)}+\RR\VV^{(h)}$, where $\RR=1-\LL$. The iterative 
integration procedure will use such splitting, see \sec(4.3) below.
 
$\LL$ will be non zero only if 
acting on a kernel $\widehat W_{2n,\aa,\oo}^{(h)}$ with $n=1,2$. In this case  
$\LL$ will be the combination of four different operators:
$\LL_j$, $j=0,1$, 
whose effect on a function of $\kk$ 
will be essentially to extract the term of order $j$
from its Taylor series in $\kk$; and $\PP_j$, $j=0,1$, 
whose effect on a functional of the sequence
$\s_h(\kk),\m_h(\kk),\ldots,\s_1,\m_1$ will be essentially 
to extract the term of order $j$ from its power series 
in $\s_h(\kk),\m_h(\kk),\ldots,\s_1,\m_1$.

The action of $\LL_j$, $j=0,1$, on the kernels 
$\widehat W_{2n,\aa,\oo}^{(h)}(\kk_1,\ldots,\kk_{2n})$
is defined as follows.\\
\\
\01) If $n=1$, 
$$\eqalign{&\LL_0\widehat W_{2,\aa,\oo}^{(h)}(\kk,\a_1\a_2\kk)=\fra14 
\sum_{\h,\h'=\pm 1}\widehat W_{2,\aa,\oo}
^{(h)}(\bk\h{\h'},\a_1\a_2\bk\h{\h'})
\cr
&\LL_1\widehat W_{2,\aa,\oo}^{(h)}(\kk,\a_1\a_2\kk)=\fra14 
\sum_{\h,\h'=\pm 1}\widehat W_{2,\aa,\oo}
^{(h)}(\bk\h{\h'},\a_1\a_2\bk\h{\h'})
\big[\h {\sin k\over \sin{\p\over M}}  +
\h'{\sin k_0\over \sin{\p\over M}}\big]\;,\cr}\Eq(4.7)$$
where $\bk\h{\h'} = \left(\h{\p\over M},\h'{\p\over
M}\right)$ are the smallest momenta allowed by
the antiperiodic boundary conditions.\\
\\
\02) If $n=2$, $\LL_1\widehat W_{4,\aa,\oo}^{(h)}=0$ and
$$\LL_0 \widehat W_{4,\aa,\oo}^{(h)}(\kk_1,\kk_2,\kk_3,\kk_4)\defin
\widehat W_{4,\aa,\oo}^{(h)}(\bk++,\bk++,\bk++,\bk++)\;.\Eq(4.8)$$
\03) If $n>2$, $\LL_0\widehat W_{2n,\aa,\oo}=\LL_1\widehat W_{2n,\aa,\oo}
=0$.\\

The action of $\PP_j$, $j=0,1$, on the kernels $\widehat W_{2n,\aa,\oo}$,
thought as functionals of the sequence $\s_h(\kk),\m_h(\kk),\ldots,\s_1,\m_1$
is defined as follows. 
$$\eqalign{&
\PP_0 \widehat W_{2n,\aa,\oo}\defin
\widehat W_{2n,\aa,\oo}\Big|_{\ss^{(h)}=\underline\m^{(h)}=0}\cr
&\PP_1 \widehat W_{2n,\aa,\oo}\defin\sum_{k\ge h,\kk}\Big[
\s_{k}(\kk){\dpr \widehat W_{2n,\aa,\oo}\over \dpr\s_k(\kk)}
\Big|_{\ss^{(h)}=\underline\m^{(h)}=0}+
\m_{k}(\kk){\dpr \widehat W_{2n,\aa,\oo}\over \dpr\m_k(\kk)}
\Big|_{\ss^{(h)}=\underline\m^{(h)}=0}\Big]\;.\cr}\Eq(4.9)$$
Given $\LL_j,\PP_j$, $j=0,1$ as above, we define the action of $\LL$
on the kernels $\widehat W_{2n,\aa,\oo}$ as follows.\\
\\
\01) If $n=1$, then
$$\LL \widehat W_{2,\aa,\oo}\defin\cases{
\LL_0(\PP_0+\PP_1)
\widehat W_{2,\aa,\oo} & if $\o_1+\o_2=0$ and $\a_1+\a_2=0$,\cr
\LL_0\PP_1
\widehat W_{2,\aa,\oo} & if $\o_1+\o_2=0$ and $\a_1+\a_2\not=0$,\cr
\LL_1\PP_0\widehat W_{2,\aa,\oo} & if $\o_1+\o_2\not=0$ and 
$\a_1+\a_2=0$,\cr
0 & if $\o_1+\o_2\not=0$ and $\a_1+\a_2\not=0$.}$$
\02) If $n=2$, then
$\LL \widehat W_{4,\aa,\oo}\defin \LL_0\PP_0\widehat W_{4,\aa,\oo}$.\\
\03) If $n>2$, then $\LL \widehat W_{2n,\aa,\oo}=0$.\\ 

Finally, the effect of $\LL$ on $\VV^{(h)}$ is, by definition,
to replace on the r.h.s. of \equ(4.6) $\widehat W_{2n,\aa,\oo}$ with 
$\LL\widehat W_{2n,\aa,\oo}$. Note that $\LL^2\VV^{(h)}=\LL\VV^{(h)}$.

Using the previous definitions we get the following result, proven in Appendix
\secc(A1.2). We use the notation $\ss^{(h)}=\{\s_k(\kk)\}^{k=h,\ldots,1}_{
\kk\in D_{-,-}}$ and $\underline\m^{(h)}=\{\m_k(\kk)\}^{k=h,\ldots,1}_{
\kk\in D_{-,-}}$.
\*
{\bf Lemma 3.1.} {\it Let the action of $\LL$ on $\VV^{(h)}$ 
be defined as above. Then
$$\LL\VV^{(h)}(\psi^{(\le h)})=(s_h+\g^h n_h) F_\s^{(\le h)}
+m_h F^{(\le h)}_{\m}+l_h
F_\l^{(\le h)} +z_{h} F_\z^{(\le h)}
\;,\Eq(4.10) $$
where $s_h,n_h,m_h,l_h$ and $z_h$ are real constants and:
$s_h$ is linear in $\ss^{(h)}$ and independent of $\underline\m^{(h)}$; 
$m_h$ is linear in $\underline\m^{(h)}$ and independent of $\ss^{(h)}$;
$n_h,l_h,z_h$ are independent of $\ss^{(h)},\underline\m^{(h)}$;
moreover, 
if $D_h\defin D_{-,-}\cap\{\kk:C_h^{-1}(\kk)>0\}$,
$$\eqalign{
F_\s^{(\le h)}(\psi^{(\le h)})&=
{1\over 2 M^2}\sum_{\kk\in D_h}\sum_{\o=\pm 1} (-i\o)
\widehat\psi^{+(\le h)}_{\o,\kk}
\widehat\psi^{-(\le h)}_{-\o,\kk}\ \ \defin\ \ 
{1\over M^2}\sum_{\kk\in D_h}
\widehat F_\s^{(\le h)}(\kk)\;,\cr
F_\m^{(\le h)}(\psi^{(\le h)})&=
{1\over 4M^2}\sum_{\kk\in D_h}\sum_{\a,\o=\pm 1} i\o
\widehat\psi^{\a(\le h)}_{\o,\kk}
\widehat\psi^{\a(\le h)}_{-\o,-\kk}\ \ \defin\ \ 
{1\over M^2}\sum_{\kk\in D_h}
\widehat F_\m^{(\le h)}(\kk)\;,\cr
F_\l^{(\le h)}(\psi^{(\le h)})&={1\over M^8}\sum_{\kk_1,...,\kk_4
\in D_h}
\widehat\psi^{+(\le h)}_{1,\kk_1}
\widehat\psi^{+(\le h)}_{-1,\kk_2} \widehat\psi^{-(\le h)}_{-1,\kk_3}
\widehat\psi^{-(\le h)}_{1,\kk_4}\d(\kk_1+\kk_2-\kk_3-\kk_4)\;\cr
F_{\z}^{(\le h)}(\psi^{(\le h)})&={1\over 2M^2}
\sum_{\kk\in D_h}\sum_{\o=\pm 1}
(i\sin k+\o\sin k_0)\widehat\psi^{+(\le h)}_{\o,\kk}
\widehat\psi^{-(\le h)}_{\o,\kk}\ \ \defin\ \ 
{1\over M^2}\sum_{\kk\in D_h}
\widehat F_\z^{(\le h)}(\kk)\;.\cr}\Eq(4.11)$$
where $\d(\kk)=M^2\sum_{{\bf n}\in\zzz^2}\d_{\kk,2\p{\bf n}}$.}\\

{\bf Remark.}
The application of $\LL$ to the kernels of the effective potential generates
the sum in \equ(4.10), \ie a linear combination of the Grassmannian
monomials in \equ(4.11) which, in the renormalization group language,
are called ``{\it relevant}'' (the first two) or ``{\it marginal}''
operators (the two others).\\

We now consider the operator $\RR\defin 1-\LL$. The following result
holds, see Appendix \secc(a1) for the proof. We use the notation
$\RR_1=1-\LL_0$, $\RR_2=1-\LL_0-\LL_1$, $\SS_1=1-\PP_0$, $\SS_2=1-\PP_0-
\PP_1$.
\*
{\bf Lemma 3.2.} {\it The action of $\RR$ on $\widehat W_{2n,\aa,\oo}$
for $n=1,2$ is the following.\\
\01) If $n=1$, then
$$\RR \widehat W_{2,\aa,\oo}=\cases{
[\SS_2+\RR_2(\PP_0+\PP_1)]
\widehat W_{2,\aa,\oo} & if $\o_1+\o_2=0$,\cr 
[\RR_1\SS_1+\RR_2\PP_0]
\widehat W_{2,\aa,\oo} & if 
$\o_1+\o_2\not=0$ and $\a_1+\a_2=0$,\cr
\RR_1\SS_1\widehat W_{2,\aa,\oo} & if 
$\o_1+\o_2\not=0$ and $\a_1+\a_2\not=0$,\cr}$$
\02) If $n=2$, then
$\RR \widehat W_{4,\aa,\oo}= [\SS_1+\RR_1\PP_0]
\widehat W_{4,\aa,\oo}$.}\\
\\
{\bf Remark.}
The effect of $\RR_j$, $j=1,2$ on $\widehat W_{2n,\aa,\oo}^{(h)}$ 
consists in  
extracting the rest of a Taylor series in $\kk$ of order $j$. 
The effect of $\SS_j$, $j=1,2$ on $\widehat W_{2n,\aa,\oo}^{(h)}$ 
consists in extracting the rest of a
power series in $(\ss^{(h)},\underline\m^{(h)})$ 
of order $j$. The definitions are given in such a way that 
$\RR\widehat W_{2n,\aa,\oo}$ is at least quadratic in $\kk,\ss^{(h)},
\underline\m^{(h)}$ if $n=1$ and at least linear in $\kk,\ss^{(h)},
\underline\m^{(h)}$ when $n=2$. This will give dimensional gain
factors in the bounds for $\RR\widehat W_{2n,\aa,\oo}^{(h)}$ w.r.t. 
the bounds for $\widehat W_{2n,\aa,\oo}^{(h)}$, $n=1,2$, as we shall see in 
details in Appendix \secc(4a).\\

\*
\sub(4.3) {\it Renormalization.}
Once that the above definitions are given we can describe our
integration procedure for $h\le 0$.
\*
We start from \equ(4.3) and we rewrite it as
$$\int P_{Z_h,\s_h,\m_h,C_h}(d\psi^{(\le h)}) \, e^{-\LL\VV^{(h)}
(\sqrt{Z_h}\psi^{(\le h)})-\RR\VV^{(h)}
(\sqrt{Z_h}\psi^{(\le h)}) -M^2 E_h}\;,\Eq(4.3a)$$
with $\LL\VV^{(h)}$ as in \equ(4.10). 
Then we include the quadratic part of  
$\LL\VV^{(h)}$ (except the term proportional to $n_h$)
in the fermionic integration, so obtaining
$$\int P_{\widehat Z_{h-1},\s_{h-1},\m_{h-1},C_h}(d\psi^{(\le h)}) \, e^{-l_h 
F_\l(\sqrt{Z_h}\psi^{(\le h)})-\g^h n_h F_\s(\sqrt{Z_h}\psi^{(\le h)})-
\RR\VV^{(h)}
(\sqrt{Z_h}\psi^{(\le h)}) -M^2 E_h}\;,\Eq(4.3aa)$$
where $\widehat Z_{h-1}(\kk)\defin Z_h (1+z_h C_h^{-1}(\kk))$ and
$$\eqalign{
&\s_{h-1}(\kk)\defin{Z_h\over\widehat Z_{h-1}(\kk)}
(\s_h(\kk)+s_hC_h^{-1}(\kk))\virg
\m_{h-1}(\kk)\defin {Z_h\over \widehat Z_{h-1}(\kk)}
(\m_h(\kk)+m_h C_h^{-1}(\kk))\cr
&a^\o_{h-1}(\kk)\defin{Z_h\over \widehat Z_{h-1}(\kk)}
a^\o_h(\kk)\virg
b^\o_{h-1}(\kk)\defin {Z_h\over \widehat Z_{h-1}(\kk)}
b^\o_h(\kk)\cr
&c_{h-1}(\kk)\defin{Z_h\over\widehat Z_{h-1}(\kk)}c_h(\kk)\virg
d_{h-1}(\kk)\defin{Z_h\over\widehat Z_{h-1}(\kk)}d_h(\kk)\;.\cr}\Eq(4.17a)$$
The integration in \equ(4.3aa) differs from the one in \equ(4.3)
and \equ(4.3a): $P_{\widehat Z_{h-1},
\s_{h-1},\m_{h-1},C_h}$\\ is defined by \equ(2.38)
with $Z_1$ and 
$A^{(1)}_\psi$ replaced by $\widehat Z_{h-1}(\kk)$ and 
$A^{(h-1)}_\psi$.

\*
Now we can perform the integration of the $\psi^{(h)}$ field.
It is convenient to rescale the fields:
$${\widehat\VV}^{(h)}(\sqrt{Z_{h-1}}\psi^{(\le h)})\ \defin\ 
\l_h F_\l(\sqrt{Z_{h-1}}\psi^{(\le h)})+
\g^h\n_h F_\s(\sqrt{Z_{h-1}}\psi^{(\le h)})
+\RR\VV^{(h)}(\sqrt{Z_h}\psi^{(\le h)})
\;,\Eq(4.18)$$
where $\l_h=\big({Z_h\over Z_{h-1}}\big)^2l_h$, $\n_h={Z_h\over Z_{h-1}}n_h$
and $\RR\VV^{(h)}=(1-\LL)\VV^{(h)}$ is the irrelevant part of $\VV^{(h)}$, and
rewrite \equ(4.3aa) as
$$e^{-M^2(t_h+E_h)}\int
P_{Z_{h-1},\s_{h-1},\m_{h-1},C_{h-1}}(d\psi^{(\le h-1)}) \, 
\int P_{Z_{h-1},\s_{h-1},\m_{h-1},
{\widetilde f}^{-1}_h}
(d\psi^{(h)})\, e^{-\widehat
\VV^{(h)}(\sqrt{Z_{h-1}}\psi^{(\le h)})}
\Eq(4.20)$$
where we used the decomposition $\psi^{(\le h)}=\psi^{(\le h-1)}+\psi^{(h)}$
(and $\psi^{(\le h-1)},\psi^{(h)}$ are independent) and 
${\widetilde f}_h(\kk)$ is defined by the relation $C_h^{-1}(\kk)\widehat 
Z_{h-1}^{-1}(\kk)=C_{h-1}^{-1}(\kk)Z_{h-1}^{-1}+\widetilde f_h(\kk)
Z_{h-1}^{-1}$, namely:
$$\widetilde f_h(\kk)\ \defin\ Z_{h-1}\Bigl[
{C_h^{-1}(\kk)\over \widehat Z_{h-1}
(\kk)}-{C_{h-1}^{-1}(\kk)\over Z_{h-1}}\Bigr]=f_h(\kk)\Bigl[
1+ {z_hf_{h+1}(\kk)\over
1+z_hf_h(\kk)}
\Bigr]\;.\Eq(4.21)$$
Note that $\widetilde f_h(\kk)$ has the same support as $f_h(\kk)$.
Moreover 
$P_{Z_{h-1},\s_{h-1},\m_{h-1},{\widetilde f}^{-1}_h}
(d\psi^{(h)})$ is defined in the same way as 
$P_{\widehat Z_{h-1},\s_{h-1},\m_{h-1},C_h}
(d\psi^{(h)})$, with $\widehat Z_{h-1}(\kk)$ resp. $C_h$
replaced by $Z_{h-1}$ resp. $\widetilde f^{-1}_h$. 
The {\it single scale} propagator is
$$\int P_{Z_{h-1},\s_{h-1},\m_{h-1},\widetilde f_h^{-1}}(d\psi^{(h)})\,
\psi^{\a(h)}_{\xx,\o}\psi^{\a'(h)}_{\yy,\o'} ={1\over Z_{h-1}}
g^{(h)}_{\underline a,\underline a'}(\xx-\yy)\virg 
\underline a=(\a,\o)\virg \underline a'=(\a',\o')\;,\Eq(4.210a)$$
where 
$$g^{(h)}_{\underline a,\underline a'}(\xx-\yy)={1\over 2M^2}
\sum_{\kk}e^{i\a\a'\kk(\xx-\yy)}
\widetilde f_h(\kk)[A_\psi^{(h-1)}(\kk)]^{-1}_{j(\underline a),
j'(\underline a')}\Eq(4.210b)$$
with $j(-,1)=j'(+,1)=1$, $j(-,-1)=j'(+,-1)=2$, $j(+,1)=j'(-,1)=3$ and
$j(+,-1)=j'(-,-1)=4$. One finds that 
$g^{(h)}_{\underline a,\underline a'}(\xx)=g_{\o,\o'}^{(1,h)}(\xx)-\a\a'
g_{\o,\o'}^{(2,h)}(\xx)$, where 
$g_{\o,\o'}^{(j,h)}(\xx)$, $j=1,2$ are defined in Appendix \secc(a3),
see \equ(4.24).

The long distance behaviour of the propagator is given by the following 
Lemma, proved in Appendix \secc(a3).

\*
{\bf Lemma 3.3.} {\it Let $\s_h\defin\s_h({\bf 0})$ and $\m_h\defin
\m_h({\bf 0})$ and assume $|\l|\le \e_1$ for a small constant $\e_1$. 
Suppose that for $h>\bar h$
$$|z_h|\le {1\over 2}\virg |s_h|\le {1\over 2}|\s_h|\virg 
|m_h|\le {1\over 2}|\m_h|\;,\Eq(4.40yz)$$
that there exists $c$ s.t. 
$$e^{-c|\l|}\le \Big|{\s_h\over\s_{h-1}}\Big|\le e^{c|\l|} 
\virg
e^{-c|\l|}\le \Big|{\m_h\over\m_{h-1}}\Big|\le e^{c|\l|}\virg
e^{-c|\l|^2}\le \Big|{Z_h\over Z_{h-1}}\Big|\le e^{c|\l|^2}\;,\Eq(4.40z)$$
and that, for some constant $C_1$, 
$${|\s_{\bar h}|\over\g^{\bar h}}\le C_1\virg
{|\m_{\bar h}|\over\g^{\bar h}}\le C_1\;;\Eq(4.40a)$$
then, for all $h\ge \bar h$, 
given the positive integers $N, n_0,n_1$ and putting $n=n_0+n_1$,
there exists a constant $C_{N,n}$ s.t. 
$$|\dpr^{n_0}_{x_0}\dpr^{n_1}_{x}g_{\underline a,\underline a'}^{(h)}
(\xx-\yy)|\le C_{N,n}{\g^{(1+n)h}\over 1+(\g^{h}
|\dd(\xx-\yy)|)^N}\virg{\it where}\quad
\dd(\xx)={M\over\p}\big(\sin{\p x\over M},\sin{\p x_0\over M})\;.
\Eq(4.400)$$ 
Furthermore, if $\PP_0$, $\PP_1$ are 
defined as in \equ(4.9) and $\SS_1$, $\SS_2$ are defined as in
Lemma 3.2, 
we have that $\PP_jg^{(h)}_{\underline a,
\underline a'}$, $j=0,1$ and $\SS_jg^{(h)}_{\underline a,
\underline a'}$, $j=1,2$, satisfy the same bound \equ(4.400),
times a factor $\big({|\s_{h}|+|\m_{h}|\over \g^{h}}\big)^j$.
The bounds for $\PP_0g^{(h)}_{\underline a,
\underline a'}$ and $\PP_1g^{(h)}_{\underline a,
\underline a'}$ hold even without hypothesis \equ(4.40a).}\\ 

After the integration of the field on scale $h$ we are left with an 
integral involving the fields 
$\psi^{(\le h-1)}$ and the new effective interaction
$\VV^{(h-1)}$, defined as 
$$e^{-\VV^{(h-1)}(\sqrt{Z_{h-1}}\psi^{(\le h-1)})-\tilde E_h M^2}=
\int P_{Z_{h-1},\s_{h-1},\m_{h-1},\widetilde f_h}(d\psi^{(h)})
e^{-\widehat \VV^{(h)}(\sqrt{Z_{h-1}}\psi^{(\le h)})}\;.\Eq(4.20a)$$
It is easy to see that $\VV^{(h-1)}$ is of the form \equ(4.6) and that 
$E_{h-1}=E_h+t_h+\tilde E_h$. It is sufficient to use the well known identity
$$M^2\tilde E_h+\VV^{(h-1)}(\sqrt{Z_{h-1}}\psi^{(\le h-1)})=
\sum_{n\ge 1}{1\over n!}(-1)^{n+1}\EE^T_h(\widehat\VV^{(h)}(\sqrt{Z_{h-1}}
\psi^{(\le h)});n)\;,\Eq(4.20az)$$
where $\EE^T_h(X(\psi^{(h)});n)$ 
is the truncated expectation of order $n$ w.r.t. the 
propagator $Z_{h-1}^{-1}g^{(h)}_{\underline a,\underline a'}$, defined
as
$$\EE^T_h(X(\psi^{(h)});n)={\dpr\over\dpr\l^n}\log\int P_{Z_{h-1},
\s_{h-1},\m_{h-1},\widetilde f_h}(d\psi^{(h)})e^{\l X(\psi^{(h)})}\Big|_{\l=
0}\;.\Eq(eth)$$

Note that the above procedure allow us to write the 
{\it running coupling constants} $\vec v_{h-1}=(\l_{h-1},\n_{h-1})$, $h\le 1$, 
in terms of $\vec v_k$, $h\le k\le 1$, namely
$\vec v_{h-1}=\b_h(\vec v_h,\ldots,\vec v_1)$,
where $\b_h$ is the so--called {\it Beta function}.
\\

\sub(4.6){\it Analiticity of the effective potential}.
We have expressed the effective potential $\VV^{(h)}$
in terms of the {\it running coupling constants} $\l_k,\n_k$, $k\ge h$,
and of the {\it renormalization constants} $Z_k,\m_k(\kk),\s_k(\kk)$, $k\ge h$.

In Appendix \secc(4a) we will prove the following result.
\*
{\bf Theorem 3.1.} {\it Let $\s_h\defin\s_h({\bf 0})$ and $\m_h\defin
\m_h({\bf 0})$ and assume
$|\l|\le \e_1$ for a small constant $\e_1$. 
Suppose that for $h> \bar h$ the hypothesis \equ(4.40yz),
\equ(4.40z) and \equ(4.40a) hold.
If, for some constant $c$,
$$\max_{h> \bar h}\{|\l_h|,|\n_h|\}\le c|\l|\;,\Eq(4.40)$$
then there exists $C>0$ s.t. the kernels in \equ(4.6) satisfy
$$\int d\xx_1\cdots d\xx_{2n}|W^{(\bar h)}_{2n,\ss,\underline j,\aa,\oo}
(\xx_1,\ldots,\xx_{2n})|
\le M^2 \g^{-\bar h D_k(n)} \,(C\,|\l|)^{max(1,n-1)}\Eq(4.45)$$
where $D_k(n)=-2+n+k$
and $k=\sum_{i=1}^{2n}\s_i$.

Moreover $|\tilde E_{\bar h+1}|+|t_{\bar h+1}|\le c|\l|\g^{2\bar h}$ and
the kernels of $\LL\VV^{(\bar h)}$ satisfy
$$|s_{\bar h}|\le C|\l||\s_{\bar h}|\virg
|m_{\bar h}|\le C|\l||\m_{\bar h}|\Eq(4.45y)$$
and
$$|n_{\bar h}|\le C|\l|\virg
|z_{\bar h}|\le C|\l|^2\virg|l_{\bar h}|\le C|\l|^2\;.\Eq(4.45yz)$$
The bounds \equ(4.45y) holds even if \equ(4.40a) does not hold.
The bounds \equ(4.45yz) holds even if \equ(4.40a) and the 
first two of \equ(4.40z) do not hold.}\\
 
{\bf Remarks.}\\
1) The above result immediately implies analyticity of the effective potential
of scale $h$
in the running coupling constants $\l_k,\n_k$, $k\ge h$, under the assumptions 
\equ(4.40yz), \equ(4.40z), \equ(4.40a) and \equ(4.40).\\
%2) From the result of the Theorem (in particular from \equ(4.45y))
%it is clear that {\it a posteriori}
%the hypothesis \equ(4.40yz) is not necessary. In fact, using \equ(4.45y),
%one can prove \equ(4.40yz) inductively (\equ(4.40yz) is true for $h=1$, 
%as it follows from Theorem 2.1).\\ 
2) The assumptions \equ(4.40z) and \equ(4.40) 
will be proved in \sec(5) and Appendix \secc(a5) below, solving the 
{\it flow equations} for $\vec v_h=(\l_h,\n_h)$ and $Z_h,\s_h,\m_h$,
given by $\vec v_{h-1}=\b_h(\vec v_h,\ldots,\vec v_1)$, $Z_{h-1}=Z_h(1+z_h)$
and \equ(4.17a). They will be proved to 
be true up to $h=-\io$.\\ 
%3) 
%Using \equ(4.40z)
%we see that the assumptions 
%\equ(4.40a) will hold for all the scales $\bar h$ s.t.
%
%$${|\s_{\bar h}|\over \g^{\bar h}}\le |\s_1|\g^{-(1+c|\l|)\bar h}\le C\virg
%{|\m_{\bar h}|\over \g^{\bar h}}\le |\m_1|\g^{-(1+c|\l|)\bar h}\le C\;,$$
%
%namely for all the scales 
%$\bar h\ge O(\max\{\log_\g|\s_1|^{1\over 1+c|\l|},
%\log_\g|\m_1|^{1\over 1+c|\l|}\})$. 
%Then we can apply
%Theorem 3.1 up to a scale $h^*_1$ of 
%the form $h^*_1=O(\max\{\log_\g|\s_1|^{1+O(\l)},
%\log_\g|\m_1|^{1+O(\l)}\})$, which will be defined more precisely in 
%\sec(6.100) below. The property of $h^*_1$ that we stress now is the 
%following.\\
%{\it $h^*_1$ is a scale such that 
%$C_2 \g^{h^*_1}\le |\s_{h^*_1}|+|\m_{h^*_1}|\le 
%C_1 \g^{h^*_1}$, with $C_1,C_2$ independent from $\l,\m_1,\s_1$}.\\
%or scales smaller than $h^*_1$,
%he construction described above and leading to Theorem 3.1 is no longer
%onvenient. We shall integrate the fields on scale $\le h^*_1$ with
% different iterative procedure, described in \sec(6) below. 

\*
\section(5, The flow of the running coupling constants.)

The convergence of the expansion for the effective potential
is proved by Theorem 3.1 under the hypothesis that
the running coupling constants are small, see \equ(4.40),
and that the bounds \equ(4.40yz), \equ(4.40z) and \equ(4.40a) 
are satisfied. We now want to show that, choosing $\l$ small enough and 
$\n$ as a suitable function of $\l$, such hypothesis are indeed verified.
In order to prove this, we will solve the flow equations for the 
renormalization constants (following 
from \equ(4.17a) and preceding line):
$${Z_{h-1}\over Z_h} = 1+ z_h\virg
{\s_{h-1}\over \s_h} = 1+{s_h/\s_h-z_h\over 1+z_h}\virg
{\m_{h-1}\over \m_h} = 1+{m_h/\m_h-z_h\over 1+z_h}\;,\Eq(5.4z)$$
together with those for the running coupling constants:
$$\eqalign{&\l_{h-1}=\l_h+\b^h_\l(\l_h,\n_h;\ldots;\l_1,\n_1)\cr
&\n_{h-1}=\g\n_h+\b^h_\n(\l_h,\n_h;\ldots;\l_1,\n_1)\;.\cr}\Eq(5.1)$$
The functions $\b^h_\l, \b^h_\n$ are called
the $\l$ and $\n$ components of the Beta function, see the comment after 
\equ(eth), and, by construction, are
{\it independent} of $\s_k,\m_k$, so that their convergence 
follow just from \equ(4.40) and the last of \equ(4.40z), \ie 
without assuming \equ(4.40a), see Theorem 3.1.
While for a general kernel we will
apply Theorem 3.1 just up to a finite scale $h^*_1$ (in order
to insure the validity of \equ(4.40a) with $\bar h=h^*_1$), 
we will inductively study the flow generated by 
\equ(5.1) up to scale $-\io$, and we shall prove that it is bounded
for all scales. The main result on the flows of $\l_h$ and $\n_h$,
proven in Appendix \secc(a5), is the following.\\

{\bf Theorem 4.1.} {\it If $\l$ is small enough, there exists an 
analytic function $\n^*(\l)$ independent of $t,u$
such that the running coupling constants 
$\{\l_h,\n_h\}_{h\le 1}$ with $\n_1=\n^*(\l)$
verify
$|\n_h|\le c|\l|\g^{(\th/2)h}$ and $|\l_h|\le c|\l|$.
Moreover the kernels $z_h,s_h$ and $m_h$ satisfy
\equ(4.40yz) and the solutions of the flow equations \equ(5.4z) 
satisfy \equ(4.40z).}
\\
%The key property we use to achieve 
%Theorem 4.1 is that the $\l$--component of
%the Beta function
%can be written as the sum of the Luttinger model Beta 
%function, which is vanishing (see [BGPS][GS][BM1] or [BeM1] for
%a simplified proof)
%plus a small rest, which is summable w.r.t. $h$. See Lemma A5.1 
%in Appendix \secc(a5).\\

Once that $\n_1$ is conveniently
chosen as in Theorem 4.1, one can study in more detail the flows of the 
renormalization constants. In Appendix \secc(a5) we prove the following.
\\

{\bf Lemma 4.1.} {\it If $\l$ is small enough and $\n_1$ is 
chosen as in Theorem 
4.1, the solution of \equ(5.4z) can be written as:
$$Z_h=\g^{\h_z(h-1)+F^h_\z}\virg \mu_h=\m_1\g^{\h_\m(h-1)+F^h_\m}\virg
\s_h=\s_1\g^{\h_\s(h-1)+F^h_\s}\Eq(lem4.1)$$
where $\h_z,\h_\m,\h_z$ and $F^h_\z,F^h_\m,F^h_\s$ are $O(\l)$
functions, independent of $\s_1,\m_1$. 

Moreover $\h_\s-\h_\m=-b\l+O(|\l|^2)$, $b>0$.}\\

\sub(6.100) {\it The scale $h^*_1$}.
The integration described in \sec(4) is iterated
until a scale $h^*_1$ defined in the following way: 
$$h^*_1\defin\cases{\min\big\{1,
\big[\log_\g|\s_1|^{1\over 1-\h_\s}\big]\big\} 
& if $|\s_1|^{1\over 1-\h_\s}
>2|\m_1|^{1\over 1-\h_\m}$,\cr 
\min\big\{1,\big[\log_\g|u|^{1\over 1-\h_\m}\big]\big\} & if 
$|\s_1|^{1\over 1-\h_\s}\le 2|\m_1|^{1\over 1-\h_\m}$.\cr}\Eq(hhh)$$

From \equ(hhh) it follows that 
$$C_2\g^{h^*_1}\le |\s_{h^*_1}|+|\m_{h^*_1}|
\le C_1\g^{h^*_1}\;,\Eq(5.20)$$ 
with $C_1,C_2$ independent of $\l,\m_1,\s_1$. 

This is obvious in the case $h^*_1=1$. If $h^*_1<1$ and 
$|\s_1|^{1\over 1-\h_\s}
>2|\m_1|^{1\over 1-\h_\m}$, then $\g^{h^*_1-1}=c_\s
|\s_1|^{1\over 1-\h_\s}$, with $1\le c_\s<\g$, 
so that, using the third of \equ(lem4.1), we see that
$C_2\g^{h^*_1}\le |\s_{h^*_1}|\le C_1'g^{h^*_1}$, for some $C_1',C_2=O(1)$. 
Furthermore, using also the second of \equ(lem4.1), we find
$${|\m_{h^*_1}|\over|\s_{h^*_1}|}=c_\s^{\h_\m-\h_\s}
|\m_1||\s_1|^{-{1-\h_\m\over 1-\h_\s}}
\g^{F^{h^*_1}_\m-F^{h^*_1}_\s}<1\Eq(1/3)$$ 
and \equ(5.20) follows.

If $h^*_1<1$ and $|\s_1|^{1\over 1-\h_\s}\le 2|\m_1|^{1\over 1-\h_\m}$, then
$\g^{h^*_1-1}=c_u
|u|^{1\over 1-\h_\m}$, with $1\le c_\m<\g$, 
so that, using the second of \equ(lem4.1) and $|\m_1|=O(|u|)$, we see that
$C_2\g^{h^*_1}\le |\m_{h^*_1}|\le C_1'\g^{h^*_1}$. Furthermore, 
using the third \equ(lem4.1), we find
$${|\s_{h^*_1}|\over|\m_{h^*_1}|}=c_u^{\h_\s-\h_\m}
|\s_1||u|^{-{1-\h_\s\over 1-\h_\m}}
\g^{F^{h^*_1}_\s-F^{h^*_1}_\m}<C_1''\;,\Eq(1/0)$$ 
for some $C_1''=O(1)$, and \equ(5.20) again follows.\\

\0{\bf Remark.} The specific value of $h^*_1$ is not crucial:
if we change $h^*_1$ in $h^*_1+n$, $n\in\ZZZ$, 
the constants $C_1,C_2$ in \equ(5.20) 
are replaced by different $O(1)$ constants
and the estimates below
are not qualitatively modified. Of course, the specific values 
of $C_1,C_2$ (then, the specific value of $h^*_1$) can affect 
the convergence radius of the pertubative series 
in $\l$. The optimal value of $h^*_1$ should be chosen by maximizing the
corresponding convergence radius. Since here we are not interested in 
optimal estimates, we find the choice in \equ(hhh) convenient.

Note also that $h^*_1$ is a non analytic function of $(\l,t,u)$ 
(in particular for small $u$ we have $\g^{h^*_1}\sim |u|^{1+O(\l)}$). 
As a consequence, the asymptotic expression for the specific heat
near the critical points (that we shall obtain in next section)
will contain non analytic functions of $u$ (in fact it will contain
terms depending on $h^*_1$). 
However, as explained in Remark (3) after the Main Theorem, this does not
imply that $C_v$ is non analytic: it is clear that in this case
the non analyticity 
is introduced ``by hands'' by our specific choice of $h^*_1$.
\\

From the results of Theorem 4.1 and Lemma 4.1, together with \equ(hhh) and 
\equ(5.20), it follows that the assumptions of Theorem 3.1
are satisfied for any $\bar h\ge h^*_1$. The integration of the scales 
$\le h^*_1$ must be performed in a different way, as discussed in 
next section.

\*
\section(6,Integration of the {$\psi$} variables: second regime)

\sub(6.aaa){\it Integration of the $\psi^{(1)}$ field.}
If $h^*_1$ is fixed as in \sec(6.100), we can use Theorem 3.1 up to
the scale $\bar h=h^*_1+1$.  

Once that all the scales $> h^*_1$ are integrated out,
it is more convenient to describe the system in terms of
the fields $\psi^{(1)}_{\o},\psi^{(2)}_{\o}$, $\o=\pm 1$,  
defined through the following
change of variables:
$$\hat\psi^{\a(\le h^*_1)}_{\o,\kk}= {1\over\sqrt{2}} (\hat\psi^{(1,
\le h^*_1)}_{\o,-\a\kk}-i\a \hat\psi^{(2,\le h^*_1)}_{\o,-\a\kk})\;,
\quad \psi^{(j)}_{\o,\xx}={1\over M^2}\sum_{\kk}e^{-i\kk\xx}\hat\psi^{
(j)}_{\o,\kk}\;.\Eq(4.21b)$$
If we perform this change of variables, we find 
$P_{Z_{h^*_1},\s_{h^*_1},\m_{h^*_1},C_{h^*_1}}
=
\prod_{j=1}^2 P^{(j)}_{
Z_{h^*_1},m_{h^*_1}^{(j)},C_{h^*_1}}$ where, if 
$\Psi^{(j,\le h^*_1),T}_\kk\defin(\psi^{(j,\le h^*_1)}_{1,\kk},\ 
\psi^{(j,\le h^*_1)}_{-1,\kk})$,
$$\eqalign{& P^{(j)}_{Z_{h^*_1},m_{h^*_1}^{(j)},C_{h^*_1}}
(d\psi^{(j,\le h^*_1)})\defin\cr
&\defin{1\over N_{h^*_1}^{(j)}}\prod_{\kk,\o} d\psi^{(j,\le h^*_1)}_{\o,\kk}
\exp\Big\{-{Z_{h^*_1}\over 4M^2}\sum_{\kk\in
D_{h^*_1}} 
C_{h^*_1}(\kk)\Psi^{(j,\le h^*_1),T}_{\kk}
A^{(h^*_1)}_j(\kk)\Psi^{(j,\le h^*_1)}_{-\kk}
\Big\}
\cr
&A^{(h^*_1)}_j(\kk)\defin\pmatrix
{(-i\sin k-\sin k_0)+a^{+(j)}_{h^*_1}(\kk)&
-i \big(m_{h^*_1}^{(j)}(\kk)+c_{h^*_1}^{(j)}(\kk)\big)\cr i
\big(m_{h^*_1}^{(j)}(\kk)+c_{h^*_1}^{(j)}(\kk)\big)&
(-i\sin k+\sin k_0)+a^{-(j)}_{h^*_1}(\kk)\cr}\cr}\Eq(4.22)$$
and $a^{\o(j)}_{h^*_1}$, $m_{h^*_1}^{(j)}$, $c_{h^*_1}^{(j)}$
are given by \equ(4.24y) with $h=h^*+1$.

The propagators $g^{(j,\le h^*_1)}_{\o_1,\o_2}$ 
associated with the fermionic integration \equ(4.22) 
are given by \equ(4.24) with $h=h^*_1+1$.
Note that, by \equ(5.20), $\max\{|m_{h^*_1}^{(1)}|,|m_{h^*_1}^{(2)}|\}=
|\s_{h^*_1}|+|\m_{h^*_1}|=O(\g^{h^*_1})$ (see \equ(4.24y) for the definition
of $m_{h^*_1}^{(1)}$, $m_{h^*_1}^{(2)}$).
From now on, for definiteness we shall suppose that 
$\max\{|m_{h^*_1}^{(1)}|,|m_{h^*_1}^{(2)}|\}\=|m_{h^*_1}^{(1)}|$. 
Then, it is easy to realize that the propagator
$g_{\o_1,\o_2}^{(1,\le h^*_1)}$ is bounded as follows.
$$|\dpr^{n_0}_{x_0}\dpr^{n_1}_{x}g_{\o_1,\o_2}^{(1,\le h^*_1)}
(\xx)|\le C_{N,n}{\g^{(1+n)h^*_1}\over 1+(\g^{h^*_1}
|\dd(\xx)|)^N}\virg n=n_0+n_1\;,\Eq(4.4*)$$
namely $g_{\o_1,\o_2}^{(1,\le h^*_1)}$ satisfies the same bound as
the single scale propagator on scale $h=h^*_1$. This suggests to 
integrate out $\psi^{(1,\le h^*_1)}$, without any
other scale decomposition. We find the following result.\\

{\bf Lemma 5.1} {\it If $|\l|\le \e_1$, 
$|\s_1|,|\m_1|\le c_1$ ($c_1,\e_1$ being the same as in Theorem 2.1)
and $\n_1$ is fixed as in Theorem 4.1,
we can rewrite the partition function as
$$\Xi_{AT}^-=\int P^{(2)}_{Z_{h^*_1}, \widehat m^{(2)}_{h^*_1}, C_{h^*_1}}
(d\psi^{(2,\le h^*_1)})e^{-\lis \VV^{(h^*_1)}({\sqrt Z_{h^*_1}}\psi^{(2,
\le h^*_1)})-M^2\lis E_{h^*_1}}
\;,\Eq(zh*)$$
where: $\widehat m^{(2)}_{h^*_1}(\kk)=m^{(2)}_{h^*_1}(\kk)-
\g^{h^*_1}\p_{h^*_1}C_{h^*_1}^{-1}(\kk)$,
with $\p_{h^*_1}$ a free parameter, s.t. $|\p_{h^*_1}|\le c|\l|$; 
$|\lis E_{h^*_1}-E_{h^*_1}|\le c|\l|\g^{2h^*_1}$;
and
$$\eqalign{\lis\VV^{(h^*_1)}(\psi^{(2)})-
\g^{h^*_1}\p_{h^*_1}F_\s^{(2,\le h^*_1)}(
\psi^{(2\le h^*_1)})
&=\sum_{n=1}^{\io}\sum_{\oo}\prod_{i=1}^{2n}
\hat\psi^{(2)}_{\o_i,\kk_i}
\lis W^{(h^*_1)}_{2n,\oo}(\kk_1,\ldots,\kk_{2n-1})
\d(\sum_{i=1}^{2n}\kk_i)=\cr
&=\sum_{n=1}^{\io}
\sum_{\ss,\underline j,\oo}\prod_{i=1}^{2n}
\dpr^{\s_i}_{j_i}\psi^{(2)}_{\o_i,\xx_i}
\lis W^{(h^*_1)}_{2n,\ss,\underline j,\oo}(\xx_1,\ldots,\xx_{2n})
\;,\cr}\Eq(4.50)$$
`with $F_\s^{(2,\le h)}$ given by the first of
\equ(4.11) with $\hat\psi^{(2,\le h)}_{\o,\kk}\hat\psi^{(2,\le h)}_{\o',-\kk}$
replacing $\hat\psi^{+(\le h)}_{\o,\kk}\hat\psi^{-(\le h)}_{\o',\kk}$; and
$\lis W^{(h^*_1)}_{2n,\ss,\underline j,\oo}$ satisfying the same 
bound \equ(4.45) as $W^{(\bar h)}_{2n,\ss,\underline j,\aa,\oo}$ 
with $\bar h=h^*_1$.}\\

In order to prove the Lemma it 
is sufficient to consider \equ(4.3) with $h=h^*_1$
and rewrite 
$P_{Z_{h^*_1},\s_{h^*_1},\m_{h^*_1},C_{h^*_1}}$ as
the product $\prod_{j=1}^2 P^{(j)}_{
Z_{h^*_1},m_{h^*_1}^{(j)},C_{h^*_1}}$. Then the integration
over the $\psi^{(1,\le h^*_1)}$ field is done as the integration
of the $\chi$'s in Appendix \secc(a1), recalling the bound \equ(4.4*). 

Finally we rewrite $m^{(2)}_{h^*_1}(\kk)$
as $\widehat m^{(2)}_{h^*_1}(\kk)+\g^{h^*_1}\p_{h^*_1}C_{h^*_1}^{-1}
(\kk)$, where 
$\p_{h^*_1}$ is a parameter to be suitably fixed below as a function 
of $\l,\s_1,\m_1$.\\

\sub(6.2aaa){\it The localization operator.}
The integration of the r.h.s. of \equ(zh*)
is done in an iterative way similar to the one described
in \sec(4).
 now we shall perform an iterative integration of the
field $\psi^{(2)}$. If $h=h^*_1,h^*_1-1,\ldots$, we shall write:
$$\Xi_{AT}^-=\int P^{(2)}_{Z_{h}, \widehat m^{(2)}_{h}, C_{h}}
(d\psi^{(2,\le h)})e^{-\lis\VV^{(h)}
(\sqrt{Z_h}\psi^{(2,\le h)})-M^2E_h}\;,\Eq(veff)$$
where $\lis\VV^{(h)}$ is given by an expansion similar to \equ(4.50), with
$h$ replacing $h^*_1$ and $Z_{h}, \widehat m^{(2)}_{h}$
are defined recursively in the following way. We first
introduce a {\it localization operator} $\LL$. As in \sec(4.2), 
we define $\LL$ as a combination of four operators $\LL_j$ 
and $\lis\PP_j$, $j=0,1$. $\LL_j$ are defined as in \equ(4.7) and \equ(4.8),
while $\lis\PP_0$ and $\lis\PP_1$,
in analogy with \equ(4.9), are defined as the operators extracting from a 
functional of $\widehat m^{(2)}_h(\kk)$, $h\le h^*_1$, the contributions
independent and linear in $\widehat m^{(2)}_h(\kk)$.
Note that inductively the kernels $\lis W_{2n,\oo}^{(h)}$ can be thought
as functionals of $\widehat m_k(\kk)$, $h\le k\le h^*_1$.
Given $\LL_j,\lis\PP_j$, $j=0,1$ as above, we define the action of $\LL$
on the kernels $\lis W_{2n,\oo}^{(h)}$ as follows.\\
\\
\01) If $n=1$, then
$$\LL \lis W_{2,\oo}^{(h)}\defin\cases{
\LL_0(\lis\PP_0+\lis\PP_1)
\lis W_{2,\oo}^{(h)} & if $\o_1+\o_2=0$,\cr
\LL_1\lis\PP_0\lis W_{2,\oo}^{(h)} & if $\o_1+\o_2\not=0$.}$$
\02) If $n>2$, then $\LL \lis W_{2n,\oo}^{(h)}=0$.\\ 

It is easy to prove the analogue of Lemma 3.1: 
$$\LL\lis\VV^{(h)}=(s_h+\g^h p_h)F_\s^{(2,\le h)}+z_h 
F_\z^{(2,\le h)}\;,\Eq(4.54)$$
where $s_h,p_h$ and $z_h$ are real constants and: $s_h$ is linear
in $\widehat m_k^{(2)}(\kk)$, $h\le k\le h^*_1$; $p_h$ and $z_h$ are
independent of $\widehat m_k^{(2)}(\kk)$. Furthermore 
$F_\s^{(2,\le h)}$ and $F_\z^{(2,\le h)}$ are given by the first and the last 
of
\equ(4.11) with $\hat\psi^{(2,\le h)}_{\o,\kk}\hat\psi^{(2,\le h)}_{\o',-\kk}$
replacing $\hat\psi^{+(\le h)}_{\o,\kk}\hat\psi^{-(\le h)}_{\o',\kk}$.\\

{\bf Remark.} Note that the action of $\LL$ on the quartic 
terms is trivial. The reason of such a choice is that in the present case
no quartic local term can appear, because of Pauli principle:
$\psi^{(2,h)}_{1,\xx}\psi^{(2,h)}_{1,\xx}\psi^{(2,h)}_{-1,\xx}
\psi^{(2,h)}_{-1,\xx}\=0$,
so that $\LL_0\lis W_{4,\oo}=0$.\\

Using the symmetry properties exposed in Appendix \secc(A1.2), we can 
prove the analogue of Lemma 3.2: if $n=1$, then
$$\RR \lis W_{2,\oo}=\cases{
[\lis\SS_2+\RR_2(\lis\PP_0+\lis\PP_1)]
\lis W_{2,\oo} & if $\o_1+\o_2=0$,\cr 
[\RR_1\lis\SS_1+\RR_2\lis\PP_0]
\lis W_{2,\aa,\oo} & if 
$\o_1+\o_2\not=0$,\cr}\Eq(rr*)$$
where $\lis\SS_1=1-\lis\PP_0$ and $\lis\SS_2=1-\lis\PP_0-\lis\PP_1$;
if $n=2$, then
$\lis W_{4,\oo}= \RR_1\lis W_{4,\oo}$.\\

\sub(6.3aaa){\it Renormalization for $h\le h^*_1$.}
If $\LL$ and $\RR=1-\LL$ are defined as in previous subsection, we can rewrite
\equ(veff) as:
$$\int P^{(2)}_{Z_{h}, \widehat m^{(2)}_{h}, C_{h}}
(d\psi^{(2,\le h)})e^{-\LL\lis\VV^{(h)}(\sqrt{Z_h}\psi^{(2,\le h)})
-\RR\lis\VV^{(h)}(\sqrt{Z_h}\psi^{(2,\le h)})-M^2E_h}\;.\Eq(veff1)$$
Furthermore, using \equ(4.54) and defining:
$$\widehat Z_{h-1}(\kk)\defin
Z_h(1+C_h^{-1}(\kk) z_h)\virg
\widehat m_{h-1}^{(2)}(\kk)\defin{Z_h\over \widehat Z_{h-1}(\kk)}
\left(\widehat m_h^{(2)}(\kk)+C_h^{-1}(\kk) s_h\right)\;,\Eq(flow)$$
we see that \equ(veff1) is equal to
$$\int P^{(2)}_{\widehat Z_{h-1},\widehat m^{(2)}_{h-1}, C_{h}}
(d\psi^{(2,\le h)})e^{-\g^h p_h F_\s^{(2,\le h)}(\sqrt{Z_{h}}
\psi^{(2),\le h})-
\RR\lis\VV^{h}(\sqrt{Z_{h}}\psi^{(2),\le h})-M^2 (E_h+t_h)}\Eq(4.58*)$$
Again, we rescale the potential: 
$$\widetilde\VV^{(h)}(\sqrt{Z_{h-1}}
\psi^{(\le h)})\defin \g^h\p_h F_\s^{(2,\le h)}(\sqrt{Z_{h-1}}
\psi^{(2,\le h)})+
\RR\lis\VV^{h}(\sqrt{Z_{h}}\psi^{(2,\le h)})\;,\Eq(recsale)$$ 
where $Z_{h-1}=\widehat Z_{h-1}({\bf 0})$ and 
$\p_h=( Z_h/ Z_{h-1})p_h$; we
define $\widetilde f_h^{-1}$ as in \equ(4.21), we perform
the single scale integration and we define the new effective potential as
$$e^{-\lis\VV^{(h-1)}(\sqrt{Z_{h-1}}\psi^{(2,\le h-1)})-M^2 
\tilde E_{h}}\defin \int 
P^{(2)}_{Z_{h-1},\widehat m^{(2)}_{h-1}, \widetilde f^{-1}_h}
(d\psi^{(2,h)})
e^{-\widetilde\VV^{h}(\sqrt{Z_{h}}\psi^{(2,\le h)})}
\;.\Eq(4.58)$$
Finally we pose $E_{h-1}=E_h+t_h+\tilde E_h$.
Note that the above procedure allow us to write the 
$\p_h$ in terms of $\p_k$, $h\le k\le h^*_1$, namely
$\p_{h-1}=\g^h\p_h+\b^h_\p(\p_h,\ldots,\p_{h^*_1})$,
where $\b^h_\p$ is the {\it Beta function}.

Proceeding as in \sec(4) we can inductively show that $\lis\VV^{(h)}$ has 
the structure of \equ(4.50), with $h$ replacing $h^*_1$ and that 
the kernels of $\lis\VV^{(h)}$ are bounded as follows.\\

{\bf Lemma 5.2.} {\it Let the hypothesis of Lemma 5.1 be satisfied
and suppose that, for $\bar h< h\le  h^*_1$ and some constants $c,\th>0$ 
$$e^{-c|\l|}\le{\widehat m^{(2)}_h\over \widehat m^{(2)}_{h-1}}\le 
e^{c|\l|}\virg
e^{-c|\l|^2}\le{Z_h\over Z_{h-1}}\le e^{c|\l|^2}
\virg |\p_h|\le c|\l|\virg |\widehat m^{(2)}_{\bar h}|\le 
\g^{\bar h}\;.\Eq(lem5.2)$$
Then the partition 
function can be rewritten as in \equ(veff)
and there exists $C>0$ s.t. the kernels of  
$\lis\VV^{(h)}$ satisfy:
$$\int d\xx_1\cdots d\xx_{2n}|\lis W^{(\bar h)}_{2n,\ss,\underline j,\oo}
(\xx_1,\ldots,\xx_{2n})|
\le M^2 \g^{-\bar h D_k(n)} \,(C\,|\l|)^{max(1,n-1)}\Eq(5.45*)$$
where $D_k(n)=-2+n+k$ and $k=\sum_{i=1}^{2n}\s_i$.
Finally $|E_{\bar h+1}|+|t_{\bar h+1}|\le c|\l|\g^{2\bar h}$.}\\

The proof of Lemma 5.2 is essentially identical to the proof
of Theorem 3.1 and we do not repeat it here.\\

%{\bf Remarks}\\
%1) The hypothesis \equ(lem5.2) can be proven by solving the flow equations
%for $\p_h, Z_h$ and $\widehat m^{(2)}_h$ in a way similar (but much simpler)
%to the one followed in \sec(5) to study the flow equations for $\l_h,\n_h,
%Z_h,\s_h$ and $\m_h$. We shall sketch the proof of \equ(lem5.2) in 
%next subsection and in Appendix \secc(pi).\\
%2) In order to insure the condition $\widehat m^{(2)}_{\bar h}\le 
%\g^{\bar h}$, we shall iterate the procedure described above up to 
%the scale $\bar h=h^*_2$ with $h^*_2$ defined by the condition: 
%$|\widehat m_k^{(2)}|\le\g^{k-1}$ for any
%$h^{*}_2\le k\le h^{*}_1$ and 
%$|\widehat m_{h^{*}_2-1}^{(2)}|> \g^{h^*_2-2}$.\\

It is possible to fix $\p_{h^*_1}$
so that the first three assumptions
in \equ(lem5.2) are valid for any $h\le h^*_1$. More precisely,
the following result holds, see Appendix \secc(pi).\\

{\bf Lemma 5.3. } {\it If $|\l|\le \e_1$, 
$|\s_1|,|\m_1|\le c_1$ and $\n_1$ is fixed as in Theorem 4.1,
there exists  
$\p^*_{h^*_1}(\l,\s_1,\m_1)$ such that, if we fix $\p_{h^*_1}=
\p^*_{h^*_1}(\l,\s_1,\m_1)$, for $h\le h^*_1$ we have: 
$$|\p_h|\le c|\l|\g^{(\th/2)(h-h^*_1)}\virg
\widehat m^{(2)}_h=\widehat m^{(2)}_{h^*_1}\g^{ F^h_m}\virg 
Z_h=Z_{h^*_1}\g^{\lis F^h_\z}\;,\Eq(4.58b)$$
where $F^h_m$ and $\lis F^h_\z$ are $O(\l)$. Moreover: 
$$\Big|\p^*_{h^*_1}(\l,\s_1,\m_1)-\p^*_{h^*_1}(\l,\s_1',\m_1')\Big|
\le c|\l|\left(\g^{(\h_\s-1)h^*_1}|\s_1-\s_1'|+
\g^{(\h_\m-1)h^*_1}|\m_1-\m_1'|\right)\;.\Eq(ph)$$}

\sub(6.101){\it The integration of the scales $\le h^*_2$.}
In order to insure that the last assumption in \equ(lem5.2) holds, 
we iterate the preceding construction up to the scale
$h^*_2$ defined as the scale s.t. $|\widehat m_k^{(2)}|\le\g^{k-1}$ for any
$h^{*}_2\le k\le h^{*}_1$ and 
$|\widehat m_{h^{*}_2-1}^{(2)}|> \g^{h^*_2-2}$.

Once we have integrated all the fields $\psi^{(>h^*_2)}$, we can integrate
$\psi^{(2,\le h^*_2)}$ without any further multiscale 
decomposition. Note in fact that by definition the propagator
satisfies the same bound \equ(4.4*) with $h^*_2$ replacing $h^*_1$.
Then, if we define
$$e^{-M^2\tilde E_{\le h^*_2}}\defin \int P_{Z_{h^*_2-1},
\widehat m^{(2)}_{h^*_2-1},C_{h^*_2}}e^{-\widetilde \VV^{(h^*_2)}(\sqrt{Z_{
h^*_2-1}}\psi^{(2,\le h^*_2)})}\;,\Eq(e*2)$$
we find that $|\tilde E_{\le h^*_2}|
\le c|\l|\g^{2h^*_2}$ (the proof is a repetition
of the estimates on the single scale integration).

Combining this bound with the results of Theorem 3.1, Lemma 5.1, Lemma 5.2
and Lemma 5.3,
together with the results of \sec(5) we finally find
that the free energy associated to $\Xi_{AT}^-$ is given by the following
{\it finite} sum, uniformly convergent with the size of $\L_M$:
$$\lim_{M\to\io}
{1\over M^2}\log \Xi_{AT}^-=E_{\le h^*_2}+(\lis E_{h^*_1}-E_{h^*_1})
+\sum_{h=h^*_2+1}^1 (\tilde E_h+t_h)\;,\Eq(oh)$$
where $E_{\le h^*_2}=\lim_{M\to\io}\tilde E_{\le h^*_2}$ and it is easy to
see that $E_{\le h^*_2}$, for any finite $h^*_2$, exists and satisfies 
the same bound of $\tilde E_{h^*_2}$.\\

\sub(an){\it Keeping $h^*_2$ finite.}
From the discussion of previous subsection, it follows that, for
any finite $h^*_2$, \equ(oh) is an analytic function of $\l,t,u$,
for $|\l|$ sufficiently small, uniformly in $h^*_2$
(this is an elementary consequence of Vitali's convergence theorem). 
Moreover, repeating the discussion of Appendix G in [M1], it can be 
proved that,
for any $\g^{h^*_2}>0$ (here $\g^{h^*_2}$ plays the
role of $|t-t_c|$ in Appendix G of [M1]), the limit \equ(oh) coincides with 
$\lim_{M\to\io}1/M^2 \log \Xi_{AT}^{\g_1,\g_2}$ for any choice 
$\g_1,\g_2$ of boundary conditions; hence this limit coincides
with $-2\log\cosh \l$ plus 
the free energy in \equ(cv), see also \equ(2.10). We can state the result 
as follows.\\

{\bf Lemma 5.4.} {\it There exists $\e_1>0$ such that, if $|\l|\le\e_1$ 
and $t\pm u\in D$ (the same as in Main Theorem), 
the free energy $f$ defined in \equ(cv) 
is real analytic in $\l,t,u$, except
possibly for the choices of $\l,t,u$ such that $\g^{h^*_2}=0$.}\\

We shall see in \sec(7) below that the specific heat is logarithmically 
divergent as $\g^{h^*_2}\to 0$. So the 
critical point is really given by the condition $\g^{h^*_2}=0$.
We shall explicitely solve the equation for the critical point
in next subsection. 

\*
\sub(6.1) {\it The critical points}.
In the present subsection we check that, if $t\pm u\in D$, $D$ being a suitable
interval centered around $\sqrt2-1$, see Main Theorem, 
there are precisely two critical points, of the form \equ(1.5). More precisely,
keeping in mind that the equation 
for the critical point is simply $\g^{h^*_2}=0$ (see the end of 
previous subsection), we prove the following.\\

{\bf Lemma 5.5. } {\it Let 
$|\l|\le\e_1$, $t\pm u\in D$ and $\p_{h^*_1}$
be fixed as in Lemma 5.3.  
Then $\g^{h^*_2}=0$ only if $(\l,t,u)=
(\l,t_c^\pm(\l,u),u)$, where $t_c^\pm(\l,u)$ is given by \equ(1.5).}\\

{\bf Proof\ }
From the definition of $h^*_2$ given above, see \sec(6.101),
it follows that $h^*_2$ satisfies the following equation:
$$\g^{h^*_2-1}=c_m\g^{F^{h^*_2}_m}\Big|
|\s_{h^*_1}|-|\m_{h^*_1}|-\a_\s\g^{h^*_1}\p_{h^*_1}\Big|\;,
\Eq(4.61a)$$
for some $1\le c_m<\g$ and $\a_\s=\sign\s_1$. 
Then, the equation $\g^{h^*_2}=0$ can be rewritten as: 
$$|\s_{h^*_1}|-|\mu_{h^*_1}|-\a_\s \g^{h^*_1}\p_{h^*_1}=0\;.
\Eq(4.61aa)$$
First note that the result of Lemma 5.5 is trivial when $h^*_1=1$.
If $h^*_1<1$, \equ(4.61aa)
cannot be solved when $|\s_1|^{1\over 1-\h_\s}>2
|\m_1|^{1\over 1-\h_\m}$. In fact, 
$$\eqalign{&|\s_1|\g^{\h_\s(h^*_1-1)+F^{h^*_1}_\s}-|\m_1|\g^{\h_\m(h^*_1-1)+
F^{h^*_1}_\m}
-\a_\s\g^{h^*_1}\p_{h^*_1}=\cr
&=|\s_1|^{1+{\h_\s\over 1-\h_\s}}c_1-
\Big(|\m_1||\s_1|^{-{1-\h_\m\over 1-\h_\s}}\Big)|\s_1|^{{1-\h_\m\over 1-\h_\s}
-{\h_\m\over 1-\h_\s}}c_1'-\a_\s\g^{h^*_1}\p_{h^*_1}\ge {\g^{h^*_1-1}
\over 3\g}\;,\cr}\Eq(4.61aaa)$$
where $c_1,c_1'$ are constants $=1+O(\l)$,
$\p_{h^*_1}=O(\l)$ and $\g^{h^*_1-1}=c_\s
|\s_1|^{1\over 1-\h_\s}$,
with $1\le c_\s< \g$. Now, if $|\m_1|>0$, the r.h.s. of \equ(4.61aaa)
equation is strictly positive.\\

So, let us consider the case $h^*_1<1$ and $|\s_1|^{1\over 1-\h_\s}\le 2
|\m_1|^{1\over 1-\h_\m}$ 
(s.t. $\g^{h^*_1}=c_u\log_\g|u|^{1\over 1-\h_\m}$, with $1\le c_u\le \g$). 
In this case
\equ(4.61aa) can be easily solved to find:
$$|\s_1|=|\m_1||u|^{\h_\m-\h_\s\over 1-\h_\m}c_u^{\h_\m-\h_\s}\g^{F^{h^*_1}_\m
-F^{h^*_1}_\s}+|u|^{1-\h_\s\over 1-\h_\m}
c_u^{1-\h_\s}\a_\s\g^{1-F^{h^*_1}_\s}\p_{h^*_1}\;.
\Eq(6.102)$$
Note that $c_u^{\h_\m-\h_\s}\g^{F^{h^*_1}_\m
-F^{h^*_1}_\s}=1+O(\l)$ is just a function of $u$,
(it does not depend on $t$), because
of our definition of $h^*_1$. 
Moreover $\p_{h^*_1}$ is a smooth
function of $t$: if we call $\p_{h^*_1}(t,u)$ resp. 
$\p_{h^*_1}(t',u)$
the correction corresponding 
to the initial data $\s_1(t,u),\m_1(t,u)$ resp. $\s_1(t',u),\m_1(t',u)$, 
we have
$$|\p_{h^*_1}(t,u)-\p_{h^*_1}(t',u)|\le c|\l|
|u|^{\h_\s-1\over 1-\h_\m}|t-t'|\;,\Eq(6.106)$$
where we used \equ(ph) and the bounds $|\s_1-\s_1'|\le c|t-t'|$ and
$|\m_1-\m_1'|\le c|u||t-t'|$, following from 
the definitions of $(\s_1,\m_1)$ in terms of $(\s,\m)$ and
of $(t,u)$, see \sec(2).

Using the same definitions we also realize that \equ(6.102) can be rewritten as
$$t=\Big[\sqrt2-1+{\n(\l)\over 2}\pm|u|^{1+\h}\Big(1+\l f(t,u)\Big)
\Big]{1+\hat\l(t^2-u^2)\over 1+\hat\l}\;,\Eq(6.105)$$
where 
$$1+\h\defin{1-\h_\s\over 1-\h_\m}\;,\Eq(eta)$$
and the crucial property is that $\h=-b\l+O(\l^2)$, $b>0$, see Lemma 4.1
and Appendix \secc(a5). We also recall that both $\h$ and $\n$ are
functions of $\l$ and are independent of $t,u$. 
Moreover $f(t,u)$ is a suitable bounded function s.t. 
$|f(t,u)-f(t',u)|\le c|u|^{-(1+\h)}|t-t'|$, as it follows from the Lipshitz 
property of $\p_{h^*_1}$ \equ(6.106). 
The r.h.s. of \equ(6.105) is
Lipshitz in $t$ with constant $O(\l)$, so that \equ(6.105) can be inverted 
w.r.t. $t$ by contractions and, for both choices of the sign, 
we find a unique solution
$$t=t_c^\pm(\l,u)=\sqrt2-1+\n^*(\l)\pm|u|^{1+\h}\big(1+F^\pm(\l,u)\big)\;,
\Eq(6.103)$$
with $|F^\pm(\l,u)|\le c\big|\l|$, for some $c$.\qed\\

\sub(6.2) {\it Computation of $h^*_2$.}
Let us now solve \equ(4.61a) in the general case of 
$\g^{h^*_2}\ge0$. 
Calling $\e\defin\g^{h^*_2-h^*_1-F^{h^*_2}_m}/c_m$, we find:
$$\eqalign{\e&=\left||\s_1|\g^{(\h_\s-1)(h^*_1-1)+F^{h^*_1}_\s}-|\m_1|
\g^{(\h_\m-1)(h^*_1-1)+F^{h^*_1}_\m}
-\a_\s\g\p_{h^*_1}\right|=\cr
&=\g^{(\h_\s-1)(h^*_1-1)+F^{h^*_1}_\s}\left||\s_1|-|\m_1|
\g^{(\h_\m-\h_\s)(h^*_1-1)+F^{h^*_1}_\m-F^{h^*_1}_\s}
-\a_\s\g^{1+(1-\h_\s)(h^*_1-1)-F^{h^*_1}_\s}\p_{h^*_1}\right|\;.
\cr}
\Eq(6.103)$$
If $|\s_1|^{1/(1-\h_\s)}\le 2|\m_1|^{1/(1-\h_\m)}$,
we use $\g^{h^*_1-1}=c_u|u|^{1/(1-\h_\m)}$ and,
from the second row of \equ(6.103), we find: $\e=C{\left||\s_1|-
|\s_{1,c}^{\a_\s}|\right||u|^{-(1+\h)}}$, where 
$\s_{1,c}^\pm=\s_1(\l,t_c^\pm,u)$
and $C=C(\l,t,u)$ is bounded above and below by $O(1)$ constants; defining
$\D$ as in \equ(1.6), we can rewrite:
$$\e=C{\left||\s_1|-
|\s_{1,c}^{\a_\s}|\right|\over|u|^{1+\h}}=C'{\left|\s_1^2-
(\s_{1,c}^{\a_\s})^2\right|\over \D|u|^{1+\h}}=C''{|t-t_c^+|\cdot|t-t_c^-|\over
\D^2}\;,\Eq(epsilon)$$
where $C'=C'(\l,t,u)$ and $C''=C''(\l,t,u)$ 
are bounded above and below by $O(1)$ constants.

In the opposite case ($|\s_1|^{1/(1-\h_s)}> 2|\m_1|^{1/(1-\h_\m)}$),
we use $\g^{h^*_1-1}=c_\s|\s_1|^{1/(1-\h_\s)}$ and, from the first row
of \equ(6.103), we find
$\e=\tilde C(1-{|\m_1||\s_1|^{-1/(1+\h)}}-\a_\s\g\p_{h^*_1})=\bar C$,
where $\tilde C$ and $\bar C$ are bounded above and below by $O(1)$ constants.
Since in this region of parameters $|t-t_c^\pm|\D^{-1}$ 
is also bounded above and below
by $O(1)$ constants, we can in both cases write 
$$\e=C_\e(\l,t,u) {|t-t_c^+|\cdot|t-t_c^-|\over
\D^2}\virg C_{1,\e}\le C_\e(\l,t,u) \le C_{2,\e}\Eq(eps2)$$
and $C_{j,\e}$, $j=1,2$, are suitable positive $O(1)$ constants.

\\
\section(7, The specific heat)

Consider the specific heat defined in \equ(cv).
The correlation function $<H^{AT}_{\xx}H^{AT}_{\yy}>_{\L_M,T}$
can be conveniently written as
$$<H^{AT}_{\xx}H^{AT}_{\yy}>_{\L,T}=
{\dpr^2\over \dpr\phi_\xx\dpr\phi_\yy}\log \Xi_{AT}(\phi)\Big|_{\phi=0}\virg
\Xi_{AT}(\phi)\defin\sum_{\s^{(1)},\s^{(2)}}e^{-\sum_{\xx\in\L}(1+\phi_\xx)
H^{AT}_\xx}\Eq(7.2)$$
where $\phi_\xx$ is a real commuting auxiliary field (with periodic boundary 
conditions).\\
\\
Repeating the construction of \sec(2), we see that $\Xi_{AT}(\phi)$
admit a Grassmanian representation similar to the one of $\Xi_{AT}$,  
and in particular, if $\xx\not =\yy$:
$$\eqalign{&{\dpr^2\over \dpr\phi_\xx\dpr\phi_\yy}\log
\Xi_{AT}(\phi)\Big|_{\phi=0}={\dpr^2\over \dpr\phi_\xx\dpr\phi_\yy}
\log\sum_{\g_1,\g_2}(-1)^{\d_{\g_1}+\d_{\g_2}}
\widehat \Xi_{AT}^{\g_1,\g_2}(\phi)\Big|_{\phi=0}\cr
&\widehat 
\Xi_{AT}^{\g_1,\g_2}(\phi)=\int\prod_{\xx\in\L_M}^{j=1,2} dH^{(j)}_\xx d
\lis H^{(j)}_\xx d V^{(j)}_\xx
d\lis V^{(j)}_\xx\, e^{S^{(1)}_{\g_1}(t^{(1)})+S^{(2)}_{\g_2}(t^{(2)})+
V_\l+\BB(\phi)}\cr}\Eq(7.3)$$
where $\d_\g$, $S^{(j)}(t^{(j)})$ and $V_\l$ where defined in \sec(2)
(see \equ(2.2) and previous lines, and \equ(2.15)), 
the apex $\g_1,\g_2$ attached to $\widehat \Xi_{AT}$ refers to the
boundary conditions assigned to the Grassmanian fields, as in \sec(2)
and finally $\BB(\phi)$ is defined as:
$$\eqalign{\BB(\phi)=
\sum_{\xx\in\L}&\phi_\xx\Big\{a^{(1)}\big(\lis H^{(1)}_{\xx} H^{(1)}_{
\xx+\hat e_1}+\lis V^{(1)}_{\xx} V^{(1)}_{\xx+\hat e_0}\big)+
a^{(2)}\big(\lis H^{(2)}_{\xx} H^{(2)}_{
\xx+\hat e_1}+\lis V^{(2)}_{\xx} V^{(2)}_{\xx+\hat e_0}\big)+\cr
&+\l \widetilde a
\big(\lis H^{(1)}_{\xx} H^{(1)}_{
\xx+\hat e_1}\lis H^{(2)}_{\xx} H^{(2)}_{\xx+\hat e_1}+
\lis V^{(1)}_{\xx} V^{(1)}_{
\xx+\hat e_0}\lis V^{(2)}_{\xx} V^{(2)}_{\xx+\hat e_0}\big)\Big\}\defin
\sum_{\xx\in\L}\phi_\xx A_\xx\;,\cr}
\Eq(7.4)$$
where $a^{(1)}$, $a^{(2)}$ and $\widetilde a$ are $O(1)$ constants, with
$a^{(1)}-a^{(2)}=O(u)$. Using \equ(7.3) and \equ(7.4) we can rewrite:
$$<H^{AT}_{\xx}H^{AT}_{\yy}>_{\L,T}={1\over 4}(\cosh J)^{2M^2}
\sum_{\g_1,\g_2}(-1)^{\d_{\g_1}+\d_{\g_2}}
{\Xi_{AT}^{\g_1,\g_2}\over \Xi_{AT}}<A_\xx A_\yy>_{\L_M,T}^{\g_1,\g_2}\;,
\Eq(axay)$$
where $<\cdot>_{\L_M,T}^{\g_1,\g_2}$ is the average w.r.t. the
boundary conditions $\g_1,\g_2$. Proceeding as in Appendix G of [M1]
one can show that, if $\g^{h^*_2}>0$, 
$<A_\xx A_\yy>_{\L_M,T}^{\g_1,\g_2}$ is exponentially
insensitive
to boundary conditions and $\sum_{\g_1,\g_2}(-1)^{\d_{\g_1}+\d_{\g_2}}
{\Xi_{AT}^{\g_1,\g_2}/ \Xi_{AT}}$ is an $O(1)$ constant. Then from
now on we will
study only $\Xi_{AT}^-(\phi)\defin
\widehat \Xi_{AT}^{(-,-),(-,-)}(\phi)$ and $<A_\xx A_\yy>_{\L_M,T}^{
(-,-),(-,-)}$.

As in \sec(2)
we integrate out the $\c$ fields and, proceeding as in Appendix A, we find:
$$\Xi_{AT}^-(\phi)=
\int P_{Z_1,\s_1,\m_1,C_1}(d\psi) e^{\VV^{(1)}+\BB^{(1)}}\;,\Eq(7.5)$$
where
$$\BB^{(1)}(\psi,\phi)=\sum_{m,n=1}^\io
\sum^{\underline\s,\underline j,
\underline\a,\oo}_{\xx_1\cdots\xx_m\atop\yy_1\cdots \yy_{2n}}
B^{(1)}_{m,2n;\underline\s,\underline j,
\underline\a,\oo}(\xx_1,\ldots,\xx_m;\yy_1,
\ldots,\yy_{2n})
\Big[\prod_{i=1}^m\phi_{\xx_i}\Big] \Big[\prod_{i=1}^{2n}
\partial^{\s_i}_{j_i}\psi^{\a_i}_{\yy_i,\o_i}\Big]\;.\Eq(6.6)$$
We proceed as for the partition function, namely as described in \sec(4)
above. We introduce
the scale decomposition described in \S 3
and we perform iteratively
the integration of the single scale fields, starting from the field of
scale $1$.
After the integration of the fields $\psi^{(1)},\ldots,\psi^{(h+1)}$, 
$h^*_1<h\le 0$, we are left with
$$\Xi_{AT}^-(\phi)=e^{-M^2 E_h+S^{(h+1)}(\phi)}\int P_{Z_h,\s_h, 
\m_h,C_h}(d\psi^{\le
h})e^{-\VV^{(h)}(\sqrt{Z_h}\psi^{(\le h)})+\BB^{(h)}
(\sqrt{Z_h}\psi^{(\le h)},\phi)}\;,\Eq(7.6)$$
where $P_{Z_h,\s_h,\m_h m_h,C_h}(d\psi^{(\le h)})$
and $\VV^{(h)}$ are the same as in \S 3, $S^{(h+1)}$ $(\phi)$ 
denotes the sum of the contributions dependent on $\phi$ but independent of
$\psi$, and finally $\BB^{(h)}(\psi^{(\le h)},\phi)$ denotes the
sum over all terms containing at least one $\phi$ field and two $\psi$
fields. $S^{(h+1)}$ and $\BB^{(h)}$ can be represented as
$$\eqalign{&S^{(h+1)}(\phi)=\sum_{m=1}^\io\sum_{\xx_1\cdots\xx_m}
S^{(h+1)}_m(\xx_1,\ldots,\xx_m)
\prod_{i=1}^m\phi_{\xx_i}\cr
&\BB^{(h)}(\psi^{(\le h)},\phi)=
\sum_{m,n=1}^\io
\sum_{\xx_1\cdots\xx_m\atop \yy_1\cdots\yy_{2n}}^{\ss,\underline j,\aa,\oo}
B^{(h)}_{m,2n;\ss,\underline j,\aa,\oo}(\xx_1,
\ldots,\xx_m;\yy_1,\ldots,\yy_{2n})
\Big[\prod_{i=1}^m\phi_{\xx_i}\Big] \Big[\prod_{i=1}^{2n}
\partial^{\s_i}\psi^{(\le h)\a_i}_{\yy_i,\o_i}\Big]\;.\cr}\Eq(7.7)$$
Since the field $\phi$ is equivalent, as regarding dimensional
bounds, to two $\psi$ fields (see Theorem 6.1 below for a more precise 
statement), the only terms in the 
expansion for $\BB^{(h)}$ 
which are not irrelevant are those with $m=n=1$, $\s_1=\s_2=0$ 
and they are marginal. 
Hence we extend the definition of the localization operator $\LL$,
so that its action on $\BB^{(h)}(\psi^{(\le
h)},\phi)$ is defined by its action on the kernels
$\widehat B^{(h)}_{m,2n;\aa,\oo}
(\qq_1,\ldots,\qq_m;\kk_1,\ldots,\kk_{2n})$:\\
\01) if $m=n=1$ and $\a_1+\a_2=\o_1+\o_2=0$, then
$\LL \widehat B^{(h)}_{1,2;\ss,\aa,\oo}(\qq_1;\kk_1,\kk_2)\defin
\PP_0\widehat B^{(h)}_{1,2;\aa,\oo}
(\kk_+;\kk_+,\kk_+)$,
where $\PP_0$ is defined as in \equ(4.9);\\
\02) in all other cases $\LL
\widehat B^{(h)}_{m,2n;\aa,\oo}=0$.
\*

Using the symmetry considerations of Appendix B together with the remark that
$\phi_\xx$ is invariant under {\it Complex conjugation}, {\it 
Hole--particle} and $(1)\otto(2)$, while under {\it Parity} 
$\phi_\xx\to\phi_{-\xx}$ and under 
{\it Rotation} $\phi_{(x,x_0)}\to\phi_{(-x_0,
-x)}$,
we easily realize that $\LL\BB^{(h)}$ has necessarily the following form:
$$\LL\BB^{(h)}(\psi^{(\le h)},\phi)={\lis Z_h\over Z_h}
\sum_{\xx,\o}{(-i\o)\over 2}\phi_\xx\psi^{
(\le h)+}_{\o,\xx}\psi^{(\le h)-}_{-\o,\xx}\;,\Eq(7.10)$$
where $\lis Z_h$ is real and $\lis Z_1=a^{(1)}|_{\s=\m=0}\=
a^{(2)}|_{\s=\m=0}$.

Note that apriori a term $\sum_{\xx,\o,\a}
\phi_\xx\psi^{(\le h)\a}_{\o,\xx}\psi^{(\le h)\a}_{-\o,\xx}$ is allowed
by symmetry but, using $(1)\otto(2)$ symmetry, 
one sees that its kernel is proportional
to $\m_k$, $k\ge h$. So, with our definition of localization,
such term contributes to $\RR\BB^{(h)}$.\\

Now that the action of $\LL$ on $\BB$ is defined, 
we can describe the single scale integration, for $h> h^*_1$. 
The integral in the r.h.s. of \equ(7.6) can be rewritten as:
$$\eqalign{&e^{-M^2 t_h}\int P_{Z_{h-1},\s_{h-1}, 
\m_{h-1},C_{h-1}}(d\psi^{\le
h-1})\cdot\cr
&\qquad\qquad\cdot\int P_{Z_{h-1},\s_{h-1}, 
\m_{h-1},\widetilde f^{-1}_{h}}(d\psi^{(h)})
e^{-\widehat \VV^{(h)}(\sqrt{Z_{h-1}}\psi^{(\le h)})+\widehat \BB^{(h)}
(\sqrt{Z_{h-1}}\psi^{(\le h)},\phi)}\;,\cr}\Eq(7.11)$$
where $\widehat \VV^{(h)}$ was defined in \equ(4.18) and 
$$\widehat \BB^{(h)}
(\sqrt{Z_{h-1}}\psi^{(\le h)},\phi)\defin \BB^{(h)}
(\sqrt{Z_h}\psi^{(\le h)},\phi)\;.\Eq(7.12)$$
Finally we define 
$$\eqalign{&
e^{-\widetilde E_h M^2+\widetilde S^{(h)}(\phi)-\VV^{(h-1)}
(\sqrt{Z_{h-1}}\psi^{(\le h-1)})+\BB^{(h-1)}(\sqrt{Z_{h-1}}
\psi^{(\le h-1)},\phi)}\defin\cr
&\qquad\defin  \int P_{Z_{h-1},\s_{h-1}, 
\m_{h-1},\widetilde f^{-1}_{h}}(d\psi^{(h)})
e^{-\widehat \VV^{(h)}(\sqrt{Z_{h-1}}\psi^{(\le h)})+\widehat \BB^{(h)}
(\sqrt{Z_{h-1}}\psi^{(\le h)},\phi)}\;,\cr}\Eq(7.13)$$
and
$$E_{h-1}\defin E_h+t_h+\widetilde E_h\virg S^{(h)}(\phi)\defin S^{(h+1)}(\phi)
+\widetilde S^{(h)}(\phi)\;.\Eq(7.14)$$
With the definitions above, it is easy to verify that $\lis Z_{h-1}$
satisfies the equation $\lis Z_{h-1}=\lis Z_h(1+\lis z_h)$,
where $\lis z_h=\lis b\l_h+O(\l^2)$, for some $\lis b\not =0$. 
Then, for some $c>0$,
$\lis Z_1e^{-c|\l|h}\le \lis Z_h\le\lis Z_1 e^{c|\l|h}$. 
The analogous of Theorem 3.1 for the kernels of $\BB^{(h)}$ holds:\\
\\
{\bf Theorem 6.1.} {\it Suppose that the hypothesis of Lemma 5.1 are satisfied.
Then, for $h^*_1\le \bar h\le 1$ and a suitable constant $C$, 
the kernels of $\BB^{(h)}$ satisfy
$$\int d\xx_1\cdots d\xx_{2n}|B^{(\bar h)}_{2n,m;\ss,\underline j,\aa,\oo}
(\xx_1,\ldots,\xx_m;\yy_1,\ldots,\yy_{2n})|
\le M^2 \g^{-\bar h (D_k(n)+m)} \,(C\,|\l|)^{max(1,n-1)}\;,\Eq(7.17)$$
where $D_k(n)=-2+n+k$ and $k=\sum_{i=1}^{2n}\s_i$.}\\

Note that, consistently with our definition of localization,
the dimension of $B^{(h)}_{2,1;(0,0),(+,-),(\o,-\o)}$ is $D_0(1)+1=0$.

Again, proceeding as in \sec(5), 
we can study the flow of $\lis Z_h$ up to $h=-\io$ and prove that
$\lis Z_h=\lis Z_1\g^{\lis\h(h-1)+F^h_{\bar z}}$,
where $\lis \h$ is a non trivial analytic function of $\l$ (its linear part is 
non vanishing) and $F^h_{\bar z}$ is a suitable
$O(\l)$ function (independent of $\s_1,\m_1$). 
We recall that $\lis Z_1=O(1)$.

We proceed as above up to the scale $h^*_1$. Once that the scale $h^*_1$
is reached we pass to the $\psi^{(1)},\psi^{(2)}$ variables,
we integrate out (say) the $\psi^{(1)}$ fields and we get
$$\int P^{(2)}_{Z_{h^*_1},\widehat m^{(2)}_{h^*_1},C_{h^*_1}}
(d\psi^{(2)(\le h^*_1)})e^{-\lis\VV^{(h^*_1)}
(\sqrt{Z_{h^*_1}}\psi^{(2,\le h^*_1)})+\lis
B^{(h^*_1)}(\sqrt{Z_{h^*_1}}
\psi^{(2,\le h^*_1)})}\;,\Eq(7.19)$$
with $\LL\lis B^{h^*_1}(\sqrt{Z_{h^*_1}}\psi^{(2),\le h^*_1})=
\lis Z_{h^*_1}\sum_\xx i\phi_{\xx}
\psi^{(2,\le h^*_1)}_{1,\xx}\psi^{(2,\le h^*_1)}_{-1,\xx}$.

The scales $h^*_2\le h\le h^*_1$
are integrated as in \sec(6) and
one finds that the flow of $\lis Z_{h}$ in this regime is trivial, \ie
if $h^*_2\le h\le h^*_1$, $\lis Z_h=\lis Z_{h^*_1}\g^{F^h_{z}}$,
with $F^h_{z}=O(\l)$.

The result is that the correlation function $<H^{AT}_\xx H^{AT}_\yy>_{\L_M,T}$
is given by a convergent power series in $\l$, uniformly in $\L_M$.
Then, the leading behaviour of the specific heat is given by the sum
over $\xx$ and $\yy$ of the lowest order contributions to 
$<H^{AT}_\xx H^{AT}_\yy>_{\L_M,T}$, namely
by the diagrams in Fig 3. Absolute convergence of the power series
of $<H^{AT}_\xx H^{AT}_\yy>_{\L_M,T}$ implies that the rest is 
a small correction. 

%%%%%%%%%%%%%%%%%%%%%%%%%%%%%%%%%%%%%%%%%%%%%%%%%%%%%%%%%%%%%%%%%%%%%%%%%%%%%
%%%%%%%%%%%%%%%%%%%%%%%%%%%%%%%%FIGURA 3%%%%%%%%%%%%%%%%%%%%%%%%%%%%%%%%%%%%%
%%%%%%%%%%%%%%%%%%%%%%%%%%%%%%%%%%%%%%%%%%%%%%%%%%%%%%%%%%%%%%%%%%%%%%%%%%%%%

\insertplot{300pt}{50pt}%
{\ins{-3pt}{8pt}{$\sum_{h=h^*_2}^{h^*_1}$}
\ins{53pt}{-3pt}{$\xx$}
\ins{91pt}{-3pt}{$\yy$}
\ins{73pt}{25pt}{$h$}
\ins{73pt}{-15pt}{$h$}
\ins{130pt}{4pt}{$+$}
\ins{150pt}{8pt}{$\sum_{h=h^*_1}^1$}
\ins{215pt}{-3pt}{$\xx$}
\ins{253pt}{-3pt}{$\yy$}
\ins{233pt}{25pt}{$h$}
\ins{233pt}{-15pt}{$h$}
}%
{fig1}{}
\vskip1.5cm
\line{\vtop{\line{\hskip2.5truecm\vbox{\advance\hsize by -5.1 truecm
\0{\ottorm FIG 3. 
The lowest order diagrams contributing to ${\scriptstyle 
<H^{AT}_\xx H^{AT}_\yy>_{\L_M,T}}$.
The wavy lines ending in the points labeled ${\scriptstyle \xx}$ and 
${\scriptstyle \yy}$ represent
the fields ${\scriptstyle \phi_\xx}$ and ${\scriptstyle \phi_\yy}$
respectively. 
The solid lines labeled by ${\scriptstyle h}$
and going from ${\scriptstyle \xx}$ to ${\scriptstyle \yy}$ 
represent the propagators ${\scriptstyle g^{(h)}(\xx-\yy)}$.
The sums are over the scale indeces and, even if not explicitly written, over
the indexes ${\scriptstyle \aa,\oo}$ 
(and the propagators depend on these indexes). 
}
} \hfill} }}
\vskip.5cm

%%%%%%%%%%%%%%%%%%%%%%%%%%%%%%%%%%%%%%%%%%%%%%%%%%%%%%%%%%%%%%%%%%%%%%%%%%%%%%
%%%%%%%%%%%%%%%%%%%%%%%%%%%%%%%%%%%%%%%%%%%%%%%%%%%%%%%%%%%%%%%%%%%%%%%%%%%%%%

The conclusion is that $C_v$, for $\l$ small and
$|t-\sqrt2+1|,|u|\le (\sqrt2-1)/4$, is given by:
$$\eqalign{&C_v={1\over |\L|}\sum_{\xx,\yy\in\L_M}\sum_{\o_1,\o_2=\pm1}
\sum_{h,h'=h^*_2}^{1}{(Z_{h\vee h'}^{(1)})^2\over Z_{h-1} Z_{h'-1}}
\Bigg[G^{(h)}_{(+,\o_1),(+,\o_2)}(\xx-\yy)
G^{(h')}_{(-,-\o_2),(-,-\o_1)}(\yy-\xx)+\cr
&+
G^{(h)}_{(+,\o_1),(-,-\o_2)}(\xx-\yy)
G^{(h')}_{(-,-\o_1),(+,\o_2)}(\xx-\yy)\Bigg]+{1\over |\L|}\sum_{\xx,\yy\in\L_M}
\sum_{h^*_2}^1
\Big({\lis Z_h\over Z_h}\Big)^2\O_{\L_M}^{(h)}(\xx-\yy)\;,\cr}\Eq(wow)$$
where $h\vee h'=\max\{h,h'\}$ and $G^{(h)}_{(\a_1,\o_1),(\a_2,\o_2)}(\xx)$
must be interpreted as
$$G^{(h)}_{(\a_1\o_1),(\a_2,\o_2)}(\xx)=\cases{
g^{(h)}_{(\a_1\o_1),(\a_2,\o_2)}(\xx) & if $h>h^*_1$,\cr
g^{(1,\le h^*_1)}_{\o_1,\o_2}(\xx)+g^{(2,h^*_1)}_{\o_1,\o_2}(\xx)
& if $h=h^*_1$,\cr
g^{(2,h)}_{\o_1,\o_2}(\xx) & if $h^*_2<h<h^*_1$,\cr
g^{(2,\le h^*_2)}_{\o_1,\o_2}(\xx) & if $h=h^*_2$.\cr}$$
Moreover, if $N,n_0,n_1\ge 0$ and $n=n_0+n_1$, 
$|\dpr_{x}^{n_0}\dpr_{x_0}\O_{\L_M}^{(h)}(\xx)|\le C_{N,n}|\l|{\g^{(2+n)h}
\over 1+(\g^h|\dd(\xx)|)^N}$.
Now, calling $\h_c$ the exponent associated to $\lis Z_h/Z_h$, from \equ(wow)
we find:
$$\eqalign{&C_v=-C_1 \g^{2\h_c h^*_1}\log_\g\g^{h^*_1-h^*_2}
\big(1+\O_{h^*_1,h^*_2}^{(1)}(\l)\big)
+C_2{1-\g^{2\h_c(h^*_1-1)}\over 2\h_c}
\big(1+\O^{(2)}_{h^*_1}(\l)\big)\;,\cr}\Eq(7.22)$$
where $|\O^{(1)}_{h^*_1,h^*_2}(\l)|,|\O^{(2)}_{h^*_1}(\l)|
\le c|\l|$, for some $c$.
Note that, defining $\D$ as in \equ(1.6), $\g^{(1-\h_\s)h^*_1}\D^{-1}$
is bounded above and below by $O(1)$ constants. Then, using \equ(eps2),
\equ(1.6) follows.
\\

\appendix(0, Proof of {\equ(2.1)})

We start from eq. (V.2.12) in [MW], expressing the partition
function of the Ising model with periodic boundary condition
on a lattice with an even number of sites
as a combination of the Pfaffians of four matrices with
different boundary conditions, defined by (V.2.10) and (V.2.11) in [MW].
In the general case (\ie $M^2$ not necessarily even), the 
(V.2.12) of [MW] becomes:
$$Z_I=\sum_{\s}e^{-\b J H_I(\s)}=(-1)^{M^2}{1\over 2}(2\cosh\b J)^{M^2}
\Big(-\Pf \lis A_1+\Pf \lis A_2+\Pf \lis A_3+\Pf \lis A_4\Big)\;,\Eqa(0.1)$$
where
$\lis A_i$ are matrices with elements $(\lis A_i)_{\xx,j;\yy,k}$, with 
$\xx,\yy\in \L_M$, $j,k=1,\ldots,6$, given by:
$$\eqalign{&({\lis A_i})_{\xx;\xx}=\pmatrix{0&0&-1&0&0&1\cr
0&0&0&-1&1&0\cr
1&0&0&0&0&-1\cr
0&1&0&0&-1&0\cr
0&-1&0&1&0&1\cr
-1&0&1&0&-1&0\cr}\cr
%&({\lis A_i})_{\xx;\xx+\hat e_1}=\pmatrix{
%0&t&0&0&0&0\cr
%0&0&0&0&0&0\cr
%0&0&0&0&0&0\cr
%0&0&0&0&0&0\cr
%0&0&0&0&0&0\cr
%0&0&0&0&0&0\cr}=-({\lis A_i}^T)_{\xx+\hat e_1;\xx}\cr
%&({\lis A_i})_{\xx;\xx+\hat e_0}=\pmatrix{
%0&0&0&0&0&0\cr
%0&0&0&0&0&0\cr
%0&0&0&t&0&0\cr
%0&0&0&0&0&0\cr
%0&0&0&0&0&0\cr
%0&0&0&0&0&0\cr}=-({\lis A_i}^T)_{\xx+\hat e_0;\xx}\cr
}\Eqa(0.2)$$
and $\big(({\lis A_i})_{\xx;\xx+\hat e_1}\big)_{i,j}=t\d_{i,1}\d_{j,2}$,
$\big(({\lis A_i})_{\xx;\xx+\hat e_0}\big)_{i,j}=t\d_{i,2}\d_{j,1}$,
$({\lis A_i})_{\xx;\xx+\hat e_1}=-({\lis A_i}^T)_{\xx+\hat e_1;\xx}$,
$({\lis A_i})_{\xx;\xx+\hat e_0}=-({\lis A_i}^T)_{\xx+\hat e_0;\xx}$; 
moreover
$$\eqalign{& (\lis A_i)_{(M,x_0);(1,x_0)}=-(\lis A_i^T)_{(1,x_0);(M,x_0)}=
(-1)^{[{i-1\over 2}]}(\lis A_i)_{(1,x_0);(2,x_0)}\cr
& (\lis A_i)_{(x,M);(x,1)}=-(\lis A_i^T)_{(x,1);(x,M)}=
(-1)^{i-1}(\lis A_i)_{(x,1);(x,2)}\;,\cr}\Eqa(0.3)$$
where $[{i-1\over 2}]$ is the bigger integer $\le {i-1\over 2}$;
in all the other cases the matrices $(\lis A_i)_{\xx,\yy}$ are 
identically zero.

Given a $(2n)\times(2n)$ antisymmetric matrix $A$, it is well--known
that $\Pf A=(-1)^n\int d\psi_1\cdots d\psi_{2n}\cdot$ $\cdot\exp\{{1\over 2}
\sum_{i,j}\psi_i A_{ij}\psi_j\}$,  
where $\psi_1,\ldots,\psi_{2n}$ are Grassmanian variables. 
Then, we can rewrite \equ(0.1) as:
$${1\over 2}(2\cosh\b J)^{M^2}
\sum_{\g}(-1)^{\d_\g}\int \prod_{\xx\in\L_M}d\lis H^\g_\xx d H^\g_\xx 
d\lis V^\g_\xx 
d V^\g_\xx d\lis T^\g_\xx d T^\g_\xx e^{S^\g(t;H,V,T)}\;,\Eqa(0.5)$$
where: $\g=(\e,\e')$; $\e,\e'=\pm 1$; $\d_\g$ is defined after \equ(2.1); 
$\lis H^\g_\xx, H^\g_\xx,
\lis V^\g_\xx, V^\g_\xx$
are Grassmanian variables with $\e$--periodic resp. $\e'$--periodic
boundary conditions in vertical resp. horizontal direction, see \equ(2.3) 
and following lines. Furthermore:
$$\eqalign{&S^\g(t;H,V,T)=t\sum_\xx\Big[\lis H^\g_\xx 
H^\g_{\xx+\hat e_1}
+\lis V^\g_\xx V^\g_{\xx+\hat e_0}\Big]+\cr
&+\sum_\xx\Big[\lis V^\g_\xx \lis H^\g_\xx+\lis H^\g_\xx T^\g_\xx
+V^\g_\xx H^\g_\xx+H^\g_\xx \lis T^\g_\xx
+T^\g_\xx \lis V^\g_\xx+\lis T^\g_\xx V^\g_\xx
+\lis T^\g_\xx T^\g_\xx\Big]\;.\cr}\Eqa(0.6)$$
The $T$--fields appear only in the diagonal elements and they can be easily
integrated out:
$$\eqalign{&
\prod_{\xx\in\L_M}\int d\lis T^\g_\xx d T^\g_\xx \exp\Big\{
\lis H^\g_\xx T^\g_\xx
+H^\g_\xx \lis T^\g_\xx
+T^\g_\xx \lis V^\g_\xx+\lis T^\g_\xx V^\g_\xx
+\lis T^\g_\xx T^\g_\xx\Big\}=\cr
&=\prod_{\xx\in\L_M}(-1-\lis H^\g_\xx H^\g_\xx-\lis V^\g_\xx V^\g_\xx-
V^\g_\xx\lis H^\g_\xx-V^\g_\xx\lis H^\g_\xx)=\cr
&=(-1)^M\exp\sum_{\xx\in\L_M}\Big[
\lis H^\g_\xx H^\g_\xx+\lis V^\g_\xx V^\g_\xx+
V^\g_\xx\lis H^\g_\xx+H^\g_\xx\lis V^\g_\xx\Big]\;,\cr}\Eqa(0.7)$$
where in the last identity we used that $\Big[
\lis H^\g_\xx H^\g_\xx+\lis V^\g_\xx V^\g_\xx+
V^\g_\xx\lis H^\g_\xx+H^\g_\xx\lis V^\g_\xx\Big]^2=0$.
Substituting \equ(0.7) into \equ(0.5) we find \equ(2.1).\\

\appendix(a1, Integration of the heavy fermions. Symmetry properties)

\asub(A1.1){\it Integration of the $\c$ fields.}
Calling $\lis\VV(\psi,\c)=Q(\psi,\c)-\n F_\s(\psi)+
V(\psi,\c)$, we obtain
$$-\widetilde E_1 M^2-Q^{(1)}(\psi)-\VV^{(1)}(\psi)=\log
\int P(d\c)e^{\lis\VV(\psi,\c)}=\sum_{n=0}^{\io}{(-1)^n\over n!}\EE^T_\chi(
\lis\VV(\psi,\c);n)\;,\Eqa(8.1)$$
where $\widetilde E_1$ is a constant
and $\VV^{(1)}$ is at least quadratic in $\psi$
and vanishing when $\l=\n=0$.
$Q^{(1)}$ is the rest (quadratic in $\psi$).
Given $s$ set of labels $P_{v_i}$, $i=1,\ldots,s$ and 
$\widetilde\c(P_{v_i})\defin
\prod_{f\in P_{v_i}}\c^{\a(f)}_{\o(f),\xx(f)}$, the truncated expectation
$\EE^T_\c(\widetilde\chi(P_{v_1}),\ldots,\widetilde\chi(P_{v_s}))$ 
can be written as
$$\EE^T_\c(\widetilde\chi(P_{v_1}),\ldots,\widetilde\chi(P_{v_s}))=
\sum_{T}\a_T\prod_{\ell\in T}
g_\c(f^1_\ell,f^2_\ell)
\int dP_{T}(\tt) \Pf G^{T}(\tt)\Eqa(8.7)$$
where: $T$ is a set of lines forming an {\it anchored tree} between
the cluster of poins $P_{v_1},\ldots,P_{v_s}$ \ie $T$ is a set
of lines which becomes a tree if one identifies all the points
in the same clusters;
$\tt=\{t_{i,i'}\in [0,1], 1\le i,i' \le s\}$, $dP_{T}(\tt)$
is a probability measure with support on a set of $\tt$ such that
$t_{i,i'}=\uu_i\cdot\uu_{i'}$ for some family of vectors $\uu_i\in \RRR^s$ of
unit norm;
$\a_T$ is a sign (irrelevant for the subsequent bounds);
$f^1_\ell, f^2_\ell$ are the field labels associated
to the points connected by $\ell$; if 
$\underline a(f)=(\a(f),\o(f))$, the propagator 
$g_\c(f,f')$ is equal to
$$g_\c(f,f')=
g^\c_{\underline a(f),\underline a(f')}(\xx(f)-\xx(f'))\defin
<\c^{\a(f)}_{\o(f),\xx(f)}
\c^{\a(f')}_{\o(f'),\xx(f')}>\;;\Eqa(1000)$$
if $2n=\sum_{i=1}^s|P_{v_i}|$, then 
$G^{T}(\tt)$ is a $(2n-2s+2)\times (2n-2s+2)$ antisymmetrix matrix, whose
elements are given by $G^{T}_{f,f'}=t_{i(f),i(f')}g_\c(f,f')$, where:
$f, f'\not\in F_T$ and $F_T\defin\cup_{\ell\in T}\{f^1_\ell,f^2_\ell\}$;
$i(f)$ is s.t. $f\in P_{i(f)}$; finally $\Pf G^T$ is the {\it Pfaffian} 
of $G^T$. 
If $s=1$ the sum over $T$ is empty, but we can still
use the above equation by interpreting the r.h.s.
as $1$ if $P_{v_1}$ is empty, and ${\rm det} G(P_1)$ otherwise.
\\

{\bf Sketch of the proof of \equ(8.7).} Equation \equ(8.7) 
is a trivial generalization of the well--known formula expressing 
truncated fermionic expectations in terms of sums of determinants [Le]. The
only difference here is that the propagators $<\c^\a_{\o_1,\xx_1}\c^\a_{
\o_2,\xx_2}>$ are not vanishing, so that Pfaffians appear instead of 
determinants. The proof can be done along the same lines
of Appendix A3 of [GM].
The only difference here is that the identity known as 
the {\it Berezin integral}, see (A3.15) of [GM], that is the starting
point to get to \equ(8.7), must be replaced by the (more general) identity:
$$\EE_\c\Big(\prod_{j=1}^{s}\widetilde\c(P_{j}) \Big)
= \Pf G = (-1)^n\int {\cal D}\c \, \exp\Big[
{1\over 2}\left( \c,G\c \right) \Big] \;,\Eqa(A3.12) $$
where: the expectation $\EE_\c$ is w.r.t. $P(d\c)$; 
if $2m=\sum_{j=1}^{s} \left| P_{j} \right|$, 
$G$ is the $2m\times 2m$ antisymmetric matrix with entries
$G_{f,f'}=g^\c_{\underline a(f),\underline a(f')}(\xx(f)-\xx(f'))$; and
$${\cal D}\c=\prod_{j=1}^{n}
\prod_{f\in P_{j}} \der\c^{\a(f)}_{\xx(f),\o(f)} 
\quad \quad\left ( \c,G\c \right) =
\sum_{f,f'\in \cup_{i}P_i}
\c^{\a(f)}_{\xx(f),\o(f)}G_{f,f'}\c^{\a(f')}_{\xx(f'),\o(f')}\;.
\Eqa(A3.9) $$
Starting from \equ(A3.12), the proof in Appendix A3 of [GM] can be repeated 
step by step in the present case, to find finally the analogue of (A.3.55)
of [GM]. Then, using again that $\int{\cal D}\c \exp(\c, G\c)/2$ is,
unless for a sign, the Pfaffian of $G$, we find \equ(8.7).\qed\\

We now use the
well--known property $|\Pf G^T|=\sqrt{|\det G^T|}$ and we can bound
$\det G^T$ by Gram--Hadamard (GH) inequality. 
Let $\HH\defin\RRR^s\otimes\HH_0$, where
$\HH_0$ is the Hilbert space of complex
four dimensional vectors $F(\kk)=(F_1(\kk),\ldots,F_4(\kk)$), $F_i(\kk)$
being a function on the set $\DD_{-,-}$, with scalar product
$<F,G>=\sum_{i=1}^4 {1/M^2}\sum_{\kk} F^*_i(\kk) G_i(\kk)$.
We can write the elements of $G^T$ as inner products of
vectors of $\HH$:
$$G_{f,f'}=t_{i(f),i(f')} g_\c(f,f')=
<\uu_{i(f)}\otimes A_{f},
\uu_{i(f')}\otimes B_{f'}>\;,\Eqa(3.97)$$
where $\uu_i\in \RRR^s$, $i=1,\ldots,s$, are vectors such that
$t_{i,i'}=\uu_i\cdot\uu_{i'}$, and, if 
$\hat g^\c_{\underline a,\underline a'}(\kk)$ is the Fourier transform of
$g^\c_{\underline a,\underline a'}(\xx-\yy)$, 
$A_f(\kk)$ and $B_{f'}(\kk)$ are given by
$$\eqalign{& A_f(\kk)=e^{-i\kk\xx(f)}\Big(\hat g^\c_{
\underline a(f),(-,1)}(\kk),
\hat g^\c_{\underline a(f),(-,-1)}(\kk),
\hat g^\c_{\underline a(f),(+,1)}(\kk),
\hat g^\c_{\underline a(f),(+,-1)}(\kk)\Big)\;,\cr
&B_{f'}(\kk)=e^{-i\kk\xx(f')}\cases{(1,0,0,0),& if $\underline a(f')=(-,1)$,\cr
(0,1,0,0),& if $\underline a(f')=(-,-1)$,\cr
(0,0,1,0),& if $\underline a(f')=(+,1)$,\cr
(0,0,0,1),& if $\underline a(f')=(+,-1)$,\cr}}
\Eqa(3.98m)$$
With these definitions and remembering \equ(b),
it is now clear that $|Pf G^T|\le C^{n-s+1}$, for some constant $C$.
Then, applying \equ(8.7) and the previous bound we find the second of
\equ(2.40).\\

We now turn to the construction of $P_{Z_1,\s_1,\m_1,C_1}$, in order to prove 
\equ(2.38). 

We define $e^{-t_1 M^2} P_{Z_1,\s_1,\m_1,C_1}(d\psi)
\defin P_\s(d\psi)e^{-Q^{(1)}(\psi)}$,
where $t_1$ is a normalization constant.
In order to write $P_{Z_1,\s_1,\m_1,C_1}(d\psi)$ 
as an exponential of a quadratic form, it is sufficient to calculate
the correlations
$$\eqalign{<\psi^{\a_1}_{\o_1,\kk}
\psi^{\a_2}_{\o_2,-\a_1\a_2\kk}>_1&\defin
\int P_{Z_1,\s_1,\m_1,C_1}(d\psi)\psi^{\a_1}_{\o_1,\kk}
\psi^{\a_2}_{\o_2,-\a_1\a_2\kk}=\cr
&=e^{-t_1M^2}
\int P_\s(d\psi)P(d\c)e^{Q(\c,\psi)}\psi^{\a_1}_{\o_1,\kk}
\psi^{\a_2}_{\o_2,-\a_1\a_2\kk}\;.\cr}\Eqa(corr)$$
It is easy to realize that the measure
$\sim P_\s(d\psi)P(d\c)e^{Q(\c,\psi)}$ 
factorizes into the product of two
measures generated by the fields $\psi^{(j)}_{\o,\xx}$, $j=1,2$,
defined by $\psi^\a_{\o,\xx}=(\psi^{(1)}_{\o,\xx}+i(-1)^\a\psi^{(2)}_{
\o,\xx})/\sqrt2$.
In fact, using this change of variables, one finds that 
$$ P_\s(d\psi)P(d\c)e^{Q(\c,\psi)}=\prod_{j=1,2}P^{(j)}(d\psi^{(j)},d\c^{(j)})=
\prod_{j=1,2}{1\over \NN^{(j)}}\exp\{-{t_\l^{(j)}\over 4M^2}\sum_{\kk}
\x^{(j),T}_\kk C^{(j)}_\kk \x^{(j)}_{-\kk}\}\;,\Eqa(8.17)$$
for two suitable matrices $C^{(j)}_\kk$, whose determinants 
$B^{(j)}(\kk)\defin \det C^{(j)}_\kk$ are equal to
$$B^{(j)}(\kk)={16\over (t_\l^{(j)})^4}\big\{
2t_\l^{(j)}[1-(t_\l^{(j)})^2](2-\cos k-\cos k_0)+(t_\l^{(j)}-t_\psi)^2
(t_\l^{(j)}-t_\c)^2\big\}\Eqa(8.18)$$
From the explicit expression of $C^{(j)}_\kk$ one finds 
$$\eqalign{
&<\psi^{(j)}_{-\kk}\psi^{(j)}_{\kk}>_1={4M^2\over t_\l^{(j)}}
{c_{1,1}^{(j)}(\kk)\over B^{(j)}(\kk)}\virg 
<\lis\psi^{(j)}_{-\kk}\psi^{(j)}_{\kk}>_1={4M^2\over t_\l^{(j)}}
{c_{-1,1}^{(j)}(\kk)\over B^{(j)}(\kk)}\virg\cr
&<\lis\psi^{(j)}_{-\kk}\lis\psi^{(j)}_{\kk}>_1={4M^2\over t_\l^{(j)}}
{c_{-1,-1}^{(j)}(\kk)\over B^{(j)}(\kk)}\;,\cr}\Eqa(8.20)$$
where, if $\o=\pm 1$, recalling that $t_\psi=\sqrt 2 -1+\n/2$ and 
defining $t_\c=-\sqrt 2-1$,
$$\eqalign{c_{\o,\o}^{(j)}(\kk)\defin&
{4\over (t_\l^{(j)})^2}\big\{2t_\l^{(j)}t_\c(-i\sin k\cos k_0+\o 
\sin k_0\cos k)+[(t_\l^{(j)})^2+t_\c^2](i\sin k-\o\sin k_0)\big\}\cr
c^{(j)}_{\o,-\o}(\kk)\defin& -i\o{4\over (t_\l^{(j)})^2}\big\{
-t_\l^{(j)}(3t_\c+t_\psi)\cos k\cos k_0+[(t_\l^{(j)})^2+2t_\c t_\psi+
t_\c^2](\cos k+\cos k_0)-\cr
&\qquad\qquad -\big(t_\l^{(j)}(t_\psi+t_\c)+2{t_\psi t_\c^2\over 
t_\l^{(j)}}\big)\big\}\;.\cr}\Eqa(8.21)$$
It is clear that, for any monomial $F(\psi^{(j)})$,
$\int P(d\psi^{(j)},d\c^{(j)})F(\psi^{(j)})=\int
P^{(j)}(d\psi^{(j)})F(\psi^{(j)})$, with
$$\eqalign{&
P^{(j)}(d\psi^{(j)})\defin{1\over N_j}\prod_\kk d\psi^{(j)}_\kk 
d\lis\psi^{(j)}_\kk\cdot\cr
&\cdot\exp\Bigl\{-{t_\l^{(j)}B^{(j)}(\kk)\over 4M^2\det c^{(j)}_\kk}
(\psi^{(j)}_\kk, \lis\psi^{(j)}_{\kk})\left(\matrix{c^{(j)}_{-1,-1}(\kk)
&-c^{(j)}_{1,-1}(\kk)\cr-c^{(j)}_{-1,1}(\kk)&c^{(j)}_{1,1}(\kk)\cr}\right)
\left(\matrix{\psi^{(j)}_{-\kk}\cr \lis\psi^{(j)}_{-\kk}\cr}\right)
\Bigr\}\;,\cr}\Eqa(8.23)$$
where $\det c^{(j)}_\kk=c^{(j)}_{1,1}(\kk)c^{(j)}_{-1,-1}(\kk)-
c^{(j)}_{1,-1}(\kk)c^{(j)}_{-1,1}(\kk)$.
If we now use the identity $t_\l^{(j)}=t_\psi(2+(-1)^j\m)/(2-\s)$
and rewrite
the measure $P^{(1)}(d\psi^{(1)})P^{(2)}(d\psi^{(2)})$
in terms of $\psi^{\pm}_{\o,\kk}$ we find:
$$P^{(1)}(d\psi^{(1)})P^{(2)}(d\psi^{(2)})=
{1\over \NN^{(1)}}\prod_{\kk,\o} d\psi^{+}_{\o,\kk}d\psi^{-}_{\o,\kk}
\exp\{-{Z_1 C_1(\kk)
\over 4M^2}\Psi^{+,T}_\kk A_\psi^{(1)}\Psi^-_\kk\}=
P_{Z_1,\s_1,\m_1,C_1}(d\psi)\;,\Eqa(8.25)$$
with $C_1(\kk)$, $Z_1$, $\s_1$ and $\m_1$
defined as after \equ(2.36), and
$A^{(1)}_\psi(\kk)$ as in \equ(2.38), with
$$\eqalign{& M^{(1)}(\kk)={2\over 2-\s}\left(\matrix{-c^+_{-1,-1}(\kk)&
c^+_{-1,1}(\kk)\cr
c^+_{1,-1}(\kk) & -c^+_{1,1}(\kk)\cr}\right)\virg
N^{(1)}(\kk)={2\over 2-\s}\left(\matrix{-c^-_{-1,-1}(\kk)&
c^-_{-1,1}(\kk)\cr
c^-_{1,-1}(\kk) & -c^-_{1,1}(\kk)\cr}\right)\;,\cr}
\Eqa(8.27)$$
where $c^\a_{\o_1,\o_2}(\kk)\defin 
[(1-\m/2)B^{(1)}(\kk)c^{(1)}_{\o_1,\o_2}(\kk)/\det c^{(1)}_\kk+\a(1+\m/2)
B^{(2)}(\kk)c^{(2)}_{\o_1,\o_2}(\kk)/\det c^{(2)}_\kk]/2$.
It is easy to verify that $A_\psi^{(1)}(\kk)$
has the form \equ(2.38).
In fact, computing the functions in \equ(8.27), one finds that, for $\kk$,
$\s_1$ and $\m_1$ small,
$$\eqalign{&M^{(1)}(\kk)=\left(\matrix{\big(1+{\s_1\over 2}
\big)(i\sin k+\sin k_0)+O(\kk^3) &
-i\s_1+O(\kk^2)\cr
i\s_1+O(\kk^2)
&\big(1+{\s_1\over 2}
\big)(i\sin k-\sin k_0)+O(\kk^3)\cr}\right)\cr
&N^{(1)}(\kk)=\left(\matrix{-{\m_1\over 2}(i\sin k+\sin k_0)+O(\kk^3)&
i\m_1+O(\m_1\kk^2)\cr
-i\m_1+O(\m_1\kk^2) & 
-{\m_1\over 2}(i\sin k-\sin k_0)+O(\kk^3)\cr}\right)\;,\cr}
\Eqa(8.28)$$
where the higher order terms in $\kk$, $\s_1$ and $\m_1$ 
contribute to the corrections $a_1^\pm(\kk)$, $b_1^{\pm}(\kk)$,
$c_1(\kk)$ and $d_1(\kk)$. They have the reality and parity properties 
described after \equ(2.38) and it is appearent that $a_1^\pm(\kk)=O(\s_1\kk)
+O(\kk^3)$, $b_1^{\pm}(\kk)=O(\m_1\kk)+O(\kk^3)$,
$c_1(\kk)=O(\kk^2)$ and $d_1(\kk)=O(\m_1\kk^2)$.\\

\asub(A1.2){\it Symmetry properties.}
In this section we identify some symmetries of 
model \equ(2.11) and we prove that the quadratic and quartic terms
in $\VV^{(1)}$ have the structure described in \equ(8.29), \equ(8.30)
and \equ(8.30a).

The formal action appearing in \equ(2.11) (see also
\equ(2.2) and \equ(2.15) for an explicit form)
is invariant under the following transformations.\\  
\\
\01) {\it Parity}: $H_\xx^{(j)}\to\lis H_{-\xx}^{(j)}$, $\lis H_\xx^{(j)}\to 
-H_{-\xx}^{(j)}$ (the same for $V$ and $\lis V$).
In terms of the variables $\hat\psi^{\a}_{\o,\kk}$, this transformation
is equivalent to $\hat\psi^\a_{\o,\kk}\to i\o\hat\psi^\a_{\o,-\kk}$ 
(the same for $\c$) and we shall call it {\it parity}.\\
\\
\02) {\it Complex conjugation}:
$\hat\psi^\a_{\o,\kk}\to\hat\psi^{-\a}_{-\o,\kk}$ (the same for $\c$)
and $c\to c^*$,
where $c$ is a generic constant appearing in the formal action and
$c^*$ is its complex conjugate. Note that \equ(2.19) is left invariant by
this transformation, that we shall 
call {\it complex conjugation}.\\
\\
\03) {\it Hole-particle}:
$H_\xx^{(j)}\to(-1)^{j+1}H_{\xx}^{(j)}$ (the same for $\lis H,V,\lis V$).
This transformation is equivalent to $\hat\psi^\a_{\o,\kk}\to\hat
\psi^{-\a}_{\o,-\kk}$ (the same for $\c$) and we shall 
call it {\it hole-particle}.\\

\04) {\it Rotation}:
$H_{x,x_0}^{(j)}\to i\lis V_{-x_0,-x}^{(j)}$, 
$\lis H_{x,x_0}^{(j)}\to i V_{-x_0,-x}^{(j)}$,
$V_{x,x_0}^{(j)}\to i\lis 
H_{-x_0,-x}^{(j)}$, $\lis V_{x,x_0}^{(j)}\to i H_{-x_0,-x}^{(j)}$.
This transformation is equivalent to 
$$\hat\psi^\a_{\o,(k,k_0)}\to-\o e^{-i\o\p/4}\hat
\psi^{\a}_{-\o,(-k_0,-k)}\virg\hat\c^\a_{\o,(k,k_0)}\to\o e^{-i\o\p/4}\hat
\c^{\a}_{-\o,(-k_0,-k)}\Eqa(rot)$$ 
and we shall call it {\it rotation}.\\

\05) {\it Reflection}:
$H_{x,x_0}^{(j)}\to i\lis H_{-x,x_0}^{(j)}$, $\lis H_{x,x_0}^{(j)}\to i 
H_{-x,x_0}^{(j)}$,
$V_{x,x_0}^{(j)}\to -iV_{-x,x_0}^{(j)}$, $\lis V_{x,x_0}^{(j)}\to 
i \lis V_{-x,x_0}^{(j)}$.
This transformation is equivalent to $\hat\psi^\a_{\o,(k,k_0)}\to i\hat
\psi^{\a}_{-\o,(-k,k_0)}$
(the same for $\c$) and we shall call it {\it reflection}.\\

\06) {\it The $(1)\otto (2)$ symmetry:}
$H_{\xx}^{(1)}\otto H_{\xx}^{(2)}$, 
$\lis H_{\xx}^{(1)}\otto\lis H_{\xx}^{(2)}$, 
$V_{\xx}^{(1)}\otto V_{\xx}^{(2)}$, 
$\lis V_{\xx}^{(1)}\otto\lis V_{\xx}^{(2)}$, $u\to -u$.
This transformation is equivalent to
$\hat\psi^\a_{\o,\kk}\to-i\a\hat\psi^{-\a}_{\o,-\kk}$ (the same for $\c$)
together with $u\to -u$ and we shall call it {\it 
$(1)\otto(2)$ symmetry}.\\

It is easy to verify that the quadratic forms $P(d\c)$, $P(d\psi)$ 
and $P_{Z_1,\s_1,\m_1,C_1}(d\psi)$
are separately invariant
under the symmetries above. Then the effective action $\VV^{(1)}(\psi)$
is still invariant under the same symmetries.
Using the invariance of $\VV^{(1)}$ under transformations (1)--(6), we now
prove that the structure of its quadratic and quartic terms is the one
described in Theorem 2.1, see in particular \equ(8.29), \equ(8.30)
and \equ(8.30a).\\

\0{\it Quartic term.}
The term $\sum_{\kk_i}W(\kk_1,\kk_2,\kk_3,\kk_4)
\hat\psi^+_{1,\kk_1}\hat\psi^+_{-1,\kk_2}
\hat\psi^-_{-1,\kk_3}\hat\psi^-_{1,\kk_4}\d(\kk_1+\kk_2-\kk_3-\kk_4)$
under complex conjugation becomes equal to
$\sum_{\kk_i}W^*(\kk_1,\kk_2,\kk_3,\kk_4)
\hat\psi^-_{-1,\kk_1}\hat\psi^-_{1,\kk_2}
\hat\psi^+_{1,\kk_3}\hat\psi^+_{-1,\kk_4}\d(\kk_3+\kk_4-\kk_1-\kk_2)$,
so that $W(\kk_1,\kk_2,\kk_3,\kk_4)=W^*(\kk_3,\kk_4,\kk_1,\kk_2)$.
Then, defining $L_1=W(\bk++,\bk++,\bk++,\bk++)$, where
$\bk++=(\p/M,\p/M)$,
and $l_1=\PP_0 L_1\defin L_1\big|_{\s_1=\m_1=0}$,
we see that $L_1$ and $l_1$ are
real. From the explicit computation of the 
lower order term we find $l_1=\l+O(\l^2)$.\\

\0{\it Quadratic terms.} 
We distinguish 4 cases (items (a)--(d) below).\\
{\it \0a)} Let $\a_1=-\a_2=+$ and $\o_1=-\o_2=\o$ and
consider the expression $\sum_{\o,\kk}W_\o(\kk;\m_1)\hat\psi^+_{\o,\kk}
\hat\psi^-_{-\o,\kk}$.
Under parity it becomes 
$\sum_{\o,\kk}W_\o(\kk;\m_1)
(i\o)\hat\psi^+_{\o,-\kk}(-i\o)\hat\psi^-_{-\o,-\kk}=
\sum_{\o,\kk}
W_\o(-\kk;\m_1)\hat\psi^+_{\o,\kk}\hat\psi^-_{-\o,\kk}$,
%\;,\Eqa(B1)$$
%
so that $W_\o(\kk;\m_1)$ is even in $\kk$.\\
Under complex conjugation it becomes 
$\sum_{\o,\kk}W_\o(\kk;\m_1)^*
\hat\psi^-_{-\o,\kk}\hat\psi^+_{\o,\kk}=-\sum_{\o,\kk}W_\o(\kk;\m_1)^*
\hat\psi^+_{\o,\kk}\hat\psi^-_{-\o,\kk}$,
%\;,\Eqa(B2)$$
%
so that $W_\o(\kk;\m_1)$ is purely imaginary.\\
Under hole-particle it becomes 
$\sum_{\o,\kk}W_\o(\kk;\m_1)\hat\psi^-_{\o,-\kk}
\hat\psi^+_{-\o,-\kk}=-\sum_{\o,\kk}W_{-\o}(\kk;\m_1)\hat\psi^+_{\o,\kk}
\hat\psi^-_{-\o,\kk}$,
%\;,\Eqa(B3)$$
%
so that $W_\o(\kk;\m_1)$ is odd in $\o$.\\
Under $(1)\otto(2)$ it becomes
$\sum_{\o,\kk}W_\o(\kk;-\m_1)(-i)
\hat\psi^-_{-\o,-\kk}(i)\hat\psi^+_{\o,-\kk}=
\sum_{\o,\kk}W_\o(\kk;-\m_1)\hat\psi^+_{\o,\kk}\hat\psi^-_{-\o,\kk}$,
%\;,\Eqa(B4)$$
%
so that $W_\o(\kk;\m_1)$ is even in $\m_1$. 
Let us define $S_1=i\o/2\sum_{\h,\h'=\pm 1}W_\o(\bk\h{\h'})$,
where $\bk\h{\h'}=(\h\p/M,\h'\p/M)$, and $\g n_1=\PP_0 S_1$,
$s_1=\PP_1 S_1=\s_1\dpr_{\s_1}S_1\big|_{\s_1=\m_1=0}+
\m_1\dpr_{\m_1}S_1\big|_{\s_1=\m_1=0}$. From the 
previous discussion we see that
$S_1, s_1$ and $n_1$ are real and $s_1$ is independent
of $\m_1$. From the computation of the lower order terms
we find $s_1=O(\l\s_1)$ and $\g n_1=\n+c^\n_1\l+O(\l^2)$,
for some constant $c^\n_1$ independent of $\l$. 
Note that, since $W_\o(\kk;\m_1)$ is even in $\kk$ 
(so that in particular no linear terms in $\kk$ appear) 
in real space no terms of the form
$\psi^+_{\o,\xx}\dpr\psi^-_{-\o,\xx}$ can appear.\\
{\it \0b)} Let $\a_1=\a_2=\a$ and $\o_1=-\o_2=\o$ and 
consider the expression $\sum_{\o,\a,\kk}W_\o^\a(\kk;\m_1)
\hat\psi^\a_{\o,\kk}
\hat\psi^\a_{-\o,-\kk}$. We proceed as in item (a) and, 
by using parity, we see that $W_\o^\a(\kk;\m_1)$ is even in $\kk$ 
and odd in $\o$.\\ 
By using complex conjugation, we see that 
$W_\o^\a(\kk;\m_1)=-W^{-\a}_{\o}(\kk;\m_1)^*$.\\
By using hole-particle, we see that 
$W_\o^\a(\kk;\m_1)$ is even in $\a$ and $W_\o^\a(\kk;\m_1)=-
W^{-\a}_{\o}(\kk;\m_1)^*$ implies that $W_\o^\a(\kk;\m_1)$ 
is purely imaginary.\\
By using $(1)\otto(2)$ we see that 
$W_\o^\a(\kk;\m_1)$ is odd in $\m_1$.

If we define 
$M_1=-i\o/2\sum_{\h,\h'}W_\o^\a(\bk\h{\h'};\m_1)$ 
and $m_1=\PP_1 M_1$, from the previous properties follows
that $M_1$ and $m_1$ are real, $m_1$ is independent of $\s_1$
and, from the computation of its lower order, $m_1=O(\l\m_1)$.
Note that, since $W_\o^\a(\kk;\m_1)$ is even in $\kk$ 
(so that in particular no linear terms in $\kk$ appear) 
in real space no terms of the form
$\psi^\a_{\o,\xx}\dpr\psi^\a_{-\o,\xx}$ can appear.\\
{\it \0c)} Let $\a_1=-\a_2=+$, $\o_1=\o_2=\o$ and consider 
the expression $\sum_{\o,\kk}W_\o(\kk;\m_1)\hat\psi^+_{\o,\kk}
\hat\psi^-_{\o,\kk}$. By using parity we see that
$W_\o(\kk;\m_1)$ is odd in $\kk$.\\
By using reflection we see that $W_\o(k,k_0;\m_1)=W_{-\o}(k,-k_0;\m_1)$.\\
By using complex conjugation we see that
$W_\o(k,k_0;\m_1)=W^*_{\o}(-k,k_0;\m_1)$.\\
By using rotation we find $W_\o(k,k_0;\m_1)=-i\o W_\o(k_0,-k;\m_1)$.\\
By using $(1)\otto(2)$ we see that
$W_{\o}(\kk;-\m_1)$ is even in $\m_1$.

If we define 
$$G_1(\kk)={1\over 4}\sum_{\h,\h'}W_\o(\bk\h{\h'};\m_1)(\h
{\sin k\over \sin\p/M}+\h'{\sin k_0\over \sin\p/M})\= a_\o\sin k+ 
b_\o\sin k_0\;,\Eqa(g1)$$ 
it can be easily verified that the previous properties imply
that
$$a_\o=a_{-\o}=-a^*_\o=i\o b_\o\defin ia\virg
b_\o=-b_{-\o}=b^*_\o=-i\o a_\o\defin \o b=-i\o i a\Eqa(4.38)$$
with $a=b$ real and independent of $\o$. As a consequence,
$G_1(\kk)=G_1(i\sin k+\o\sin k_0)$ for some real constant $G_1$. If $z_1
\defin\PP_0 G_1$ and we compute the lowest order contribution to $z_1$,
we find $z_1=O(\l^2)$.\\
{\it\0d)} Let $\a_1=\a_2=\a$, $\o_1=\o_2=\o$ and consider the expression
$\sum_{\a,\o,\kk}W_\o^\a(\kk;\m_1)\hat\psi^\a_{\o,\kk}\hat\psi^\a_{\o,-\kk}$.
Repeating the proof in item (c) we see that 
$W_\o^\a(\kk;\m_1)$ is odd in $\kk$ and in $\m_1$ and, if we define
$F_1(\kk)={1\over 4}\sum_{\h,\h'}W_\o^\a(\bk\h{\h'};\m_1)(\h
{\sin k\over \sin\p/M}+\h'{\sin k_0\over \sin\p/M})$, 
we can rewrite $F_1(\kk)=F_1(i\sin k+\o\sin k_0)$. Since 
$W_\o^\a(\kk;\m_1)$ is odd in $\m_1$, we find $F_1=O(\l\m_1)$.\\

Note that, with the definition of $\LL$ introduced in \sec(4.2), the 
result of the previous discussion is the following:
$$\LL\VV^{(1)}(\psi)=(s_1+\g n_1) F_\s^{(\le 1)}
+m_1 F^{(\le 1)}_{\m}+l_1
F_\l^{(\le 1)} +z_1 F_\z^{(\le 1)}
\;,\Eqa(BB.10) $$
where $s_1,n_1,m_1,l_1$ and $z_1$ are real constants and:
$s_1$ is linear in $\s_1$ and independent of $\m_1$; 
$m_1$ is linear in $\m_1$ and independent of $\s_1$;
$n_1,l_1,z_1$ are independent of $\s_1,\m_1$;
moreover $F_\s^{(\le 1)}$, $F_\m^{(\le 1)}$, $F_\l^{(\le 1)}$,
$F_\z^{(\le 1)}$ are defined by \equ(4.11) with $h=1$.\\

\0{\bf Proof of Lemma 3.1.}
The symmetries (1)--(6) discussed above are preserved by the iterative 
integration procedure. In fact it is easy to verify that
$\LL\VV^{(h)}$, $\RR\VV^{(h)}$ and $P_{Z_{h-1},\s_{h-1},\m_{h-1},\widetilde 
f_h}(d\psi^{(h)})$ are, step by step, separately
invariant under the transformations (1)--(6). Then Lemma 3.1 can be proven 
exactly in the same way \equ(BB.10) was proven above.\\

\0{\bf Proof of Lemma 3.2.}
It is sufficient to note that the symmetry 
properties discussed above imply that: $\LL_1 W_{2,\aa,\oo}=0$ if 
$\o_1+\o_2=0$; $\LL_0 W_{2,\aa,\oo}=0$ if 
$\o_1+\o_2\not=0$; $\PP_0 W_{2,\aa,\oo}=0$ if 
$\a_1+\a_2\not=0$; and use the definitions of $\RR_i$, $\SS_i$, $i=1,2$.

\*
\appendix(a3, Proof of Lemma 3.3)

The propagators $g_{\underline a,\underline a'}^{(h)}(\xx)$
can be written in terms of the propagators
$g_{\o,\o'}^{(j,h)}(\xx)$, $j=1,2$,
see \equ(4.210b) and following lines; $g_{\o,\o'}^{(j,h)}(\xx)$
are given by
$$\eqalign{&g^{(j,h)}_{\o,\o}(\xx-\yy)=\cr
&\qquad\qquad={2\over M^2}
\sum_\kk e^{-i\kk(\xx-\yy)}\widetilde f_h(\kk) 
{-i\sin k+\o\sin k_0+a^{-(j)}_{h-1}(\kk)\over 
\sin^2 k+\sin^2 k_0+\big(\lis m^{(j)}_{h-1}(\kk)\big)^2+
\d B^{(j)}_{h-1}(\kk)}\cr 
&g^{(j,h)}_{\o,-\o}(\xx-\yy)=\cr
&\qquad\qquad={2\over M^2}
\sum_\kk e^{-i\kk(\xx-\yy)} 
\widetilde f_h(\kk)
{-i\o \lis m^{(j)}_{h-1}(\kk)
\over \sin^2 k+\sin^2 k_0+\big(\lis m^{(j)}_{h-1}(\kk)
\big)^2+\d B^{(j)}_{h-1}(\kk)}
\;,\cr}\Eqa(4.24)$$
where
$$\eqalign{&a^{\o(j)}_{h-1}(\kk)\defin -a^\o_{h-1}(\kk)+(-1)^j
b^\o_{h-1}(\kk)\virg
c^{(j)}_{h-1}(\kk)\defin c_{h-1}(\kk)+(-1)^jd_{h-1}(\kk)\cr
&m_{h-1}^{(j)}(\kk)\defin \s_{h-1}(\kk)+(-1)^j\m_{h-1}(\kk)\virg
\lis m_{h-1}^{(j)}(\kk)\defin m_{h-1}^{
(j)}(\kk)+c^{(j)}(\kk)\cr
&\d B^{(j)}_{h-1}(\kk)\defin\sum_\o\big[a_{h-1}^{\o(j)}(\kk)
(i\sin k-\o\sin k_0)+a^{\o(j)}_{h-1}(\kk)
a^{-\o(j)}_{h-1}(\kk)/2\big]\;.\cr}\Eqa(4.24y)$$ 
In order to bound the propagators defined above, we need estimates
on $\s_h(\kk),\m_h(\kk)$ and on the ``corrections''
$a^\o_{h-1}(\kk)$, $b^\o_{h-1}(\kk)$, $c_{h-1}(\kk)$, $d_{h-1}(\kk)$.
As regarding $\s_h(\kk)$ and $\m_h(\kk)$, in [BM] is proved (see Proof
of Lemma 2.6) that, on the support of $f_h(\kk)$, for some $c$,
$c^{-1}|\s_h|\le|\s_{h-1}(\kk)|\le c|\s_h|$ and 
$c^{-1}|\m_h|\le|\m_{h-1}(\kk)|\le c|\m_h|$.
%\;.\Eqa(4.41)$$
%
Note also that, if $h\ge \bar h$, using the first two of \equ(4.40z), we have
${|\s_h|+|\m_h|\over \g^h}\le 2C_1$.
As regarding the corrections, 
using their iterative definition \equ(4.17a), the asymptotic estimates
near $\kk={\bf 0}$ of the corrections on scale $h=1$ (see lines after 
\equ(2.38)) and the hypothesis \equ(4.40z), we easily find that,
on the support of $f_h(\kk)$:
$$\eqalign{&
a_{h-1}^\o(\kk)= O(\s_h\g^{(1-2c|\l|)h})+O(\g^{(3-c|\l|^2)h})\virg
b_h^\o(\kk)=O(\m_h\g^{(1-2c|\l|)h})+O(\g^{(3-c|\l|^2)h})\;,\cr
& c_h(\kk)=O(\g^{(2-c|\l|^2)h})\virg d_h(\kk)=O(\m_h\g^{(2-2c|\l|)h})\;.\cr}
\Eqa(c.5)$$
The bounds on the propagators follow from the remark that, as a consequence
of the estimates discussed above,
the denominators in \equ(4.24) are $O(\g^{2h})$
on the support of $f_h$.\\

\*
\appendix(4a, Analyticity of the effective potentials)
It is possible to write $\VV^{(h)}$  \equ(4.6)
in terms of {\it Gallavotti-Nicolo' trees.}\\

%%%%%%%%%%%%%%%%%%%%%%%%%%%%%%%%%%%%%%%%%%%%%%%%%%%%%%%%%%%%%%%%%%%%
%%%%%%%%%%%%%%%%%%%%%%%%%%%%FIGURA 4%%%%%%%%%%%%%%%%%%%%%%%%%%%%%%%%
%%%%%%%%%%%%%%%%%%%%%%%%%%%%%%%%%%%%%%%%%%%%%%%%%%%%%%%%%%%%%%%%%%%%

\insertplot{300pt}{150pt}%
{\ins{30pt}{85pt}{$r$}\ins{50pt}{85pt}{$v_0$}\ins{130pt}{100pt}{$v$}%
\ins{35pt}{-2pt}{$h$}\ins{55pt}{-2pt}{$h+1$}\ins{135pt}{-2pt}{$h_v$}%
\ins{215pt}{-2pt}{$0$}\ins{235pt}{-2pt}{$+1$}\ins{255pt}{-2pt}{$+2$}}%
{fig51}{}
\vskip.7cm
\line{\vtop{\line{\hskip5.5truecm\vbox{\advance\hsize by -5.1 truecm
\0{\ottorm FIG 4. A tree with its scale labels.}
} \hfill} }}
\vskip.5cm

%%%%%%%%%%%%%%%%%%%%%%%%%%%%%%%%%%%%%%%%%%%%%%%%%%%%%%%%%%%%%%%%%%%%%%%%%
%%%%%%%%%%%%%%%%%%%%%%%%%%%%%%%%%%%%%%%%%%%%%%%%%%%%%%%%%%%%%%%%%%%%%%%%%

We need some definitions and notations.

\0 1) Let us consider the family of all trees which can be constructed
by joining a point $r$, the {\it root}, with an ordered set of $n\ge 1$
points, the {\it endpoints} of the {\it unlabeled tree},
so that $r$ is not a branching point. $n$ will be called the
{\it order} of the unlabeled tree and the branching points will be called
the {\it non trivial vertices}.
Two unlabeled trees are identified if they can be superposed by a suitable
continuous deformation, so that the endpoints with the same index coincide.
Then the number of unlabeled trees with $n$ end-points
is bounded by $4^n$.

\0 2) We associate a label $h\le 0$ with the root and we denote $\TT_{h,n}$
the corresponding set of labeled trees with $n$ endpoints. Moreover, 
we introduce
a family of vertical lines, labeled
by an integer taking values in
$[h,2]$, and we represent any
tree $\t\in\TT_{h,n}$ so that, if $v$ is an
vendpoint or a non trivial vertex, it is contained in a vertical line with
index $h_v>h$, to be called the {\it scale} of $v$, while the root is on the
line with index $h$. There is the constraint that, if $v$ is an endpoint,
$h_v>h+1$; if there is only one end-point its scale must
be equal to $h+2$,
for $h\le 0$.
%The tree will intersect in general the vertical lines in set of
%points different from the root, the endpoints and the non trivial vertices;
%these points will be called {\it trivial vertices}. The set of the {\it
%vertices} of $\t$ will be the union of the endpoints, the trivial vertices
%and the non trivial vertices.
%Note that, if $v_1$ and $v_2$ are two vertices and $v_1<v_2$, then
%$h_{v_1}<h_{v_2}$.
Moreover, there is only one vertex immediately following
the root, which will be denoted $v_0$ and can not be an endpoint;
its scale is $h+1$.

\0 3) With each endpoint $v$ of scale $h_v=+2$ we associate one of the
contributions to $\VV^{(1)}$ given by \equ(2.40);
with each endpoint $v$ of
scale $h_v\le 1$ one of the terms in
$\LL \VV^{(h_v-1)}$ defined in \equ(4.10).
Moreover, we impose the constraint that, if $v$ is an endpoint and
$h_v\le 1$,
$h_v=h_{v'}+1$, if $v'$ is the non trivial vertex immediately preceding $v$.

%\0 4) If $v$ is not an endpoint, the {\it cluster } $L_v$ with frequency $h_v$
%is the set of endpoints following the vertex $v$; if $v$ is an endpoint, it is
%itself a ({\it trivial}) cluster. The tree provides an organization of
%endpoints into a hierarchy of clusters.

\0 4) We introduce a {\it field label} $f$ to distinguish the field variables
appearing in the terms associated with the endpoints as in item 3);
the set of field labels associated with the endpoint $v$ will be called $I_v$.
Analogously, if $v$ is not an endpoint, we shall
call $I_v$ the set of field labels associated with the endpoints following
the vertex $v$; $\xx(f)$, $\s(f)$ and $\o(f)$ will denote the space-time
point, the $\s$ index and the $\o$ index, respectively, of the
field variable with label $f$.

\0 5) We associate with any vertex $v$ of the tree a subset $P_v$ of $I_v$,
the {\it external fields} of $v$. These subsets must satisfy various
constraints. First of all, if $v$ is not an endpoint and $v_1,\ldots,v_{s_v}$
are the $s_v$ vertices immediately following it, then $P_v \subset \cup_i
P_{v_i}$; if $v$ is an endpoint, $P_v=I_v$. We shall denote $Q_{v_i}$ the
intersection of $P_v$ and $P_{v_i}$; this definition implies that $P_v=\cup_i
Q_{v_i}$. The subsets $P_{v_i}\bs Q_{v_i}$, whose union will be made, by
definition, of the {\it internal fields} of $v$, have to be non empty, if
$s_v>1$, that is if $v$ is a non trivial vertex.
Given $\t\in\TT_{j,n}$, there are many possible choices of the subsets $P_v$,
$v\in\t$, compatible with the previous constraints; let us call $\bP$ one of
this choices. 
Given $\bP$, we consider the family $\cal G_\bP$ of all
connected Feynman graphs, such that, for any $v\in\t$, the internal fields of
$v$ are paired by propagators of scale $h_v$, so that the following condition
is satisfied: for any $v\in\t$, the subgraph built by the propagators
associated with all vertices $v'\ge v$ is connected. The sets $P_v$ have, in
this picture, the role of the external legs of the subgraph associated with$v$.
The graphs belonging to $\cal G_\bP$ will be called {\it compatible with
$\bP$} and we shall denote $\PP_\t$ the family of all choices of $\bP$ such
that $\cal G_\bP$ is not empty.

\06) we associate with any vertex $v$ an index $\r_v\in\{s,p\}$
and correspondingly an operator $\RR_{\r_v}$, where 
$\RR_s$ or $\RR_p$ are defined as 
$$\RR_s\defin\cases{\SS_2 & if $n=1$ and $\o_1+\o_2=0$,\cr
\RR_1\SS_1 & if $n=1$ and $\o_1+\o_2\not=0$,\cr
\SS_1 & if $n=2$,\cr
1 & if $n>2$;\cr}\Eqa(rs)$$
and 
$$\RR_p\defin\cases{\RR_2(\PP_0+\PP_1) & if $n=1$ and $\o_1+\o_2=0$,\cr
\RR_2\PP_0 & if $n=1$, $\o_1+\o_2\not=0$
and $\a_1+\a_2=0$,\cr
0 & if $n=1$, $\o_1+\o_2\not=0$
and $\a_1+\a_2\not=0$,\cr
\RR_1\PP_0 & if $n=2$,\cr
0 & if $n>2$.\cr}\Eqa(rp)$$
Note that $\RR_s+\RR_p=\RR$, see Lemma 3.1. 

\*
The effective potential can be written in the following way:
$$\VV^{(h)}(\sqrt{Z_h}\psi^{(\le h)}) + M^2 \tilde E_{h+1}=
\sum_{n=1}^\io\sum_{\t\in\TT_{h,n}}
\VV^{(h)}(\t,\sqrt{Z_h}\psi^{(\le h)});,\Eqa(3.27a)$$
where, if $v_0$ is the first vertex of $\t$ and $\t_1,\ldots,\t_s$ 
are the subtrees of $\t$ with root $v_0$,\\
$\VV^{(h)}(\t,\sqrt{Z_h}\psi^{(\le h)})$ is defined inductively by the relation
$$\eqalign{
&\qquad \VV^{(h)}(\t,\sqrt{Z_h}\psi^{(\le h)})=\cr
&{(-1)^{s+1}\over s!} \EE^T_{h+1}[\bar
V^{(h+1)}(\t_1,\sqrt{Z_{h}}\psi^{(\le h+1)});\ldots; \bar
V^{(h+1)}(\t_{s},\sqrt{Z_{h}}\psi^{(\le h+1)})]\;,\cr}\Eqa(3.28a)$$
and $\bar V^{(h+1)}(\t_i,\sqrt{Z_{h}}\psi^{(\le h+1)})$:
 
\0 a) is equal to $\RR_{\r_{v_i}}\widehat
\VV^{(h+1)}(\t_i,\sqrt{Z_{h}}\psi^{(\le h+1)})$ if
the subtree $\t_i$ with first vertex $v_i$
is not trivial (see \equ(4.18) for the definition of
$\widehat \VV^{(h)}$);
 
\0 b) if $\t_i$ is trivial and $h\le -1$, it
is equal to one of the terms in $\LL\widehat\VV^{(h+1)}$,
see \equ(4.18), or,
if $h=0$, to one of the terms contributing to 
$\widehat\VV^{(1)}(\sqrt{Z_1}\psi^{\le 1})$.\\

\asub(a.0) The explicit 
expression for the kernels of $\VV^{(h)}$ can be found from 
\equ(3.27a) and \equ(3.28a) by writing the truncated expectations 
of monomials of $\psi$ fields using the analogue of \equ(8.7): if 
$\widetilde\psi(P_{v_i})=\prod_{f\in P_{v_i}}\psi^{\a(f)(h_v)}_{\xx(f),
\o(f)}$, the following identity holds:
$$\EE^T_{h_v}(\widetilde\psi(P_{v_1}),\ldots,\widetilde\psi(P_{v_s}))=
\Big({1\over Z_{h_v-1}}\Big)^n\sum_{T_v}\a_{T_v}\prod_{\ell\in T_v}
g^{(h_v)}(f^1_\ell,f^2_\ell)
\int dP_{T_v}(\tt) \Pf G^{T_v}(\tt)\Eqa(a.1)$$
where $g^{(h)}(f,f')=g_{\underline a(f),\underline a(f')}(\xx(f)-\xx(f'))$
and the other symbols in \equ(a.1) have the same meaning as those in 
\equ(8.7).

Using iteratively \equ(a.1) we can express the kernels
of $\VV^{(h)}$ as sums of products of propagators of the fields (the ones
associated to the anchored trees $T_v$) and Pfaffians of matrices $G^{T_v}$.
%
%We begin to the describe the effect of the iterative construction in a 
%case simpler than the one we are really interested in.\\

\asub(a.1) {\it If the $\RR$ operator were not applied to 
the vertices $v\in\t$} then the result of the iteration would lead to the 
following relation:
$$\VV^*_h(\t,\sqrt{Z_h}\psi^{(\le h)})=\sqrt{Z_h}^{|P_{v_0}|}\sum_{
\bP\in\PP_\t}\sum_{T\in\bT}\int d\xx_{v_0}W^*_{\t,\bP,\bT}
(\xx_{v_0})\Big\{\prod_{f\in P_{v_0}}
\psi^{\a(f)(\le h)}_{
\xx(f),\o(f)}\Big\}\;,\Eqa(a.2)$$
where $\xx_{v_0}$ is the set of integration variables
asociated to $\t$ and $T=\bigcup_v T_v$; 
$W^*_{\t,\bP,\bT}$ is given by
$$\eqalign{&W^*_{\t,\bP,\bT}(\xx_{v_0})=
\Big[\prod_{v\,\hbox{\ottorm not e.p.}}
\Big({Z_{h_v}\over Z_{h_v-1}}\Big)^{|P_v|\over 2}\Big]
\Big[\prod_{i=1}^n K^{h_i}_{v^*_i}(\xx_{v^*_i})\Big]
\Big\{\prod_{v\,\hbox{\ottorm not e.p.}}{1\over s_v!} \int
dP_{T_v}(\tt_v) \;\cdot\cr
&\cdot\; \Pf G^{h_v,T_v}(\tt_v)
\Big[\prod_{l\in T_v} g^{(h_v)}(f^1_l,f^2_l)\Big]\Big\}\;,\cr}
\Eqa(a.3)$$
where: $e.p.$ is an abbreviation of ``end points'';
$v_1^*,\ldots, v_n^*$ are the endpoints of $\t$, $h_i\=h_{v_i^*}$
and $K^{h_v}_v(\xx_v)$ are the corresponding kernels (equal to 
$\l_{h_v-1}\d(\xx_v)$ or $\n_{h_v-1}\d(\xx_v)$ if $v$ is an endpoint 
of type $\l$ or $\n$ on scale $h_v\le 1$; or equal to one of the kernels
of $\VV^{(1)}$ if $h_v=2$).

We can bound \equ(a.3) 
using \equ(4.400) and the Gram--Hadamard inequality,
see Appendix \secc(a1), we would find:
$$\int d\xx_{v_0} |W^*_{\t,\bP,T}(\xx_{v_0})|\le
C^n M^2|\l|^n \g^{-h(-2+|P_{v_0}|/2 )}
\prod_{v\,\hbox{\ottorm not e.p.}} \left\{ {1\over s_v!}
\Big({Z_{h_v}\over Z_{h_v-1}}\Big)^{|P_v|\over 2}
\g^{-[-2+{|P_v|\over 2}]}\right\}\;.\Eqa(a.170)$$
We call $D_v=-2+{|P_v|\over 2}$ the {\it dimension}
of $v$, depending on the number of the external
fields of $v$. If $D_v<0$ for any $v$ one can
sum over $\t,\bP,T$ obtaining convergence
for $\l$ small enough; however 
$D_v\le 0$ when there are two or four external lines.
We will take now into account the effect of the $\RR$ operator
and we will see how the bound \equ(a.170) is improved.

\asub(a.1) 
The effect of application of $\PP_j$ and $\SS_j$ 
is to replace a kernel $W^{(h)}_{2n,\ss,\underline j,\aa,\oo}$
with $\PP_j W^{(h)}_{2n,\ss,\underline j,\aa,\oo}$ and 
$\SS_j W^{(h)}_{2n,\ss,\underline j,\aa,\oo}$. 
If inductively, starting from the end--points,
we write the kernels $W^{(h)}_{2n,\ss,\underline j,\aa,\oo}$ 
in a form similar to 
\equ(a.3), we easily realize that, eventually, 
$\PP_j$ or $\SS_j$ will act on some 
propagator of an anchored tree or on some Pfaffian $\Pf G^{T_v}$, 
for some $v$.
It is easy to realize that $\PP_j$ and $\SS_j$, when applied to Pfaffians,
do not break the Pfaffian structure. In fact the effect 
of $\PP_j$ on the Pfaffian of an antisymmetric matrix $G$ 
with elements $G_{f,f'}$, $f,f'\in J$, $|J|=2k$,
is the following (the proof is trivial):
$$\PP_0\Pf G=\Pf G^0\virg
\PP_1\Pf G=
{1\over 2 }\sum_{f_1,f_2\in J}\PP_1G_{f_1,f_2}
(-1)^\p\Pf G_1^0\;,\Eqa(a.500)$$
where $G^0$ is the matrix with elements $\PP_0 G_{f,f'}$, $f,f'\in J$;
$G_1^0$ is the matrix with elements $\PP_0 G_{f,f'}$, $f,f'\in J_1\defin
J\setminus
\{f_1\cup f_2\}$ and $(-1)^\p$ is the sign of the permutation
leading from the ordering $J$ of the labels $f$ in the l.h.s. to the 
ordering $f_1,f_2,J_1$ in the r.h.s. The effect of 
$\SS_j$ is the following, see Appendix \secc(a7) for a proof:
$$\SS_1\Pf G={1\over 2\cdot k!}\sum_{f_1,f_2\in J}\SS_1G_{f_1,f_2}
\sum_{J_1\cup J_2=
J\setminus\cup_i f_i}^*(-1)^\p k_1!\, k_2!\,
\Pf G_1^0\,\Pf G_2\;,\Eqa(a.5)$$
where: the $*$ on the sum means that $J_1\cap J_2=\emptyset$; $|J_i|=2k_i$, 
$i=1,2$; $(-1)^\p$ is the sign of the permutation leading 
from the ordering $J$ of the fields labels on the l.h.s. to the ordering 
$f_1,f_2,J_1,J_2$ on the r.h.s.;
$G_1^0$ is the matrix with elements $\PP_0 G_{f,f'}$, $f,f'\in J_1$;
$G_2$ is the matrix with elements $G_{f,f'}$, $f,f'\in J_2$.
The effect of $\SS_2$ on $\Pf G^T$ is given by a formula 
similar to \equ(a.5).
% furthermore:
%%
%$$\eqalign{&\SS_2\Pf G={1\over 2}\sum_{f_1,f_2\in J}\SS_2G_{f_1,f_2}
%(-1)^\p \Pf G_1^{0}+\cr
%&+{1\over 2^3\cdot k!}\sum_{f_1,\ldots,f_4\in J}^{**}\PP_1 G_{f_1,f_2}\,
%\PP_1 G_{f_3,f_4}\sum_{J_1\cup J_2=
%J\setminus\cup_i f_i}^*(-1)^\p k_1!\, k_2!\,
%\Pf G_1^{0}\,\Pf G_2\;,\cr}\Eqa(a.6)$$
%%
%where the $**$ on the sum over $f_1,\ldots,f_4$ means that $f_1,f_2,f_3,f_4$
%are all different. \equ(a.6) can be proven analogously to \equ(a.5), see 
%Appendix \secc(a7).
%A consequence of \equ(a.500), \equ(a.5) and \equ(a.6) is that  
%any term of the form $\PP_i\Pf G$ or $\SS_i\Pf G$ 
%can still be bounded by Gram inequality,
%in a way similar to the one explained in Appendix \secc(a1) 
%for the integration of 
%the $\c$ fields: the key remark is that
Note that the
number of terms in the sums appearing in
\equ(a.500), \equ(a.5) (and in the analogous equation for $\SS_2\Pf G^T$),
is bounded by $c^k$ for some constant $c$.
\\
\asub(a.2) It is possible to show 
that the $\RR_j$ operators   
produce derivatives applied to the propagators of the anchored 
trees and on the elements of $G^{T_v}$; and a product of ``zeros''
of the form $d_{j}^{b}(\xx(f^1_\ell)-\xx(f^2_\ell))$, $j=0,1$, $b=0,1,2$, 
associated to the lines $\ell\in T_v$.
This is a well known result, and a very detailed discussion
can be found in \S 3 of [BM]. By such analysis, and using
\equ(a.500),\equ(a.5),
we get the following expression for 
$\RR\VV^{(h)}(\t,\sqrt{Z_h}\psi^{(\le h)})$:
$$\eqalign{&\RR\VV^{(h)}(\t,\sqrt{Z_h}\psi^{(\le h)})=\cr
&\qquad=\sqrt{Z_h}^{|P_{v_0}|}\sum_{
\bP\in\PP_\t}\sum_{T\in\bT}\sum_{\b\in B_T}\int d\xx_{v_0}W_{\t,\bP,\bT,\b}
(\xx_{v_0})\Big\{\prod_{f\in P_{v_0}}
\hat\dpr^{q_\b(f)}_{j_\b(f)}\psi^{\a(f)(\le h)}_{
\xx_\b(f),\o(f)}\Big\}\;,\cr}\Eqa(a.7)$$
where: $B_T$ is a set of indeces which allows to distinguish the
different terms produced by the non trivial $\RR$ operations;
$\xx_\b(f)$ is a coordinate obtained by interpolating two points in $
\xx_{v_0}$, in a suitable way depending on $\b$; $q_\b(f)$ is a nonnegative 
integer $\le 2$; $j_\b(f)=0,1$ and $\hat\dpr^q_j$ is a suitable differential 
operator, dimensionally equivalent to $\dpr^q_j$ (see [BM] for a precise 
definition); $W_{\t,\bP,\bT,\b}$ is given by:
$$\eqalign{W_{\t,\bP,\bT,\b}(\xx_{v_0})&=
\Big[\prod_{v\,\hbox{\ottorm not e.p.}}
\Big({Z_{h_v}\over Z_{h_v-1}}\Big)^{|P_v|\over 2}\Big]
\Big[\prod_{i=1}^n d^{b_\b(v_i^*)}_{j_\b(v_i^*)}(\xx^i_\b,
\yy^i_{\b})\PP_{I_\b(v^*_i)}^{C_\b(v^*_i)}
\SS_{i_\b(v^*_i)}^{c_\b(v^*_i)}K^{h_i}_{v^*_i}(\xx_{v^*_i})\Big]\cdot\cr
&\cdot\Big\{\prod_{v\,\hbox{\ottorm not e.p.}}{1\over s_v!} \int
dP_{T_v}(\tt_v)\PP_{I_\b(v)}^{C_\b(v)}
\SS_{i_\b(v)}^{c_\b(v)}\Pf G^{h_v,T_v}_\b(\tt_v)\cdot\cr
&\cdot\Big[\prod_{l\in T_v}\hat\dpr^{q_\b(f^1_l)}_{j_\b(f^1_l)}
\hat\dpr^{q_\b(f^2_l)}_{j_\b(f^2_l)}[d^{b_\b(l)}_{j_\b(l)}(\xx_l,\yy_l) 
\PP_{I_\b(l)}^{C_\b(l)}
\SS_{i_\b(l)}^{c_\b(l)}g^{(h_v)}(f^1_l,f^2_l)]\Big]\Big\}\;,\cr}
\Eqa(a.8)$$
where: $v^*_1,\ldots,v^*_n$ are the endpoints of $\t$; 
$b_\b(v)$, $b_\b(l)$, $q_\b(f^1_l)$ and $q_\b(f^2_l)$ are
nonnegative integers $\le 2$; $j_\b(v)$, $j_\b(f^1_l)$, $j_\b(f^2_l)$
and $j_\b(l)$ can be 
$0$ or $1$; $i_\b(v)$ and $i_\b(l)$ can be $1$ or $2$;
$I_\b(v)$ and $I_\b(l)$ can be $0$ or $1$; $C_\b(v)$, $c_\b(v)$, $C_\b(l)$
and $c_\b(l)$ can be $0,1$ and $\max\{C_\b(v)+c_\b(v),C_\b(l)+ c_\b(l\})\le 1$;
$G^{h_v,T_v}_\b(\tt_v)$ is obtained from $G^{h_v,T_v}(\tt_v)$
by substituting the element $t_{i(f),i(f')}g^{(h_v)}(f,f')$ with 
$t_{i(f),i(f')}\hat\dpr^{q_\b(f)}_{j_\b(f)}
\hat\dpr^{q_\b(f')}_{j_\b(f')}g^{(h_v)}(f,f')$.\\

It would be very difficult to give a precise description
of the various contributions of the sum over $B_T$, but fortunately
we only need to know some very general properties, which easily
follows from the construction in \S 3. 

1)There is a constant $C$ such that, $\forall T\in {\bf T}_\t$,
$|B_T|\le C^n$;
for any $\b\in B_T$, the following inequality is satisfied
$$\Big[\prod_{f\in \cup_v P_v} \g^{h(f) q_\b(f)} \Big]
\Big[\prod_{l\in T} \g^{-h(l) b_\b(l)} \Big]\le
\prod_{v\,\hbox{\ottorm not e.p.}} \g^{-z(P_v)}\;,\Eqa(a.9)$$
where: $h(f)=h_{v_0}-1$ if $f\in P_{v_0}$, otherwise it is the scale of
the vertex where the field with label $f$ is contracted;
$h(l)=h_v$, if $l\in T_v$ and
$$z(P_v)=\cases{
1 & if $|P_v|=4$ and $\r_v=p\;,$ \cr
2 & if $|P_v|=2$ and $\r_v=p\;,$ \cr
1 & if $|P_v|=2$, $\r_v=s$ and $\sum_{f\in P_v}\o(f)\not=0\;,$ \cr
0 & otherwise.\cr}\Eqa(a.10)$$
2)If we define
$$\prod_{v\in\t}\Big[\Big({|\s_{h_v}|+|\m_{h_v}|\over 
\g^{h_v}}\Big)^{c_\b(v)i_\b(v)}
\prod_{\ell\in T_v}\Big({|\s_{h_v}|+|\m_{h_v}|\over 
\g^{h_v}}\Big)^{c_\b(\ell)i_\b(\ell)}\Big]
\defin \prod_{v\in V_\b}\Big({|\s_{h_v}|+|\m_{h_v}|\over 
\g^{h_v}}\Big)^{i(v,\b)}\;.\Eqa(a.11)$$
the indeces $i(v,\b)$ satisfy, for any $B_T$, the following property:
$$\sum_{w\ge v}i(v,\b)\ge z'(P_v)\;,\Eqa(a.11z)$$
where
$$z'(P_v)=\cases{
1 & if $|P_v|=4$ and $\r_v=s\;,$ \cr
2 & if $|P_v|=2$ and $\r_v=s$ and $\sum_{f\in P_v}\o(f)=0\;,$ \cr
1 & if $|P_v|=2$, $\r_v=s$ and $\sum_{f\in P_v}\o(f)\not=0\;,$ \cr
0 & otherwise.\cr}\Eqa(a.12)$$
%{\bf Remark.}
%As we shall see below, the factors $\g^{h(f) q_\b(f)}$ and 
%$\g^{-h(l) b_\b(l)}$ in \equ(a.9), unless for a 
%constant $c^n$, are the contributions to the bound
%on $W_{\t,\bP,\bT,\b}$ coming from 
%the zeros and the derivative operators
%in \equ(a.8) (in fact it is easy to realize that each 
%derivative applied on a 
%propagator on scale $h$ contributes to the bounds with 
%a factor $\sim\g^h$; each zero associated to a line of $T_v$ with a factor 
%$\g^{-h_v}$). As we said in section \secc(a.2) and as described in [BM], 
%these zeros and derivatives operators
%come from the application of $\RR_1$ and $\RR_2$ on the nodes of $\t$;
%then the r.h.s. of \equ(a.9) collects the dimensional
%gains associated to the operators $\RR_1$ and $\RR_2$, in terms of which 
%$\RR_s$ and $\RR_p$ are written, see \equ(rs) and \equ(rp).
%Note that these gain factors are not enough to
%give negative dimension to all the vertices of $\t$: 
%in fact their effect is to replace $-2+|P_v|/2$ with 
%$-2+|P_v|/2+z(P_v)$ in the analogous of \equ(a.170). Then, the
%vertices with $|P_v|=2,4$ and $\r_v=s$ still have positive dimension,
%\ie they have $2-|P_v|/2-z(P_v)\ge 0$. In order to 
%improve the power counting for such vertices we will have to use the 
%gain factors coming from 
%the $\SS_i$ operators in \equ(a.8).\\

\asub(a.6) 
%We now want to bound $\int d\xx_{v_0}|W_{\t,\bP,\bT,\b}(\xx_{v_0})|$,
%using the expression \equ(a.8). The different factors in 
%\equ(a.8) give the contributions described in items (1) and (2)
%below, easily understood by
%dimensional arguments (see [BM] for a more detailed proof).
%In particular we shall extensively use that the
%derivative operators and the zeros coming from the action 
%of $\RR_i$, $i=1,2$, can be  
%dimensionally bounded as described in the last Remark;
%and the operators $\PP_i$ or $\SS_i$, when  
%applied on a propagator or on a Pfaffian 
%on scale $h$, contribute with a factor $c[(|\s_h|+|\m_h|)/\g^h]^i$.
We can bound 
any $|\PP_{I_\b(v)}^{C_\b(v)}\SS_{i_\b(v)}^{c_\b(v)}
\Pf G^{h_v,T_v}_\b|$ in \equ(a.8), with $C_\b(v)+c_\b(v)=0,1$, 
by using \equ(a.500), \equ(a.5) and 
Gram inequality, as illustrated in Appendix \secc(a1) for the case of the 
integration of the $\c$ fields. Using that the elements of $G$ are all
propagators on scale $h_v$, dimensionally bounded as in Lemma 3.3, we find:
$$\eqalign{&
|\PP_{I_\b(v)}^{C_\b(v)}\SS_{i_\b(v)}^{c_\b(v)}
\Pf G^{h_v,T_v}_\b|\le C^{\sum_{i=1}^{s_v}|P_{v_i}|-|P_v|-
2(s_v-1)}\cdot\cr
&\qquad\cdot
\g^{{h_v\over 2}\left(\sum_{i=1}^{s_v}|P_{v_i}|-|P_v|-2(s_v-1)\right)}
\Big[\prod_{f\in J_v}
\g^{h_v q_\b(f)}\Big]\Big({|\s_{h_v}|+|\m_{h_v}|\over 
\g^{h_v}}\Big)^{c_\b(v)i_\b(v)+C_\b(v)I_\b(v)}\;,\cr}\Eqa(a.13)$$
where $J_v=\cup_{i=1}^{s_v}P_{v_i}\setminus Q_{v_i}$.
We will bound
the factors $\Big({|\s_{h_v}|+|\m_{h_v}|\over 
\g^{h_v}}\Big)^{C_\b(v)I_\b(v)}$ 
using \equ(4.40a) by a constant.\\

If we call 
$$\eqalign{
J_{\t,\bP,T,\b}&=\int d\xx_{v_0}
\Big| \Big[ \prod_{i=1}^n d_{j_\b(v^*_i)}^{b_\b(v^*_i)}(\xx^i_\b,\yy^i_\b)
\PP_{I_\b(v^*_i)}^{C_\b(v^*_i)}
\SS_{i_\b(v^*_i)}^{c_\b(v^*_i)}K^{h_i}_{v^*_i}(\xx_{v^*_i})\Big]\cdot\cr
&\cdot \Big\{ \prod_{v\,\hbox{\ottorm not e.p.}} {1\over s_v!}
\Big[\prod_{l\in T_v} \hat\partial^{q_\b(f^1_l)}_{j_\b(f^1_l)}
\hat\partial^{q_\b(f^2_l)}_{j_\b(f^2_l)} [d^{b_\b(l)}_{j_\b(l)}(\xx_l,\yy_l)
\PP_{I_\b(l)}^{C_\b(l)}\SS_{i_\b(l)}^{c_\b(l)}
g^{(h_v)}(f^1_l,f^2_l)]\Big]\Big\}\Big|\;,\cr}
\Eqa(a.14)$$
we have, under the hypothesis \equ(4.40),
$$\eqalign{&
J_{\t,\bP,T,\a}\le C^n M^2 |\l|^n\Big[\prod_{i=1}^n
\Big({|\s_{h^*_i}|+|\m_{h^*_i}|\over 
\g^{h^*_i}}\Big)^{c_\b(v^*_i)i_\b(v^*_i)}\Big]\cdot\cr
&\cdot\Big\{\prod_{v\,\hbox{\ottorm not e.p.}}{1\over s_v!} C^{2(s_v-1)}
\g^{h_v n_\n(v)}\g^{-h_v\sum_{l\in T_v}b_\b(l)}
\g^{-h_v\sum_{i=1}^n b_\b(v_i^*)}
\g^{-h_v(s_v-1)}\cdot\cr
&\cdot\g^{h_v\sum_{l\in T_v}\left[q_\b(f^1_l)+q_\b(f^2_l)\right]}
\Big\}\Big[\prod_{\ell\in T}\Big({|\s_{h_v}|+|\m_{h_v}|\over 
\g^{h_v}}\Big)^{c_\b(\ell)i_\b(\ell)}
\Big]\;,\cr} \Eqa(a.15)$$ 
where $n_\n(v)$ is the number
of vertices of type $\n$ with scale $h_v+1$. 

Now, substituting \equ(a.13), \equ(a.15) into \equ(a.8), using
\equ(a.9), we find that:
$$\eqalign{
&\int d\xx_{v_0} |W_{\t,\bP,T,\b}(\xx_{v_0})|\le
C^n M^2|\l|^n \g^{-h D_k(|P_{v_0}|)}
\prod_{v\in V_\b}\Big({|\s_{h_v}|+|\m_{h_v}|\over\g^{h_v}}\Big)^{i(v,\b)}
\;\cdot\cr
&\cdot\; \prod_{v\,\hbox{\ottorm not e.p.}} \left\{ {1\over s_v!}
C^{\sum_{i=1}^{s_v}|P_{v_i}|-|P_v|}
\Big({Z_{h_v}\over Z_{h_v-1}}\Big)^{|P_v|\over 2}
\g^{-[-2+{|P_v|\over 2}+z(P_v)]}\right\}\;,\cr}\Eqa(a.17)$$
where, if $k=\sum_{f\in P_{v_0}}q_\b(f)$, $D_k(p)=-2+p+k$ and we 
have used \equ(a.11).
Note that, given $v\in \t$ and $\t\in \TT_{h,n}$ and
using \equ(4.40a) together with the first two of \equ(4.40z), 
$$\eqalign{&{|\sigma_{h_v}|\over \g^{h_v}}={|\sigma_h|\over \g^h}
{|\sigma_{h_v}|\over |\sigma_h|}
\g^{h-h_v} \le {|\sigma_h|\over \g^h} \g^{(h-h_v)(1-c|\l|)}\le
C_1\g^{(h-h_{\bar v})(1-c|\l|)}\cr
&{|\m_{h_v}|\over \g^{h_v}}={|\m_h|\over \g^h}
{|\m_{h_v}|\over |\m_h|}
\g^{h-h_v} \le {|\m_h|\over \g^h} \g^{(h-h_v)(1-c|\l|)}\le
C_1\g^{(h-h_v)(1-c|\l|)}\cr}\Eqa(a.18)$$
Moreover  
the indeces $i(v,\b)$ satisfy, for any $B_T$, 
\equ(a.12) sso that, using \equ(a.18) and \equ(a.11z), we find
$$\prod_{v\in V_\b}\Big({|\s_{h_v}|+|\m_{h_v}|\over 
\g^{h_v}}\Big)^{i(v,\b)}\le C_1^n \prod_{v\, {\rm not}\, {\rm e.p.}}
\g^{-z'(P_v)}\;.\Eqa(esse)$$
%
%{\bf Remark.}
%The r.h.s. of \equ(esse) collects the dimensional
%gains associated to the operators $\SS_1$ and $\SS_2$, in terms of which 
%$\RR_s$ is written, see \equ(rs).
%Note that these gain factors, together with the factors 
%$\prod_{v\, {\rm not}\, {\rm e.p.}}
%\g^{-z(P_v)}$ are now enough to renormalize all the vertices of $\t$:
%the key property is 
%
%$$z(P_v)+z'(P_v)=\cases{
%1 & if $|P_v|=4$\cr
%2 & if $|P_v|=2$\cr
%0 & otherwise,\cr}\Eqa(a.12)$$
%
%so that 
%
%$$D_v\defin -2+{|P_v|\over 2}+z(P_v)+(1-c|\l|)z'(P_v)\ge {|P_v|\over 6}\;,
%\Eqa(a.20)$$
%
%which means that any vertex $v\in\t$ has negative dimension $-D_v$.\\
Substituting \equ(a.18) into \equ(a.17) and using \equ(a.11z), we find:
$$\eqalign{
&\int d\xx_{v_0} |W_{\t,\bP,T,\b}(\xx_{v_0})|\le
C^n M^2|\l|^n \g^{-h D_k(|P_{v_0}|)}
\;\cdot\cr
&\cdot\; \prod_{v\,\hbox{\ottorm not e.p.}} \left\{ {1\over s_v!}
C^{\sum_{i=1}^{s_v}|P_{v_i}|-|P_v|}
\Big({Z_{h_v}\over Z_{h_v-1}}\Big)^{|P_v|\over 2}
\g^{-[-2+{|P_v|\over 2}+z(P_v)+(1-c|\l|)z'(P_v)]}\right\}\;.\cr}\Eqa(a.19)$$
where 
$$D_v\defin -2+{|P_v|\over 2}+z(P_v)+(1-c|\l|)z'(P_v)\ge {|P_v|\over 6}\;.
\Eqa(a.20)$$
Then
\equ(4.45) in Theorem 3.1 
follows from the previous bounds and the remark that
$$\sum_{\t\in\TT_{h,n}}
\sum_{\bP\in\PP_\t}\sum_{T\in\bT}\sum_{\b\in B_T}
\prod_{v}{1\over s_v!}\g^{-{|P_v|\over 6}}\le
c^n\;,\Eqa(a.21)$$
for some constant $c$, see [BM] or [GM] for further details.\\

The bound on $\tilde E_h$, $t_h$, \equ(4.45y) and \equ(4.45yz) follow 
from a similar analysis. 
The remarks following \equ(4.45y) and \equ(4.45yz) follows from noticing 
that in the expansion for $\LL\VV^{(h)}$ appear only propagators of type 
$\PP_0 g^{(h_v)}_{\underline a,\underline a'}$ or
$\PP_1 g^{(h_v)}_{\underline a,\underline a'}$ (in order to bound 
these propagators we do not need \equ(4.40a), see 
the last statement in Lemma 3.3). 
Furthermore, by construction $l_h,n_h$ and $z_h$ are 
independent of $\s_k,\m_k$, so that, in order to prove \equ(4.45yz) 
we do not even need the first two inequalities in \equ(4.40z).
\qed
\\

\asub(short) The sum over all the trees with root scale $h$
and  with at least a $v$ with $h_v=k$ is $O(|\l|\g^{{1\over 2}(h-k)})$;
this follows from the fact that
the bound \equ(a.21) holds, for some $c=O(1)$, 
even if $\g^{-|P_v|/6}$ is replaced
by $\g^{-\k |P_v|}$, for any constant $\k>0$ independent of $\l$;
and that $D_v$, instead of using \equ(a.20), can also 
be bounded as $D_v\ge 1/2 +|P_v|/12$. This property 
is called {\it short memory property.}
\*
\appendix(a5, Proof of Theorem 4.1 and Lemma 4.1)

We consider the space $\MMM_\th$ of sequences $\un=\{\n_h\}_{h\le 1}$
such that $|\n_h|\le c|\l|\g^{(\th/2)h}$; we shall think $\MMM_\th$
as a Banach space with norm $||\cdot||_\th$, where $||\un||_\th\defin
\sup_{k\le 1}|\n_k|\g^{-(\th/2)k}$. We will proceed as follows: we 
first show that, for any sequence $\un\in\MMM_\th$, 
the flow equation for $\n_h$, the hypothesis 
\equ(4.40yz), \equ(4.40z) and the property $|\l_h(\un)|\le c|\l|$
are verified, uniformly in $\un$. Then we fix $\un\in\MM_\th$
via an exponentially convergent iterative procedure, in such a way that
the flow equation for $\n_h$ is satisfied.\\

\asub(a5.1) {\it Proof of Theorem 4.1.}
Given $\un\in\MMM_\th$,
let us suppose inductively that \equ(4.40yz), \equ(4.40z) and that,
for $k>\bar h+1$,
$$|\l_{k-1}(\un)-\l_k(\un)|\le c_0|\l|^{2}
\g^{(\th/2)k}\;,\Eqa(5.7)$$
for some $c_0>0$. Note that \equ(5.7) is certainly true for $h=1$ 
(in that case the r.h.s. of \equ(5.7)
is just the bound on $\b^1_\l$).
Note also that \equ(5.7) implies  
that $|\l_k|\le c|\l|$, for any $k> \bar h$.

Using \equ(4.45y), the second of
\equ(4.45yz) and \equ(5.4z) we find that \equ(4.40yz), \equ(4.40z)
are true with $\bar h$ replaced by $\bar h-1$. 

We now consider the equation $\l_{h-1}=\l_h+\b^h_\l(\l_h,\n_h;\ldots;
\l_1,\n_1)$, $h>\bar h$. 
The function $\b^h_\l$ can be expressed as a convergent
sum over tree diagrams, as described in Appendix \secc(4a); note that
it depends on $(\l_h,\n_h;\ldots;
\l_1,\n_1)$ directly through the end--points of the trees and indirectly
through the factors $Z_h/Z_{h-1}$.

We can write $\PP_0 g^{(h)}_{(+,\o),(-,\o)}(\xx-\yy)=
g^{(h)}_{L,\o}(\xx-\yy)+
r^{(h)}_\o(\xx-\yy)$, where 
$$g^{(h)}_{L,\o}(\xx-\yy)\defin {4\over M^2}\sum_{\kk}e^{-i\kk(\xx-\yy)}
\widetilde f_h(\kk){1\over ik+\o k_0}\Eqa(gL)$$
and $r^{(h)}_\o$ is the rest, satisfying the same bound as 
$g^{(h)}_{(+,\o),(-,\o)}$, times a factor $\g^h$. This decomposition
induces the following decomposition for $\b^h_\l$:
$$\eqalign{&\b^h_\l(\l_h,\n_h;\ldots;\l_1,\n_1)=\cr
&\qquad=\b_{\l,L}^h
(\l_h,\ldots,\l_h)+
\sum_{k=h+1}^1 
D^{h,k}_\l + r^h_\l(\l_h,\ldots,\l_1)+\sum_{k\ge h}\n_k\tilde\b_\l^{h,k}
(\l_k,\n_k;\ldots;\l_1,\n_1)\;,\cr}\Eqa(5.1aaa)$$
with 
$$\eqalign{&|\b^h_{\l,L}|\le c|\l|^2 \g^{\th h}\;,\qquad
|D^{h,k}_\l|\le c |\l| \g^{\th(h-k)} |\l_k-\l_h|\;,\qquad\cr
&|r^h_\l|\le c|\l|^2\g^{(\th/2) h}\;,\qquad |\tilde\b_\l^{h,k}|\le
c|\l|\g^{\th(h-k)}\;.\cr}\Eqa(5.1bbb)$$
The first two terms in \equ(5.1aaa)
$\b^h_{\l,L}$ collect the contributions
obtained by posing $r^{(k)}_\o=0$, $k\ge h$ and substituting
the discrete $\d$ function defined after \equ(4.11) with 
$M^2\d_{\kk,{\bf 0}}$.  
The first of \equ(5.1bbb) is called 
the {\it vanishing of the Luttinger
model Beta function} property, see [BGPS][GS][BM1] (or [BeM1] for
a simplified proof), and it is a crucial property
of interacting fermionic systems in $d=1$.

Using the decomposition \equ(5.1aaa) and the bounds \equ(5.1bbb)
we prove the following bounds for $\l_{\bar h}(\un)$, $\un\in\MMM_\th$:
$$|\l_{\bar h}(\un)-\l_1(\un)|\le c_0|\l|^2
\virg
|\l_{\bar h}(\un)-\l_{\bar h+1}(\un)|\le c_0|\l|^2\g^{(\th/2) \bar h}
\;,\Eqa(5.1cc)$$
for some $c_0>0$.
Moreover, given $\un,\un'\in\MMM_\th$, we show that:
$$|\l_{\bar h}(\un)-\l_{\bar h}(\un')|\le c|\l|
||\un-\un'||_0\;,\Eqa(5.1d)$$
where $||\un-\un'||_0\defin\sup_{h\le 1}|\n_h-\n_h'|$.\\

\0{\it Proof of \equ(5.1cc).} 
We decompose $\l_{\bar h}-\l_{\bar h+1}=\b^{\bar h+1}_\l$ 
as in \equ(5.1aaa). Using the bounds
\equ(5.1bbb) and the inductive hypothesis \equ(5.7), we find:
$$\eqalign{|\l_{\bar h}(\un)-\l_{\bar h+1}(\un)| 
&\le c|\l|^2\g^{\th (\bar h+1)}+
\sum_{k\ge \bar h+2}c|\l|\g^{\th(\bar h+1-k)}\sum_{k'=\bar h+2}^k 
c_0|\l|^{2}\g^{(\th/ 2)k'}+\cr
&+c|\l|^2\g^{(\th/2)(\bar h+1)}+\sum_{k\ge \bar h+1}
c^2|\l|^2\g^{(\th/2)k}\g^{(\th(\bar h+1-k))}
\;,\cr}
\Eqa(5.8)$$
which, for $c_0$ big enough, immediately implies the second of
\equ(5.1cc) with $h\to h-1$; from this bound and the 
hypothesis \equ(5.7) follows the first of \equ(5.1cc).\qed\\

\0{\it Proof of \equ(5.1d).} If we take two sequences 
$\un,\un'\in\MMM_\th$, we
easily find that the beta function for $\l_{\bar h}(\un)-\l_{\bar h}(\un')$
can be represented by a tree expansion similar to the one for $\b^h_\l$,
with the property that the trees giving a non vanishing contribution 
have necessarily one end--point on scale $k\ge h$ associated to a 
coupling constant $\l_k(\un)-\l_k(\un')$ or $\n_k-\n_k'$. Then we find:
$$\l_{\bar h}(\un)-\l_{\bar h}(\un')
=\l_{1}(\un)-\l_{1}(\un')+\sum_{\bar h+1\le k\le 1}
[\b^k_\l(\l_k(\un),\n_k;
\ldots;\l_1,\n_1)-\b^k_\l(\l_k(\un'),\n_k';
\ldots;\l_1,\n_1')]
\;.\Eqa(5.8aaa)$$
Note that $|\l_1(\un)-\l_1(\un')|\le c_0|\l||\n_1-\n_1'|$,
because $\l_1=\l/Z_1^2+O(\l^2/Z_1^4)$
and $Z_1=\sqrt2-1+\n/2$. 
If we inductively suppose that,
for any $k> \bar h$, $|\l_k(\un)-\l_k(\un')|\le 2c_0|\l|
||\un-\un'||_0$, we
find, by using the decomposition
\equ(5.1aaa):
$$|\l_{\bar h}(\un)-\l_{\bar h}(\un')|\le c_0|\l||\n_1-\n_1'|+
c|\l|\sum_{k\ge \bar h+1}\g^{(\th/2)k}\sum_{k'\ge k}
\g^{\th(k-k')}\Big[2c_0|\l|\,||\un-\un'||_0+
|\n_k-\n'_k|\Big]\;.\Eqa(5.8aab)$$
Choosing $c_0$ big enough, \equ(5.1d) follows.\qed\\

We are now left with fixing the sequence $\un$ in such a way that 
the flow equation for $\n$ is satisfied.
Since we want to fix $\un$ in such a way that $\n_{-\io}=0$, we must have:
$$\n_1=-\sum_{k=-\io}^1\g^{k-2}\b^k_\n(\l_k,\n_k;\ldots;\l_1,\n_1)\;.
\Eqa(5.8a)$$
If we manage to fix $\n_1$ as in \equ(5.8a), we also get:
$$\n_h=-\sum_{k\le h}\g^{k-h-1}\b^k_\n(\l_k,\n_k;\ldots;\l_1,\n_1)\;.
\Eqa(5.8b)$$
We look for a fixed point 
of the operator $\bT:\MMM_\th\to\MMM_\th$ defined as:
$$(\bT\un)_h=-\sum_{k\le h}\g^{k-h-1}\b^k_\n(\l_k(\un),\n_k;\ldots;\l_1,\n_1)
\,.\Eqa(5.8c)$$
where $\l_k(\un)$ is the solution of the first line of \equ(5.1), 
obtained as a function of the {\it parameter} $\un$, as described above.

If we find a fixed point $\un^*$ of \equ(5.8c), 
the first two lines in \equ(5.1) 
will be simultaneously solved by $\ul(\un^*)$ and $\un^*$ respectively, 
and the solution will have the desired smallness properties for $\l_h$ and 
$\n_h$.

First note that, if $|\l|$ is sufficiently small, then $\bT$ leaves 
$\MMM_\th$ invariant: in fact, as a consequence of parity
cancellations, the $\n$--component
of the Beta function satisfies:
$$
\b^h_\n(\l_h,\n_h;\ldots;\l_1,\n_1)=
\b^h_{\n,1}(\l_h;\ldots;\l_1)+\sum_{k}\n_k
\tilde\b^{h,k}_\n(\l_h,\n_h;\ldots;\l_1,\n_1)\Eqa(gg)$$
where, if $c_1, c_2$ are suitable constants
$$|\b^h_{\n,1}|\le c_1 |\l|\g^{\th h}\;\quad
|\tilde\b^{h,k}_\n|\le c_2|\l|\g^{\th(h-k)}\;.\Eqa(kmn)$$
by using \equ(gg)
and choosing $c=2 c_1$ we obtain
$$|(\bT \n)_h|\le \sum_{k\le h} 2 c_1|\l|\g^{(\th/2) k}\g^{k-h}\le 
c|\l|\g^{(\th/2)h}\;,\Eqa(5.8d)$$
Furthermore, using \equ(gg) and \equ(5.1d), we find that 
$\bT$ is a contraction on $\MMM_\th$:
$$\eqalign{&|(\bT \n)_h-(\bT\un')_h|\le \sum_{k\le h}
\g^{k-h-1}|\b^k_\n(\l_k(\un),\n_k;
\ldots;\l_1,\n_1)-\b^k_\n(\l_k(\un'),\n_k';
\ldots;\l_1,\n_1')|\le\cr
&\le c\sum_{k\le h}\g^{k-h-1}
\left[\g^{\th k}\sum_{k'=k}^1 |\l_{k'}(\un)-
\l_{k'}(\un')|+\sum_{k'=k}^1\g^{\th(k-k')}|\l||\n_{k'}-\n_{k'}'|
\right]\le \cr
&\le c'\sum_{k\le h}\g^{k-h-1}\Big[|k|\g^{\th k}|\l|\,||\un-\un'||_0
+\sum_{k'=k}^1\g^{\th(k-k')}
|\l|\g^{(\th/2)k'}||\un-\un'||_\th
\le\cr
&\le c''|\l|\g^{(\th/2)h}
||\un-\un'||_\th\;.\cr}\Eqa(5.8e)$$
hence $||(\bT \n)-(\bT\un')||_\th\le c''|\l|||\un-\un'||_\th$.
Then, a unique fixed point $\un^*$ for $\bT$ exists on $\MMM_\th$. 
Proof of 
Theorem 4.1 is concluded by noticing that $\bT$ 
is analytic (in fact 
$\b^h_\n$ and $\ul$ are analytic in $\un$ in the domain $\MMM_\th$).\qed\\

\asub(a5.2){\it Proof of Lemma 4.1} 
From now on we shall think $\l_h$ and $\n_h$ fixed, with $\n_1$ conveniently
chosen as above ($\n_1=\n_1^*(\l)$). Then we have $|\l_h|\le c|\l|$ 
and $|\n_h|\le c|\l|\g^{(\th/2)h}$, for some $c,\th>0$. Having fixed 
$\n_1$ as a convenient function 
of $\l$, we can also think $\l_h$ and $\n_h$ as functions of $\l$.\\

\0{\it The flow of $Z_h$.}
The flow of $Z_h$ is given by the first of \equ(5.4z) with $z_h$
independent of $\s_k,\m_k$, $k\ge h$. By Theorem 3.1
we have that $|z_h|\le c|\l|^2$, uniformly in $h$. Again, as for $\l_h$ and 
$\n_h$, we can formally study this equation up to $h=-\io$.
We define $\g^{-\h_z}
\defin\lim_{h\to-\io} 1+z_h$, so that
$$\eqalign{&
\log_\g Z_h=\sum_{k\ge h+1}\log_\g (1+z_k)=\h_z(h-1)+\sum_{k\ge h+1}
r_\z^k\virg
r^k_\z\defin\log_\g\big(1+{z_k-z_{-\io}\over 1+z_{-\io}}
\big)\;.\cr}\Eqa(5.9)$$
Using the fact that $z_{k-1}-z_k$ is necessarily proportional
to $\l_{k-1}-\l_k$ or to $\n_{k-1}-\n_k$ and that $\l_{k-1}-\l_k$
is bounded as in \equ(5.7), we easily find:
$|r^k_\z|\le c\sum_{k'\le k}|z_{k'-1}-z_{k'}|\le c'|\l|^2\g^{(\th/2)k}$. 
So, if $F^h_\z\defin \sum_{k\ge h+1}
r_\z^k$ and 
$F^1_\z=0$, then $F^h_\z=O(\l)$ and $Z_h=\g^{\h_z(h-1)+F^h_\z}$.
Clearly, by definition, $\h_z$ and $F^h_\z$ only depend on $\l_k$, $\n_k$, 
$k\le 1$, so they are independent of $t$ and $u$.\\

\0{\it The flow of $\m_h$.}
The flow of $\m_h$ is given by the last of \equ(5.4z). 
One can easily show inductively 
that $\m_k(\kk)/\m_h$, $k\ge h$, is independent of $\m_1$, so that
one can think that $\m_{h-1}/\m_h$ is just a function of $\l_h$, $\n_h$.
Then, again we can study the flow equation for $\m_h$ up to $h\to -\io$.
We define $\g^{-\h_\m}\defin \lim_{h\to-\io} 1+(m_h/\m_h-z_h)/(1+z_h)$,
so that, proceeding as for $Z_h$, we see that 
$$\mu_h=\m_1\g^{\h_\m(h-1)+F^h_\m}\;,\Eqa(mu)$$ 
for a suitable $F^h_\m=O(\l)$.
Of course $\h_\m$ and $F^h_\m$ are independent of $t$ and $u$.\\

\0{\it The flow of $\s_h$.}
The flow of $\s_h$ can be studied as the one of $\m_h$. If 
we define $\g^{-\h_\s}\defin\lim_{h\to-\io}1+(s_h/\s_h-z_h)/(1+z_h)$, we find
that 
$$\s_h=\s_1\g^{\h_\s(h-1)+F^h_\s}\;,\Eqa(sigma)$$ 
for a suitable $F^h_\s=O(\l)$.
Again, $\h_\s$ and $F^h_\s$ are independent of $t,u$.\\

We are left with proving that $\h_\s-\h_\m\not=0$. It is sufficient to note 
that, by direct computation of the lowest order terms, 
for some $\th>0$, \equ(5.4z) can
be written as:
$$\eqalign{&
z_h=b_1 \l_h^2+O(|\l|^2\g^{\th h})+O(|\l|^3)\virg b_1>0\cr
&{s_h/ \s_h}=-b_2\l_h+O(|\l|\g^{\th h})+O(|\l|^2)\virg b_2>0\cr
&{m_h/ \m_h}=b_2\l_h+O(|\l|\g^{\th h})+O(|\l|^2)\virg b_2>0\;,\cr}
\Eqa(5.5z)$$
where $b_1,b_2$ are constants independent of $\l$ and $h$.
Using \equ(5.5z) and the definitions of $\h_\m$ and $\h_\s$ we
find: $\h_\s-\h_\m=(2b_2/\log\g)\l+O(\l^2)$.\qed\\

\appendix(pi, Proof of Lemma 5.3)

Proceeding as in 
\sec(5) and Appendix \secc(a5), we first solve the equations for
$Z_h$ and $\widehat m^{(2)}_h$ parametrically in 
$\underline\p=\{\p_h\}_{h\le h^*_1}$.
If $|\p_h|\le c|\l|\g^{(\th/2)
(h-h^*_1)}$, the first two assumptions of \equ(lem5.2) easily follow.
Now we will construct a sequence $\underline\p$ such that 
$|\p_h|\le c|\l|\g^{(\th/2)
(h-h^*_1)}$ and satisfying the 
flow equation $\p_{h-1}=\g^h\p_h+\b^h_\p(\p_h,\ldots,
\p_{h^*_1})$.\\

\asub(pi1) {\it Tree expansion for {$\b^h_\p$}.}
$\b^h_\p$ can be expressed
as sum over tree diagrams, similar to those used in Appendix
\secc(4a). The main difference
is that they
have vertices on scales $k$ between $h$ and $+2$. The vertices on scales 
$h_v\ge h^*_1+1$ are associated to the truncated expectations \equ(3.28a); 
the vertices on scale $h_v=h^*_1$ 
are associated to truncated expectations w.r.t. the propagators
$g^{(1,h^*_1)}_{\o_1,\o_2}$; the vertices on scale $h_v<h^*_1$ 
are associated to truncated expectations w.r.t. the propagators
$g^{(2,h_v+1)}_{\o_1,\o_2}$. Moreover the end--points
on scale $\ge h^*_1+1$ are associated to the couplings $\l_h$ or $\n_h$,
as in Appendix \secc(4a); 
the end--points on scales $h\le h^*_1$ are necessarily associated to the
couplings $\p_h$.\\

\asub(pi2) {\it Bounds on {$\b^h_\p$}.}
The non vanishing trees contributing to $\b^h_\p$ must have at least 
one vertex on scale $\ge h^*_1$: in fact the diagrams depending only on 
the vertices of type $\p$ are vanishing (they are chains, 
so they are vanishing, because of the compact support property
of the propagator). This means that, by the short 
memory property, see the Remark at the end of Appendix \secc(4a):
$|\b^h_\p|\le c|\l|\g^{\th(h-h^*_1)}$.\\

\asub(pi3) {\it Fixing the counterterm.}
We now proceed as in Appendix \secc(a5) 
but the analysis here is easier, because no $\l$
end--points can appear and the bound $|\b^h_\p|\le c|\l|\g^{\th(h-h^*_1)}$
holds.
As in Appendix \secc(a5), we can formally consider the flow equation 
up to $h=-\io$, even if $h^*_2$ is a finite integer. This
is because the beta function is independent of 
$\widehat m^{(2)}_k$, $k\le h^*_1$
and admits bounds uniform in $h$. If we want to fix 
the counterterm $\p_{h^*_1}$
in such a way that $\p_{-\io}=0$, we must have, for any $h\le h^*_1$:
$$\p_h=-\sum_{k\le h}\g^{k-h-1}\b^k_\p(\p_k,\ldots,\p_{h^*_1})\;.\Eqa(must)$$
Let $\tilde{\MMM}$ be the space of sequences $\underline\p=
\{\p_{-\io},\ldots,\p_{h^*_1}\}$ 
such that $|\p_h|\le c |\l| \g^{-(\th/2) (h-h^*_1)}$.
We look for a fixed point 
of the operator $\tilde{\bT}:\tilde{\MMM}
\to\tilde{\MMM}$ defined as:
$$(\tilde\bT\underline\p)_h=-\sum_{k\le h}\g^{k-h-1}
\b^k_\p(\p_k;\ldots;\p_{h^*_1})\,.\Eqa(5.8c0)$$

Using that $\b^k_\p$ is independent from $\hat m_k^{(2)}$
and
the bound on the beta function, choosing $\l$ small enough and proceeding
as in the proof of Theorem 4.1, we find that $\tilde\bT$ is a contraction
on $\tilde{\MMM}$, so that we find a unique fixed point, and the first
of \equ(4.58b) follows.\\

\asub(pi4) {\it The flows of $Z_h$ and $\widehat m^{(2)}_h$.}
Once that $\p_{h^*_1}$ is fixed via the iterative procedure of \sec(pi3),
we can study in more detail the flows of $Z_h$ and $\widehat m^{(2)}_h$ 
given by \equ(flow). Note that $z_h$ and $s_h$ can be again expressed as 
a sum over tree diagrams and, as discussed for $\b^h_\p$, see \sec(pi2),
any non vanishing diagram must have at least one vertex on scale $\ge h^*_1$. 
Then, by the short memory property, see \sec(short), we have 
$z_h=O(\l^2\g^{\th(h-h^*_1)})$ and $s_h=O(\l\widehat m^{(2)}_h\g^{\th
(h-h^*_1)})$ and, repeating the proof of Lemma 4.1,
we find the second and third of \equ(4.58b).\\

\asub(pi6){\it The Lipshitz property {\equ(ph)}.}
Clearly,  
$\p_{h^*_1}^*(\l,\s_1,\m_1)-\p^*_{h^*_1}(\l,\s_1',\m_1')$ can be expressed
via a tree expansion similar to the one discussed above;
in the trees with non vanishing value, there 
is either a difference of propagators at scale $h\ge h^*_1$
with couplings $\s_h,\m_h$ and  $\s'_h,\m'_h$, giving
in the dimensional bounds
an extra factor
$O(|\s_h-\s_h'| \g^{-h})$ or $O(|\m_h-\m_h'| \g^{-h})$; or a 
difference of propagators at scale $h\le h^*_1$
(computed by definition at $\widehat m^{(2)}_h=0$)
with the ``corrections'' $a^\o_h,c_h$ associated to $\s_1,\m_1$ or 
$\s'_1,\m'_1$, giving
in the dimensional bounds
an extra factor
$O(|\s_1-\s_1'|)$ or $O(|\m_1-\m_1'|)$. Then, 
$$\eqalign{&\Big|\p_{h^*_1}(\l,\s_1,\m_1)-\p_{h^*_1}(\l,\s_1',\m_1')\Big|
\le c|\l|\sum_{k\le h^*_1}\g^{k-h^*_1-1}\cdot\cr
&\qquad\cdot \Big[ \sum_{h\ge h^*_1}
\left({|\s_h-\s_h'|\over \g^h}+{|\m_h-\m_h'|\over \g^h}\right)+
\sum_{k\le h\le h^*_1}\big(|\s_1-\s_1'|+|\m_1-\m_1'|\big)\Big]\;,\cr}
\Eqa(last)$$
from which, using \equ(mu) and \equ(sigma), we easily get \equ(ph).

\*
\appendix(a7, Proof of {\equ(a.5)})

We have, by definition 
$\Pf G=(2^k k!)^{-1}\sum_{\pp}(-1)^\pp G_{p(1)p(2)}\cdots
G_{p(2k-1)p(2k)}$,
where $\pp=(p(1),\ldots$ $\ldots,p(|J|))$ 
is a permutation of the indeces $f\in J$ (we suppose $|J|=2k$)
and $(-1)^\pp$ its sign. 

If we apply $\SS_1=1-\PP_0$ to $\Pf G$ and we call 
$G^0_{f,f'}\defin \PP_0G_{f,f'}$, we find that $\SS_1\Pf G$
is equal to
$$\eqalign{&{1\over 2^k k!}\sum_{\pp}(-1)^\pp \Big[G_{p(1)p(2)}
\cdots
G_{p(2k-1)p(2k)}-G^0_{p(1)p(2)}\cdots
G^0_{p(2k-1)p(2k)}\Big]={1\over 2^k k!}\sum_{\pp}(-1)^\pp\sum_{j=1}^k\cdot\cr
&\cdot\Big(G^0_{p(1)p(2)}\cdots G^0_{
p(2j-3)p(2j-2)}\Big)\SS_1 G_{p(2j-1)p(2j)} \Big(G_{p(2j+1)p(2j+2)}\cdots
G_{p(2k-1)p(2k)}\Big)\;,\cr}\Eqa(a7.2)$$
where in the last sum the meaningless factors must be put equal to 1.
We rewrite the two sums over $\pp$ and $j$ in the following way:
$$\sum_\pp\sum_{j=1}^k=\sum_{j=1}^k\sum_{f_1,f_2\in J\atop f_1\not=
f_2}
\sum_{J_1,J_2}^*\sum_{\pp}^{**}\;,\Eqa(a7.3)$$
where: the $*$ on the second sum means that the sets $J_1$ and $J_2$ are s.t.
$(f_1,f_2,J_1,J_2)$ is a partition of $J$; the $**$ on the second
sum means that $p(1),\ldots,p(2j-2)$ belong to $J_1$, $(p(2j-1),p(2j))=
(f_1,f_2)$ and $p(2j+1),\ldots,p(2k)$ belong to $J_2$. 
Using \equ(a7.3) we can rewrite \equ(a7.2) as 
$$\eqalign{&
\SS_1\Pf G={1\over 2^k k!}\sum_{j=1}^k\sum_{f_1,f_2\in J\atop f_1\not=
f_2}(-1)^\p\SS_1 G_{f_1,f_2}\sum_{J_1,J_2}^{*}\cdot\cr
&\qquad\cdot\sum_{\pp_1,\pp_2}(-1)^{\pp_1+\pp_2}
\Big(G^0_{p_1(1)p_1(2)}\cdots G^0_{
p_1(2k_1-1)p(2k_1)}\Big)\Big(G_{p_2(1)p_2(2)}\cdots
G_{p_2(2k_2-1)p(2k_2)}\Big)\;,\cr}\Eqa(a7.4)$$
where: $(-1)^\p$ is the sign of the permutation leading
from the ordering $J$ to the ordering $(f_1,f_2,$ $J_1,J_2)$;
$\pp_i$, $i=1,2$ is a permutation of the labels in $J_i$ (we suppose $|J_i|=
2k_i$) and $(-1)^{\pp_i}$
is its sign. It is clear that \equ(a7.4) is equivalent to \equ(a.5).\\

{\bf Acknowledgments. } AG thanks Prof. J. L. Lebowitz
for his invitation at Rutgers University, where part of this work was done;
and acknowledges the NSF Grant DMR 
01--279--26, which partially supported his work. VM thanks
Prof. T. Spencer  for his nice invitation
at the Institute for Advanced Studies, in Princeton,
where part of this work was done. We both thank Prof. G. Gallavotti
for many important remarks and suggestions.

\baselineskip=12pt
\vskip1cm

\centerline{\titolo References}
\*
\halign{\hbox to 1.2truecm {[#]\hss} &%
\vtop{\advance\hsize by-1.25truecm\0#}\cr
AT& {J. Ashkin, E.Teller: Statistics of Two-Dimensional
Lattices with Four Components.
Phys. Rev. {\bf 64}, 178-184 (1943). }\cr
B& {R.J. Baxter:
Eight-Vertex Model in Lattice Statistics.
{\it Phys. Rev. Lett.} {\bf 26}, 832--833 (1971). }\cr
Ba&{R. Baxter, Exactly solved models in statistical mechanics,
Academic Press (1982)}\cr
BG1& {G. Benfatto, G. Gallavotti:
Perturbation Theory of the Fermi Surface in Quantum Liquid. A General
Quasiparticle Formalism and One-Dimensional Systems.
{\it J. Stat. Phys.} {\bf 59}, 541--664 (1990). }\cr
BG& {G. Benfatto, G. Gallavotti:
Renormalization group. Physics notes 1, Princeton University
Press (1995). }\cr
BGPS& {G. Benfatto, G. Gallavotti, A. Procacci, B. Scoppola:
Beta function and Schwinger functions for a Many Fermions System 
in One Dimension. {\it Comm. Math. Phys.} {\bf 160}, 93--171 (1994).}\cr
BM& {G. Benfatto, V.Mastropietro.
Renormalization group, hidden symmetries
and approximate Ward identities in the $XYZ$ model.
{\it Rev. Math. Phys.} 13 (2001), no. 11, 1323--143;
and {\it Comm. Math. Phys.} 231, 97-134 (2002)}\cr
BM1& {F. Bonetto, V. Mastropietro: Beta function and Anomaly 
of the Fermi Surface for a d=1 System of interacting Fermions in a 
Periodic Potential. {\it Comm. Math. Phys.} {\bf 172}, 57--93 (1995).}\cr
Bad& { M. Badehdah et al, Physica B, 291, 394 (2000)}\cr  
Bez& { C.G.Bezerra, A.M. Mariz. The anisotropic Ashkin-Teller model:
a renormalization Group study. Cond-mat010280, to appear on 
{\it Physica A}}\cr
Bar& { N.C. Bartelt, T.L. Einstein  et al.:
{\it Phys. Rev. B} 40, 10759 (1989)}\cr
Be& { S. Bekhechi et al 
Physica A, 264, 503 (1999)}\cr 
BeM1& {G. Benfatto, V. Mastropietro: Ward identities and Dyson equations
in interacting Fermi systems. To appear on {\it J. Stat. Phys.}}\cr
DR& {E. Domany, E.K. Riedel:
{\it Phys. Rev. Lett.} 40, 562 (1978)}\cr
GM& {G. Gentile, V. Mastropietro:
Renormalization group for one-dimensional fermions.
A review on mathematical results. {\it Phys. Rep.} 352 (2001), no. 4-6,
273--43}\cr
GS& {G. Gentile, B. Scoppola: Renormalization Group and 
the ultraviolet problem in the Luttinger model. {\it Comm. Math. Phys.}
{\bf 154}, 153--179 (1993).}\cr
K&{L.P. Kadanoff,
Connections between the Critical Behavior of the Planar Model and That of the
Eight-Vertex Model. {\it Phys. Rev. Lett.} 39, 903-905 (1977)}\cr
KW&{L.P. Kadanoff and F.J. Wegner, {\it Phys. Rev. B} 4, 3989--3993 (1971)}\cr
Ka&{P.W.Kasteleyn, Dimer Statistics and phase transitions, {\it J. Math.Phys.}
4, 287 (1963)}\cr
F&{C. Fan, On critical properties of the Ashkin-Teller model,
{\it Phis. Lett.}, 6, 136-136 (1972)}\cr
H&{C.Hurst, New approach to the Ising problem, 
{\it J.Math. Phys.} 7,2, 305-310 (1966)}\cr
ID&{ C. Itzykson, J. Drouffe, "Statistical field theory: 1," Cambridge Univ.
Press, 1989.}\cr
Le& {A. Lesniewski:
Effective action for the Yukawa 2 quantum field Theory.
{\it Comm. Math. Phys.} {\bf 108}, 437-467 (1987). }\cr
Li& {H. Lieb, Exact solution of the problem of entropy of two-dimensional
ice,
{\it Phys. Rev. Lett.}, 18, 692-694, (1967)}\cr
LP&{A.Luther, I.Peschel. Calculations
of critical exponents in two dimension from quantum
field theory in one dimension.
{\it Phys. Rev. B} 12, 3908-3917 (1975)}\cr
M1&{V. Mastropietro:
Ising
models with four spin interaction at criticality, {\it Comm. Math. Phys}
{\bf 244}, 595--642 (2004)}\cr
ML&{D. Mattis, E. Lieb:
Exact solution of a many fermion system and its associated boson field.
{\it J. Math. Phys.} {\bf 6}, 304--312 (1965). }\cr
MW&{B. McCoy, T. Wu, The two-dimensional Ising model,
Harvard Univ. Press, 1973.}\cr
MPW&{E.Montroll, R.Potts, J.Ward. Correlation
and spontaneous magnetization of the two dimensional
Ising model. {\it J. Math. Phys.} 4,308 (1963)}\cr
N& {M.P.M. den Nijs: Derivation of extended scaling relations between
critical exponents in two dimensional models from the one dimensional
Luttinger model, {\it Phys. Rev. B}, 23, 11 (1981)
6111-6125}\cr
O& {L.Onsager: Critical statistics. A two dimensional
model with an order-disorder transition.
{\it Phys. Rev.}, 56, 117-149 (1944)}\cr
PB& {A.M.M. Pruisken, A.C. Brown.
Universality fot the critical lines of the eight vertex,
Ashkin-Teller and Gaussian models, {\it Phys. Rev.} B, 23, 3 (1981)
1459-1468}\cr
PS& {H. Pinson, T.Spencer: Universality in 2D critical Ising model.
To appear in Comm. Math. Phys.}\cr
S& { S. Samuel:
The use of anticommuting variable integrals in statistical mechanics'',
{\it J. Math. Phys.} 21 (1980) 2806}\cr
Su& {S.B. Sutherland:
Two-Dimensional Hydrogen Bonded Crystals.
{\it J. Math. Phys.} {\bf 11}, 3183--3186 (1970). }\cr
Spe& {T. Spencer: A mathematical approach to universality in
two dimensions. {\it Physica A} {\bf 279}, 250-259 (2000). }\cr
SML& { T. Schultz, D. Mattis, E. Lieb: Two-dimensional
Ising model as a soluble problem
of many Fermions. {\it Rev. Mod. Phys.} 36 (1964) 856.}\cr
W& {F.J. Wegner: Duality relation between the Ashkin-Teller
and the eight-vertex model, {J. Phys. C}, 5,L131--L132
(1972)}\cr
WL& {F.Y. Wu, K.Y. Lin: Two phase transitions in
the Ashkinh-Teller model. {J. Phys. C}, {\bf 5}, L181--L184
(1974).}\cr
Wu& {F.W.Wu: The Ising model with four spin interaction.
{\it Phys. Rev.} B 4, 2312-2314 (1971).}\cr}

\bye